\RequirePackage{etoolbox}
\documentclass[11pt,a4paper]{article}

\usepackage{amsthm}
\usepackage{amsmath}
\usepackage{natbib}
\usepackage[colorlinks,citecolor=blue,urlcolor=blue,filecolor=blue,backref=page]{hyperref}
\usepackage{graphicx}
\usepackage{enumitem}
\usepackage{verbatim}
\usepackage{graphics}
\usepackage{color}
\usepackage{caption}
\usepackage{comment}
\usepackage{amssymb}
\usepackage{subcaption}
\usepackage{tikz}
\usepackage{algorithm}
\usepackage{multirow}

\title{Improved bridge constructs for stochastic differential equations}
\author{Gavin A.~Whitaker$^1$ \and\ Andrew Golightly$^1$\thanks{email: \texttt{andrew.golightly@ncl.ac.uk}} \and\ Richard J.~Boys$^1$ \and\ Chris Sherlock$^2$}
\date{\small $^{1}$School of Mathematics \& Statistics, Newcastle University,\\
  Newcastle upon Tyne, NE1 7RU, UK \\
$^{2}$Department of Mathematics and Statistics, Lancaster University, Lancaster, LA1 4YF}

\newcommand{\half}{{1/2}}
\newcommand{\EM}{\textrm{\tiny EM}}
\newcommand{\MDB}{\textrm{\tiny MDB}}
\newcommand{\LI}{\textrm{\tiny LB}}
\newcommand{\GP}{\textrm{\tiny GP}}
\newcommand{\GPS}{\textrm{\tiny GP-S}}
\newcommand{\GPN}{\textrm{\tiny GP-N}}
\newcommand{\R}{\textrm{\tiny RB}}
\newcommand{\RM}{\textrm{\tiny RB$^-$}}
\newcommand{\expect}{\textrm{E}}
\newcommand{\tr}{\textrm{tr}}

\begin{document}
\maketitle
\begin{abstract}
We consider the task of generating discrete-time realisations of a nonlinear multivariate 
diffusion process satisfying an It\^o stochastic differential equation conditional 
on an observation taken at a fixed future time-point. Such realisations are typically termed 
diffusion bridges. Since, in general, no closed form expression exists for the transition 
densities of the process of interest, a widely adopted solution works with the Euler-Maruyama 
approximation, by replacing the intractable transition densities with Gaussian approximations. 
However, the density of the conditioned discrete-time process remains intractable, necessitating 
the use of computationally intensive methods such as Markov chain Monte Carlo. Designing an 
efficient proposal mechanism which can be applied to a noisy and partially observed system 
that exhibits nonlinear dynamics is a challenging problem, and is the focus of this paper. 
By partitioning the process into two parts, one that accounts for nonlinear dynamics in 
a deterministic way, and another as a residual stochastic process, we develop a class of 
novel constructs that bridge the residual process via a linear approximation. In addition, we 
adapt a recently proposed construct to a partial and noisy observation regime. We compare 
the performance of each new construct with a number of existing approaches, using three 
applications. 
\end{abstract}


\section{Introduction}\label{intro}

Diffusion processes satisfying stochastic differential equations (SDEs) 
provide a flexible class of models for describing many continuous-time 
physical processes. Some application areas and indicative references include 
finance, e.g. \cite{Kalogeropoulos_2010}, \cite{Stramer_Bognar_Schneider2010}, 
reaction networks, e.g. \cite{Fuchs_2013}, \cite{GolightlyHS_2015} 
and population dynamics, e.g. \cite{Heydari_2014}. Fitting such models to data 
observed at discrete-times can be problematic since the transition densities of the diffusion 
process are likely to be intractable. A review of inferential methods for 
diffusions can be found in \cite{Fuchs_2013}. A widely adopted solution is to approximate 
the unavailable transition densities either analytically 
\citep{Ait-Sahalia_2002,Ait-Sahalia_2008} or numerically 
\citep{pederen_1995b,Elerian_2001,Eraker_2001,Roberts_Stramer_2001}. 
Within the Bayesian paradigm, the numerical approach can 
be seen as a data augmentation problem. The simplest implementation augments 
low-frequency data by introducing intermediate time-points between observation 
times. An Euler-Maruyama scheme is then applied by approximating 
the transition densities over the induced discretisation as Gaussian. 
Computationally intensive algorithms such as Markov chain 
Monte Carlo (MCMC) are then used to integrate over the uncertainty associated with 
the missing data. The key challenges of designing such an MCMC scheme include 
overcoming dependence between the parameters and missing data (first highlighted 
as a problem by \cite{Roberts_Stramer_2001}) and overcoming dependence between 
successive values of the missing data. Dealing with the latter requires repeatedly generating realisations 
known as \emph{diffusion bridges} from an approximation of the conditioned 
process. Methods built upon exact simulation, that avoid use of the 
Euler-Maruyama approximation and the associated discretisation error, 
have been proposed by \cite{Beskos_2006} (see also \cite{Beskos_2009}). 
However, these exact methods are limited to diffusions which can be 
transformed to have unit diffusion coefficient, known as reducible 
diffusions. 

Designing bridge constructs for irreducible, multivariate diffusions 
is a challenging problem and has received much 
attention in recent literature. The simplest approach (see e.g. 
\cite{pederen_1995b}) is based on the forward dynamics of the diffusion 
process and generates a bridge by sampling iteratively from the Euler-Maruyama 
approximation of the unconditioned SDE. This myopic approach induces a 
discontinuity at the observation time (as the discretisation gets finer) 
and is well known to lead to low Metropolis-Hastings acceptance rates. 
The modified diffusion bridge (MDB) construct of \cite{Durham_2001} (see also 
extensions to the partial and noisy observation case in 
\cite{Golightly_Wilkinson_2008}) pushes the bridge process towards the 
observation in a linear way and provides the optimal sampling method 
when the drift and diffusion coefficients of the SDE are constant 
\citep{Stramer_Yan_2006}. However, this construct is less effective 
when the process exhibits nonlinear dynamics. Several approaches 
have been proposed to overcome this problem. For example, 
\cite{Lindstrom_2012} (see also \cite{Fearnhead_2008} for a similar approach) 
combines the Pedersen and MDB approaches, 
with a tuning parameter governing the precise dynamics of the resulting sampler. 
\cite{delmoral14} (see also \cite{Lin_2010}) use a sequential Monte Carlo 
scheme to generate realisations according to the forward dynamics, pushing 
the resulting trajectories towards the observation using a sequence of 
reweighting steps. \cite{Schauer_2014} combine the ideas of \cite{Delyon_2006} 
and \cite{Clark_1990} to obtain a bridge based on the addition of a guiding term 
to the drift of the process under consideration. The guiding term is 
derived using a tractable approximation of the target process. 

\subsection{Contributions and organisation of the paper}\label{cont}

Our contribution is the development of a novel class of 
bridge constructs that are computationally and statistically 
efficient, simple to implement, and can be applied in scenarios 
where only partial and noisy measurements of the 
system are available. Essentially, the process is partitioned 
into two parts, one that accounts for nonlinear dynamics in 
a deterministic way, and another as a residual stochastic process. A 
bridge construct is obtained for the target process by 
applying the MDB sampler of 
\cite{Durham_2001} to the end-point conditioned residual 
process. We consider two implementations of this 
approach. Firstly, we use the bridge introduced by 
\cite{Whitaker_2015} that constructs the residual process 
by subtracting the solution of an ordinary differential 
equation (ODE) system based on the drift, from the target 
process. Secondly, we recognise that the intractable 
SDE governing the residual process can be approximated 
by a tractable process. We therefore extend the first 
approach by additionally subtracting the expectation 
of the approximate residual process and bridging the 
remainder with the MDB sampler. In addition, we adapt 
the guided proposal proposed by \cite{Schauer_2014} 
to a partial and noisy observation regime. 

We evaluate the performance of each bridge construct 
(as well as the constructs proposed by 
\cite{Durham_2001} and \cite{Lindstrom_2012}) using 
three examples: a simple birth-death model, a 
Lotka-Volterra system and a model of aphid growth.

The remainder of this article is organised as follows. 
Section~\ref{sde} provides a brief introduction to the 
problem of sampling conditioned SDEs and examines 
two previously proposed approaches. In Section~\ref{bridge} 
we describe a novel class of bridge constructs and adapt 
an existing approach to a more general observation regime. 
Applications are considered in Section~\ref{app} and a 
discussion is provided in Section~\ref{disc}.

\section{Sampling conditioned SDEs}\label{sde}

Consider a continuous-time $d$-dimensional It\^o process $\{X_t, t\geq 0\}$
governed by the SDE paramaterised by $\theta=(\theta_{1},\ldots,\theta_{p})'$ of the form
\begin{equation}
dX_t=\alpha(X_t,\theta)\,dt+\sqrt{\beta(X_t,\theta)}\,dW_t,
\quad X_0=x_0. \label{eqn:sdmem}
\end{equation}
Here, $\alpha$ is a $d$-vector of drift functions, the 
diffusion matrix $\beta$ is a $d \times d$ positive 
definite matrix with a square root representation 
$\sqrt{\beta}$ such that $\sqrt{\beta}\sqrt{\beta}'=\beta$ 
and $W_t$ is a $d$-vector of (uncorrelated) standard 
Brownian motion processes. We assume that $\alpha$ and 
$\beta$ are sufficiently regular so that the SDE has 
a weak non-explosive solution \citep{oksendal_2003}.

For tractability, we make the same assumption as \cite{Golightly_Wilkinson_2008}, 
\cite{Golightly_Wilkinson_2011}, \cite{Picchini_2014} and \cite{Lu_2015} 
among others, that the process is observed at $t=T$ according 
to
\begin{equation}\label{eqn:obs}
Y_T = F' X_T + \epsilon_T, \qquad \epsilon_T|\Sigma\sim N(0,\Sigma).
\end{equation}
Here, $Y_T$ is a $d_o$-vector, $F$ is a constant $d\times d_o$
matrix and $\epsilon_T$ is a random $d_o$-vector for some $d_o\le d$. This flexible setup
allows for only observing a subset of components. For 
simplicity we also assume that the process is known exactly at 
$t=0$. This is the case when a diffusion process is observed completely 
and without error. In the case of partial and/or noisy observations, 
typically the initial position is an unknown parameter in an MCMC scheme 
and a new bridge is created at each iteration conditional on the current 
parameter values, so in terms of the bridge, the initial position 
is effectively known. The complication of multiple partial 
and/or noisy observations is discussed in Section~\ref{disc}.

Our aim 
is to generate discrete-time realisations of $X_t$ conditional on 
$x_0$ and $y_T$. To this end, we partition $[0,T]$ as
\[
0=\tau_{0}<\tau_{1}<\tau_{2}<\cdots<\tau_{m-1}<\tau_{m}=T,
\]
giving $m$ intervals of equal length $\Delta\tau=T/m$. Since, in 
general, the form of the SDE in (\ref{eqn:sdmem}) will not permit 
an analytic solution, we work with the Euler-Maruyama
approximation which gives the change in the process over a small 
interval of length $\Delta\tau$ as a Gaussian random vector. Specifically, 
we have that
\[
X_{\tau_{k+1}}-X_{\tau_k}=\alpha(X_{\tau_k},\theta)\,\Delta \tau+\sqrt{\beta(X_{\tau_k},\theta)}\,\Delta W_{\tau_k}
\]
where $\Delta W_{\tau_k}\sim N(0,\Delta\tau I_d)$ and $I_d$ is the $d\times d$ identity 
matrix. The continuous-time conditioned 
process is then approximated by the discrete-time skeleton bridge, with the latent 
values \hbox{$x_{(0,T]}=(x_{\tau_1},\ldots, x_{\tau_{m}}=x_T)'$} having the (posterior) density 
\begin{equation}\label{eqn:target}
\pi(x_{(0,T]}|x_{0},y_{T},\theta,\Sigma)\propto \pi(y_T|x_T,\Sigma)
\prod\limits_{k=0}^{m-1}\pi(x_{\tau_{k+1}}|x_{\tau_{k}},\theta)
\end{equation}
where 
\[\pi(x_{\tau_{k+1}}|x_{\tau_{k}},\theta)=N\left(x_{\tau_{k+1}}\,;\,x_{\tau_{k}}+\alpha(x_{\tau_{k}},\theta)\Delta\tau,~\beta(x_{\tau_{k}},\theta)\Delta\tau\right)\] 
is the transition density under the Euler-Maruyama approximation, 
$\pi(y_T|x_T,\Sigma)= N(y_T\,;\,F'x_T,\Sigma)$ and $N(\cdot;m,V)$ 
denotes the multivariate Gaussian density with mean 
vector $m$ and variance matrix $V$. In the special case where $x_T$ is known 
(so that $y_{T}=x_{T}$ and $F=I_d$), the latent values 
$x_{(0,T)}=(x_{\tau_1},\ldots, x_{\tau_{m-1}})'$ have the density
\begin{equation}\label{eqn:target2}
\pi(x_{(0,T)}|x_{0},x_{T},\theta)\propto \prod\limits_{k=0}^{m-1}\pi(x_{\tau_{k+1}}|x_{\tau_{k}},\theta).
\end{equation}
For nonlinear forms of the drift 
and diffusion coefficients, the products in (\ref{eqn:target}) and (\ref{eqn:target2}) will be intractable 
and samples can be generated via computationally intensive algorithms 
such as Markov chain Monte Carlo or importance sampling. We focus on the former 
but note that in either case, the efficiency of the algorithm will depend on 
the proposal mechanism used to generate the bridge. A common approach to 
constructing an efficient proposal is to factorise the target in (\ref{eqn:target}) 
as
\begin{equation}\label{eqn:fact}
\pi(x_{(0,T]}|x_{0},y_{T},\theta,\Sigma)\propto \prod\limits_{k=0}^{m-1}\pi(x_{\tau_{k+1}}|x_{\tau_{k}},y_T,\theta,\Sigma).
\end{equation}
The density in (\ref{eqn:target2}) can be factorised in a similar manner. 
This suggests seeking proposal densities of the form $q(x_{\tau_{k+1}}|x_{\tau_{k}},y_T,\theta,\Sigma)$ 
which aim to approximate the intractable constituent densities in (\ref{eqn:fact}). 
In what follows, we consider some 
existing approaches for generating bridges via approximation 
of $\pi(x_{\tau_{k+1}}|x_{\tau_{k}},y_T,\theta,\Sigma)$ 
before outlining our contribution. For each bridge, the proposal densities 
take the form 
\begin{equation}\label{eqn:qprop}
q(x_{\tau_{k+1}}|x_{\tau_k},y_T,\theta,\Sigma)=
N\left(x_{\tau_{k+1}}\,;\, x_{\tau_k}+\mu(x_{\tau_k})\Delta\tau\,,\,\Psi(x_{\tau_k})\Delta\tau\right)
\end{equation}
and our focus is on the choice of $\mu(\cdot)$ and $\Psi(\cdot)$. For simplicity and where possible, 
we drop the parameters $\theta$ and $\Sigma$ from the notation as they 
remain fixed throughout. 

\subsection{Myopic simulation}

Ignoring the information in the observation $y_T$ and 
simply applying the Euler-Maruyama approximation over 
each interval of length $\Delta\tau$ leads to a proposal 
density of the form given by (\ref{eqn:qprop}) with 
$\mu_{\EM}(x_{\tau_k})=\alpha(x_{\tau_k})$ and 
$\Psi_{\EM}(x_{\tau_k})=\beta(x_{\tau_k})$. 
Sampling iteratively according to $(\ref{eqn:qprop})$ for $k=0,1,\ldots,m-1$ 
gives a proposed bridge which we denote by $x_{(0,T]}^*$. 
The Met\-ropolis-Hastings (MH) acceptance probability for a move from $x_{(0,T]}$ 
to $x_{(0,T]}^*$ is 
\[
\min\left\{1\,,\,\frac{\pi(y_T|x_T^*)}{\pi(y_T|x_T)}\right\}.
\]
This strategy is likely to work well provided that the 
observation $y_T$ is not particularly informative, that is, when the 
measurement error dominates the intrinsic stochasticity of the process. However, 
as $\Sigma$ is reduced, the MH acceptance rate decreases. A 
related approach can be found in \cite{pederen_1995b}, 
where it is assumed that $x_T$ is known. In this case, 
a move from $x_{(0,T)}$ to $x_{(0,T)}^*$ is accepted 
with probability
\[
\min\left\{1\,,\,\frac{\pi(x_T|x_{\tau_{m-1}}^*)}{\pi(x_T|x_{\tau_{m-1}})}\right\}
\]
which tends to 0 as $m\to\infty$ (or equivalently, $\Delta\tau\to 0$).

\subsection{Modified diffusion bridge}\label{mdbsect}

For known $x_T$, \cite{Durham_2001} derive a linear Gaussian approximation 
of $\pi(x_{\tau_{k+1}}|x_{\tau_{k}},x_T)$, leading to a sampler known as the modified 
diffusion bridge (MDB). Extensions to the partial and noisy 
observation regime are considered in \cite{Golightly_Wilkinson_2008}. 
In brief, the joint distribution of $X_{\tau_{k+1}}$ and 
$Y_T$ (conditional on $x_{\tau_{k}}$) is approximated by 
%\begin{align*}
%\begin{pmatrix}
%X_{\tau_{k+1}} \\
%Y_{T} \end{pmatrix}\bigg| x_{\tau_{k}}&\sim N\left\{\begin{pmatrix}
%x_{\tau_{k}}+\alpha_{k}\Delta\tau  \\[0.2em]
%F' (x_{\tau_{k}}+\alpha_{k}\Delta_k) \end{pmatrix},\begin{pmatrix}
%\beta_{k} \Delta\tau & \beta_{k} F\Delta\tau \\
%F'\beta_{k}\Delta\tau & F'\beta_{k} F\Delta_k + \Sigma \end{pmatrix}\right\}
%\end{align*} 
\[
\begin{pmatrix}
X_{\tau_{k+1}} \\
Y_{T} \end{pmatrix}\bigg| x_{\tau_{k}}\sim  N\left\{\begin{pmatrix}
x_{\tau_{k}}+\alpha_{k}\Delta\tau  \\[0.2em]
F' (x_{\tau_{k}}+\alpha_{k}\Delta_k) \end{pmatrix},\begin{pmatrix}
\beta_{k} \Delta\tau & \beta_{k} F\Delta\tau \\
F'\beta_{k}\Delta\tau & F'\beta_{k} F\Delta_k + \Sigma \end{pmatrix}\right\}
\]
where $\alpha_{k}=\alpha(x_{\tau_k})$, $\beta_{k}=\beta(x_{\tau_k})$ 
and $\Delta_{k}=T-\tau_{k}$. Conditioning on $Y_T=y_T$ 
gives
%\begin{equation}\label{eqn:mdb1}
%q_{\MDB}(x_{\tau_{k+1}}| x_{\tau_{k}},y_{T})= 
%N\left(x_{\tau_{k+1}}\,;\, x_{\tau_k}+\mu_{\MDB}(x_{\tau_k})\Delta\tau\,,\,\Psi_{\MDB}(x_{\tau_k})\Delta\tau\right)
%\end{equation}
%where
%\begin{equation}\label{eqn:mdb2}
%\mu_{\MDB}(x_{\tau_k}) =\alpha_{k}+\beta_{k} F\left(F'\beta_{k}F\Delta_k + \Sigma\right)^{-1}
%\left\{y_{T}-F'(x_{\tau_{k}}+\alpha_{k}\Delta_k)\right\}   
%\end{equation}
\begin{equation}\label{eqn:mdb2}
\mu_{\MDB}(x_{\tau_k}) =\alpha_{k}+\beta_{k} F\left(F'\beta_{k}F\Delta_k + \Sigma\right)^{-1} 
\left\{y_{T}-F'(x_{\tau_{k}}+\alpha_{k}\Delta_k)\right\}   
\end{equation}
and
\begin{equation}\label{eqn:mdb3}
\Psi_{\MDB}(x_{\tau_k})=\beta_{k}-\beta_{k} F\left(F'\beta_{k} F\Delta_k + \Sigma\right)^{-1}F'\beta_{k}\Delta\tau. 
\end{equation}

In the case of no measurement error and observation of all components 
(so that $x_T$ is known), (\ref{eqn:mdb2}) and (\ref{eqn:mdb3}) become
\[
\mu_{\MDB}^*(x_{\tau_k}) = \frac{x_T-x_{\tau_k}}{T-\tau_k} \quad \textrm{and} \quad 
\Psi_{\MDB}^*(x_{\tau_k})=\frac{T-\tau_{k+1}}{T-\tau_k}\beta(x_{\tau_k}).
\]

\subsubsection{Connection with continuous-time conditioned processes} \label{cts}

Consider the case of no measurement error and full observation of 
all components. The SDE satisfied by the conditioned process 
$\{X_t,t\in[0,T]\}$, takes the form 
\begin{equation}\label{eqn:guip1}
dX_t=\tilde{\alpha}(X_t)\,dt+\sqrt{\beta(X_t)}\,dW_t,
\quad X_0=x_0
\end{equation}
where the drift is
\begin{equation}\label{eqn:guip2}
\tilde{\alpha}(X_t)=\alpha(X_t)+\beta(X_t)\nabla_{x_t}\log p(x_T|x_t).
\end{equation}
See for example chap. IV.39 of \cite{Rogers2000} for a derivation. 
Note that $p(x_T|x_t)$ denotes the (intractable) transition density of the 
unconditioned process defined in (\ref{eqn:sdmem}). Approximating 
$\alpha(X_t)$ and $\beta(X_t)$ in (\ref{eqn:sdmem}) by the constants $\alpha(x_T)$ 
and $\beta(x_T)$ yields a process for which $p(x_T|x_t)$ is tractable. 
The corresponding conditioned process satisfies 
\begin{equation}\label{eqn:SDEbridge}
dX_t=\frac{X_{T}-X_t}{T-t}\,dt+\sqrt{\beta(X_t)}\,dW_t.
\end{equation}
Use of (\ref{eqn:SDEbridge}) as a proposal process has been justified by \cite{Delyon_2006} 
(see also \cite{Stramer_Yan_2006}, \cite{Marchand2011} and \cite{Papaspiliopoulos_2013}), who 
show that the distribution of the target process (conditional on $x_T$) 
is absolutely continuous with respect to the distribution of the 
solution to (\ref{eqn:SDEbridge}). As discussed by \cite{Papaspiliopoulos_2013}, 
it is impossible to simulate exact (discrete-time) realisations of 
(\ref{eqn:SDEbridge}) unless $\beta(\cdot)$ is constant. They also note that 
performing a local linearisation of (\ref{eqn:SDEbridge}) according to 
\cite{Shoji1998} (see also \cite{Shoji2011}) gives a tractable process 
with transition density
\[
q(x_{\tau_{k+1}}| x_{\tau_{k}},x_{T})= 
N\left(x_{\tau_{k+1}}\,;\, x_{\tau_k}+\frac{x_T-x_{\tau_k}}{T-\tau_k}\Delta\tau \,,\,\frac{T-\tau_{k+1}}{T-\tau_k}\beta(x_{\tau_k})\Delta\tau  \right), 
\]
that is, the transition density of the modified diffusion bridge discussed in the previous section. Plainly, 
taking the Euler-Maruyama approximation of (\ref{eqn:SDEbridge}) yields the MDB construct, albeit 
without the time dependent multiplier of $\beta(x_{\tau_k})$ in the variance. As observed by 
\cite{Durham_2001} and discussed in \cite{Papaspiliopoulos_2012} and 
\cite{Papaspiliopoulos_2013}, the inclusion of the time dependent multiplier can lead to improved 
empirical performance. 

Unfortunately, the MDB is only efficient when the drift of
(\ref{eqn:sdmem}) is approximately constant. When this is not the
case, so that realisations of the SDE started from the same point
exhibit strong and similar non-linearity over the inter-observation
time, the modified diffusion bridge is likely to be unsatisfactory.

\subsection{Lindstr\"{o}m bridge}\label{lind}

A bridge construct that combines the myopic sampler with the 
MDB is proposed in \cite{Lindstrom_2012}, for the special case 
of known $x_T$. Extending the sampler to the observation scenario 
in (\ref{eqn:obs}) is straightforward. Whereas the MDB approximates 
the variance of $Y_T|x_{\tau_k}$ by $F'\beta_k F\Delta_k+\Sigma$, 
the simplest version of the Lindstr\"{o}m bridge (LB) has that
 \[
\textrm{Var}(Y_T|x_{\tau_k})\simeq
	F'\{\beta_k \Delta_k +C(\Delta_{k+1})^2\}F + \Sigma,
\]
where $C(\Delta_{k+1})^2$ is the squared bias of $X_T|x_{\tau_{k+1}}$ 
using a single Euler-Maruyama time-step 
and $C$ is an unknown matrix. By assuming that the squared bias is 
a fraction $\gamma$ of the variance over an interval of length 
$\Delta\tau$, a heuristic choice of 
$C$ is given by
\[
C_{\textrm{Heur}} = \dfrac{\gamma\beta_k}{\Delta\tau},
\]
with $\gamma>0$. This particular choice of $C_{\textrm{Heur}}$ 
ensures that $\textrm{Var}(Y_T|x_{\tau_k})$ is a positive definite 
matrix. The joint distribution of $X_{\tau_{k+1}}$ and 
$Y_T$ (conditional on $x_{\tau_{k}}$) is then approximated by 
%\begin{align*}
%\begin{pmatrix}
%X_{\tau_{k+1}} \\
%Y_{T} \end{pmatrix}\bigg| x_{\tau_{k}}&\sim N\left\{\begin{pmatrix}
%x_{\tau_{k}}+\alpha_{k}\Delta\tau  \\[0.2em]
%F' (x_{\tau_{k}}+\alpha_{k}\Delta_k) \end{pmatrix},\begin{pmatrix}
%\beta_{k} \Delta\tau & \beta_{k} F\Delta\tau \\
%F'\beta_{k}\Delta\tau & F'\beta_{k} F\Delta_k^\gamma + \Sigma \end{pmatrix}\right\}
%\end{align*} 
\[
\begin{pmatrix}
X_{\tau_{k+1}} \\
Y_{T} \end{pmatrix}\bigg| x_{\tau_{k}}\sim \\
 N\left\{\begin{pmatrix}
x_{\tau_{k}}+\alpha_{k}\Delta\tau  \\[0.2em]
F' (x_{\tau_{k}}+\alpha_{k}\Delta_k) \end{pmatrix},\begin{pmatrix}
\beta_{k} \Delta\tau & \beta_{k} F\Delta\tau \\
F'\beta_{k}\Delta\tau & F'\beta_{k} F\Delta_k^\gamma + \Sigma \end{pmatrix}\right\}
\]
where $\Delta_k^\gamma=\Delta_k +\gamma(\Delta_{k+1})^2/\Delta\tau$. Conditioning on $Y_T=y_T$ 
gives
%\begin{equation}\label{eqn:lind1}
%q_{\LI}(x_{\tau_{k+1}}| x_{\tau_{k}},y_{T})= 
%N\left(x_{\tau_{k+1}}\,;\, x_{\tau_k}+\mu_{\LI}(x_{\tau_k})\Delta\tau\,,\,\Psi_{\LI}(x_{\tau_k})\Delta\tau\right)
%\end{equation}
%where
\begin{equation}\label{eqn:lind2}
\mu_{\LI}(x_{\tau_k}) =\alpha_{k}+\beta_{k} F\left(F'\beta_{k}F\Delta_k^\gamma + \Sigma\right)^{-1}
\left\{y_{T}-F'(x_{\tau_{k}}+\alpha_{k}\Delta_k)\right\}  
\end{equation}
and
\begin{equation}\label{eqn:lind3}
\Psi_{\LI}(x_{\tau_k})=\beta_{k}-\beta_{k} F\left(F'\beta_{k} F\Delta_k^\gamma + \Sigma\right)^{-1}F'\beta_{k}\Delta\tau. 
\end{equation}
In the case of no measurement error and observation of all components, 
(\ref{eqn:lind2}) and (\ref{eqn:lind3}) become
\begin{align*}
\mu_{\LI}^*(x_{\tau_k}) &= w_k^\gamma \mu_{\MDB}^*(x_{\tau_k})+(1-w_k^\gamma)\alpha(x_{\tau_k}) 
\intertext{and} 
\Psi_{\LI}^*(x_{\tau_k})&= w_k^\gamma \Psi_{\MDB}^*(x_{\tau_k})+(1-w_k^\gamma)\beta(x_{\tau_k})
\end{align*}
where
\[
w_k^\gamma=\frac{(\tau_{k+1}-\tau_k)(T-\tau_k)}{(\tau_{k+1}-\tau_k)(T-\tau_k)+\gamma(T-\tau_{k+1})^2}.
\]
The Lindstr\"{o}m bridge can therefore be seen as a convex combination of the MDB 
and myopic samplers, with $\gamma=0$ giving the MDB and $\gamma=\infty$ giving 
the myopic approach. In practice, \cite{Lindstrom_2012} suggests that 
$\gamma\in[0.01,1]$, given that these values have proved successful 
in simulation experiments. Note also that for a fixed $\gamma$, 
if $T-\tau_{k+1}\gg \Delta\tau$ then $w_k^\gamma\simeq 0$ and the myopic 
sampler dominates. However, as $\tau_{k+1}$ approaches $T$, $w_k^\gamma$ 
approaches 1 and the LB is dominated by the MDB.
 
Whilst the LB attempts to account for nonlinear dynamics 
by combining the MDB with the myopic approach, having to specify a 
model-dependent tuning parameter is unsatisfactory, since different choices 
of $\gamma$ will lead to different properties of the proposed 
bridges. Moreover, the link between the regularised sampler 
and the continuous-time conditioned process is unclear.

\section{Improved bridge constructs}\label{bridge}

In this section we describe a novel class of bridge constructs 
that require no tuning parameters, are simple to implement (even when
only a subset of components are observed with Gaussian noise) and can
account for nonlinear dynamics driven by the drift. In addition, we 
discuss the recently proposed bridging strategy of 
\cite{Schauer_2014} and describe an implementation method in the 
case of partial observation with additive Gaussian measurement error.  

\subsection{Bridges based on residual processes}

Suppose that $X_t$ is partitioned as $X_t=\zeta_t+R_t$ 
where $\{\zeta_t,t\geq 0\}$ is a deterministic 
process and $\{R_t,t\geq 0\}$ is a residual stochastic process, 
satisfying
\begin{align}
d\zeta_t &= f(\zeta_t)dt, \quad \zeta_0=x_0, \nonumber \\
dR_t&=\{\alpha(X_t)-f(\zeta_t)\}dt+\sqrt{\beta(X_t)}\,dW_t, \quad R_0=0. \label{eqn:resTarget}
\end{align}
We then aim to choose $\zeta_t$ (and therefore $f(\cdot)$) to adequately account 
for nonlinear dynamics (so that the drift in (\ref{eqn:resTarget}) is approximately 
constant), and construct the MDB of Section~\ref{mdbsect} 
for the \emph{residual stochastic process} rather than the target 
process itself. Suitable choices of $\zeta_t$ and $f(\cdot)$ can be 
found in Sections~\ref{driftsub} and \ref{LNAsub}. It should be clear from the discussion in Section~\ref{mdbsect} 
that for known $x_T$, the MDB approximates the density of $R_{\tau_{k+1}}|r_{\tau_k},r_T$ 
by
\begin{equation}\label{eqn:res1}
q(r_{\tau_{k+1}}| r_{\tau_{k}},r_{T})=
N\left(r_{\tau_{k+1}}\,;\, r_{\tau_k}+\frac{r_T-r_{\tau_k}}{T-\tau_k}\Delta\tau \,,\,\frac{T-\tau_{k+1}}{T-\tau_k}\beta(x_{\tau_k})\Delta\tau  \right). 
\end{equation}
In this case, the connection between (\ref{eqn:res1}) and the intractable continuous-time 
conditioned residual process can be established by following the arguments 
of Section~\ref{cts}. By approximating the drift and diffusion matrix in 
(\ref{eqn:resTarget}) by the constants $\alpha(x_T)-f(\zeta_T)$ and $\beta(x_T)$ 
gives a process with a tractable transition density. The corresponding conditioned 
process then satisfies
\begin{equation}\label{eqn:SDEbridgeRes}
dR_t=\frac{R_{T}-R_t}{T-t}\,dt+\sqrt{\beta(X_t)}\,dW_t.
\end{equation} 
The density in (\ref{eqn:res1}) is then obtained by a local 
linearisation of (\ref{eqn:SDEbridgeRes}). 
 
It remains for us to choose $\zeta_t$ to balance the 
accuracy and computational efficiency of the resulting construct. 
We explore two possible choices in the remainder of this section.

\subsubsection{Subtracting the drift}\label{driftsub}

In the simplest approach to account for dynamics based on the drift, 
we take $\zeta_t=\eta_t$ and $f(\cdot)=\alpha(\cdot)$ where 
\begin{align}
d\eta_t &= \alpha(\eta_t)dt, \quad \eta_0=x_0, \label{eqn:dgode_eta}
\intertext{so that}
dR_t&=\{\alpha(X_t)-\alpha(\eta_t)\}dt+\sqrt{\beta(X_t)}\,dW_t, \quad R_0=0. \label{eqn:dgode_r}
\end{align}
The MDB can be constructed for the residual process by approximating the 
joint distribution of $R_{\tau_{k+1}}$ and $Y_T-F'\eta_T$ (conditional 
on $r_{\tau_k}$), where $Y_T-F'\eta_T$ can be seen as a partial and noisy observation of $R_T$ since
\[
Y_T-F'\eta_T=F' R_T + \epsilon_T,\qquad \epsilon_T|\Sigma\sim N(0,\Sigma).
\]
As in Section~\ref{mdbsect}, we obtain the (approximate) joint distribution
%\begin{equation}\label{eqn:joint}
%\begin{pmatrix}
%R_{\tau_{k+1}} \\
%Y_T-F'\eta_T \end{pmatrix}\bigg| r_{\tau_k}\sim N\left\{\begin{pmatrix}
%r_{\tau_k}+(\alpha_k-\alpha^\eta_k)\Delta\tau  \\[0.2em]
%F'(r_{\tau_k}+(\alpha_k-\alpha^\eta_k)\Delta_k) \end{pmatrix},
%\begin{pmatrix}
%\beta_k \Delta\tau & \beta_k F\Delta\tau \\
%F'\beta_k\Delta\tau & F'\beta_k F\Delta_k + \Sigma \end{pmatrix}\right\}
%\end{equation}
\begin{equation}\label{eqn:joint}
\begin{pmatrix}
R_{\tau_{k+1}} \\
Y_T-F'\eta_T \end{pmatrix}\bigg| r_{\tau_k}\sim 
N\left\{\begin{pmatrix}
r_{\tau_k}+(\alpha_k-\alpha^\eta_k)\Delta\tau  \\[0.2em]
F'(r_{\tau_k}+(\alpha_k-\alpha^\eta_k)\Delta_k) \end{pmatrix},
\begin{pmatrix}
\beta_k \Delta\tau & \beta_k F\Delta\tau \\
F'\beta_k\Delta\tau & F'\beta_k F\Delta_k + \Sigma \end{pmatrix}\right\} 
\end{equation}
where $\alpha^\eta_k=\alpha(\eta_{\tau_k})$ and $\alpha_k$, $\beta_k$ and $\Delta_k$  
are as defined in Section~\ref{mdbsect}. Note that the mean in (\ref{eqn:joint}) 
uses the tangent $\alpha^\eta_k$ at $(\tau_k,\eta_{\tau_k})$ 
to approximate $d\eta_t/dt$ over time intervals of length $\Delta\tau$ 
and $\Delta_k$. Since $\eta_{\tau_{k+1}}$ will be available either exactly 
from the solution of (\ref{eqn:dgode_eta}) or 
from the output of a (stiff) ODE solver, we propose to approximate 
$d\eta_t/dt$ via the chord between $(\tau_k,\eta_{\tau_k})$ and 
$(\tau_{k+1},\eta_{\tau_{k+1}})$, that is, by
\[
\delta_k^\eta=\frac{\eta_{\tau_{k+1}}-\eta_{\tau_k}}{\Delta\tau}.
\]
Replacing $\alpha_k^\eta$ in (\ref{eqn:joint}) with $\delta_k^\eta$, 
conditioning on $y_{T}-F'\eta_{T}$ and using the 
partition $X_t=\eta_t+R_t$ gives $\Psi_{\R}(x_{\tau_k})=\Psi_{\MDB}(x_{\tau_k})$ and
%\begin{equation}\label{eqn:res4}
%q_{\R}(x_{\tau_{k+1}}| x_{\tau_{k}},y_{T})= 
%N\left(x_{\tau_{k+1}}\,;\, x_{\tau_k}+\Delta\eta_{\tau_k}+\mu_{\R}(x_{\tau_k})\Delta\tau\,,\,\Psi_{\R}(x_{\tau_k})\Delta\tau\right)
%\end{equation}
%where $\Delta\eta_{\tau_k}=\eta_{\tau_{k+1}}-\eta_{\tau_k}$, $\Psi_{\R}(x_{\tau_k})=\Psi_{\MDB}(x_{\tau_k})$ and 
\begin{equation}\label{eqn:res2}
\mu_{\R}(x_{\tau_k}) =\alpha_k+\beta_{k} F\left(F'\beta_{k}F\Delta_k + \Sigma\right)^{-1}
\left\{y_{T}-F'(\eta_T+r_{\tau_{k}}+(\alpha_k-\delta_k^\eta)\Delta_k)\right\}. 
\end{equation}
%\begin{equation}\label{eqn:res2}
%\mu_{\R}(x_{\tau_k}) =\alpha_k-\delta_k^\eta+\beta_{k} F\left(F'\beta_{k}F\Delta_k + \Sigma\right)^{-1}
%\left\{y_{T}-F'(\eta_T+r_{\tau_{k}}+(\alpha_k-\delta_k^\eta)\Delta_k)\right\}.   
%\end{equation}
%and
%\begin{equation}\label{eqn:res3}
%\Psi_{\R}(x_{\tau_k})=\beta_{k}-\beta_{k} F\left(F'\beta_{k} F\Delta_k + \Sigma\right)^{-1}F'\beta_{k}\Delta\tau. 
%\end{equation}
Note that in the case of known $x_T$, $\Psi_{\R}^*(x_{\tau_k})=\Psi_{\MDB}^*(x_{\tau_k})$ and
(\ref{eqn:res2}) becomes %and (\ref{eqn:res3}) become
\[
\mu_{\R}^*(x_{\tau_k}) = \delta_k^\eta+\frac{(x_T-x_{\tau_k})-(\eta_T-\eta_{\tau_k})}{T-\tau_k}. %\quad \textrm{and} \quad 
%\Psi_{\R}^*(x_{\tau_k})=\frac{T-\tau_{k+1}}{T-\tau_k}\beta(x_{\tau_k}).
\]

\subsubsection{Further subtraction using the linear noise approximation}\label{LNAsub}

Whilst the solution of the SDE governing the residual stochastic 
process in (\ref{eqn:dgode_r}) is unavailable 
in closed form, a tractable approximation can be obtained. Therefore, in situations 
where $\eta_t$ fails to adequately capture the target process dynamics, 
we propose to further subtract an approximation of the conditional 
expectation $\rho_t=\expect(R_{t}|r_{0},y_T)$, which we denote 
by $\hat{\rho}_t=\expect(\hat{R}_{t}|r_{0},y_T)$. Here, $\{\hat{R}_t, t\in[0,T]\}$ 
is obtained through the linear noise approximation (LNA) of (\ref{eqn:dgode_r}). 
The LNA can be derived in a number of more or less formal ways (see e.g. 
\cite{Kurtz_1970}, \cite{vanKampen2001} and \cite{Fearnhead_2014}). 
Here, we give a brief exposition of the LNA and refer the reader to 
\cite{Fearnhead_2014} and the references therein for a complete derivation.

By Taylor expanding $\alpha(X_t)$ 
and $\beta(X_t)$ about $\eta_t$ (the solution of (\ref{eqn:dgode_eta})), 
truncating the expansion of $\alpha$ at the first two terms and taking 
only the first term of the expansion of $\beta$, we obtain
 \[
d\hat{R}_t = H(\eta_t)\hat{R}_t\,dt + \sqrt{\beta(\eta_t)}\,dW_t,
\]
where $H(\eta_t)$ is the Jacobian matrix with $(i,j)$th element 
$(H(\eta_t))_{i,j}=\partial\alpha_i(\eta_t)/\partial\eta_{j,t}$. It should be clear 
from the truncations used in the Taylor expansions of the drift and diffusion coefficients that 
the key assumption underpinning the LNA is that the stochastic term $\beta(X_t)$ is ``small''. 
Now, for a fixed initial condition $\hat{R}_0=\hat{r}_0$, it is straightforward 
to show that 
\begin{equation}\label{eqn:lna0}
\hat{R}_t|\hat{R}_0=\hat{r}_0 \sim N\left(P_t \hat{r}_0\,,\,P_t\psi_tP_t'\right)
\end{equation}
where $P_t$ and $\psi_t$ satisfy the ODE system
\begin{align}
\frac{dP_t}{dt}&=H(\eta_t)P_t, \quad P_0=I_d,   \label{eqn:lna1}\\
\frac{d\psi_t}{dt}&=P_t^{-1}\beta(\eta_t)(P_t^{-1})',\quad \psi_0=0. \label{eqn:lna2}
\end{align}
The joint distribution of $\hat{R}_t$ and $Y_T-F'\eta_T$ (conditional on $\hat{r}_0$) 
is
%\begin{equation}\label{eqn:lnaJoint}
%\begin{pmatrix}
%\hat{R}_{t} \\
%Y_T-F'\eta_T \end{pmatrix}\bigg| \hat{r}_0\sim N\left\{\begin{pmatrix}
%P_t\hat{r}_0  \\[0.2em]
%F'P_T\hat{r}_0 \end{pmatrix},
%\begin{pmatrix}
%P_t\psi_tP_t' &P_t\psi_tP_T'F  \\
%F'P_T\psi_tP_t' &F'P_T\psi_TP_T'F  + \Sigma \end{pmatrix}\right\}.
%\end{equation}
\begin{equation}\label{eqn:lnaJoint}
\begin{pmatrix}
\hat{R}_{t} \\
Y_T-F'\eta_T \end{pmatrix}\bigg| \hat{r}_0\sim 
 N\left\{\begin{pmatrix}
P_t\hat{r}_0  \\[0.2em]
F'P_T\hat{r}_0 \end{pmatrix},
\begin{pmatrix}
P_t\psi_tP_t' &P_t\psi_tP_T'F  \\
F'P_T\psi_tP_t' & F'P_T\psi_TP_T'F  + \Sigma \end{pmatrix}\right\}. 
\end{equation}
Conditioning further on $y_T-F'\eta_T$ and noting that $\hat{r}_0=r_0=~0$ gives
\begin{align*}
\hat{\rho}_t&=\expect(\hat{R}_{t}|r_{0},y_T)\\
&=P_t\psi_tP_T'F(F'P_T\psi_TP_T'F  + \Sigma)^{-1}(y_T-F'\eta_T).
\end{align*}

Having obtained an explicit, closed-form (subject to the solution of (\ref{eqn:dgode_eta}), 
(\ref{eqn:lna1}) and (\ref{eqn:lna2})) approximation of the expected conditioned 
residual process, we adopt the partition $X_t=\eta_t+\hat{\rho}_t+R_t^-$ where 
$\{R_t^-,t\in[0,T]\}$ is the residual stochastic process resulting from the 
additional decomposition of $X_t$. Although the SDE satisfied by $R_t^-$ 
will be intractable, the joint distribution of 
$R_{\tau_{k+1}}^-$ and $Y_T-F'(\eta_T+\hat{\rho}_T)$ can be approximated (conditional 
on $r_{\tau_{k}}^-$) by 
\[
\begin{pmatrix}
R_{\tau_{k+1}}^- \\
Y_T-F'(\eta_T+\hat{\rho}_T) \end{pmatrix}\bigg| r_{\tau_k}^-\sim 
N\left\{\begin{pmatrix}
r_{\tau_k}^- +(\alpha_k-\delta_k^\eta-\delta_k^\rho)\Delta\tau  \\[0.2em]
F'(r_{\tau_k}^- +(\alpha_k-\delta_k^\eta-\delta_k^\rho)\Delta_k) \end{pmatrix},\begin{pmatrix}
\beta_k \Delta\tau & \beta_k F\Delta\tau \\
F'\beta_k\Delta\tau & F'\beta_k F\Delta_k + \Sigma \end{pmatrix}\right\}
\]
where again we use the chord 
\[
\delta_k^\rho=\frac{\hat{\rho}_{\tau_{k+1}}-\hat{\rho}_{\tau_k}}{\Delta\tau}
\]
in preference to the tangent. Hence we obtain $\Psi_{\RM}(x_{\tau_k})=\Psi_{\MDB}(x_{\tau_k})$ and
%\begin{equation}\label{eqn:resSub}
%q_{\RM}(x_{\tau_{k+1}}| x_{\tau_{k}},y_{T})= 
%N\left(x_{\tau_{k+1}}\,;\, x_{\tau_k}+\Delta\eta_{\tau_k}+\Delta\hat{\rho}_{\tau_k}+\mu_{\RM}(x_{\tau_k})\Delta\tau\,,\,\Psi_{\RM}(x_{\tau_k})\Delta\tau\right)
%\end{equation}
%where $\Delta\hat{\rho}_{\tau_k}=\hat{\rho}_{\tau_{k+1}}-\hat{\rho}_{\tau_k}$, $\Psi_{\RM}(x_{\tau_k})=\Psi_{\MDB}(x_{\tau_k})$ and
\begin{equation}\label{eqn:resSub2} 
\mu_{\RM}(x_{\tau_k}) =\alpha_k+\beta_{k} F\left(F'\beta_{k}F\Delta_k + \Sigma\right)^{-1} 
\left\{y_{T}-F'(\eta_T+\hat{\rho}_T+r_{\tau_{k}}^- +(\alpha_k-\delta_k^\eta-\delta_k^\rho)\Delta_k)\right\}.  
\end{equation}
%\begin{align}
%\mu_{\RM}(x_{\tau_k}) &=\alpha_k-\delta_k^\eta-\delta_k^\rho+\beta_{k} F\left(F'\beta_{k}F\Delta_k + \Sigma\right)^{-1} \nonumber\\
%& \qquad\qquad \times\left\{y_{T}-F'(\eta_T+\hat{\rho}_T+r_{\tau_{k}}^- +(\alpha_k-\delta_k^\eta-\delta_k^\rho)\Delta_k)\right\}.  \label{eqn:resSub2} 
%\end{align}
%and
%\begin{equation}\label{eqn:resSub3}
%\Psi_{\RM}(x_{\tau_k})=\beta_{k}-\beta_{k} F\left(F'\beta_{k} F\Delta_k + \Sigma\right)^{-1}F'\beta_{k}\Delta\tau. 
%\end{equation}
Note that in the case of known $x_T$, $\Psi_{\RM}^*(x_{\tau_k})=\Psi_{\MDB}^*(x_{\tau_k})$
and (\ref{eqn:resSub2}) becomes %and (\ref{eqn:resSub3}) become
\[
\mu_{\RM}^*(x_{\tau_k}) = \delta_k^\eta+\delta_k^\rho+\frac{(x_T-x_{\tau_k})-(\eta_T-\eta_{\tau_k})-(\hat{\rho}_T-\hat{\rho}_{\tau_k})}{T-\tau_k}. % \quad \textrm{and} \quad 
%\Psi_{\RM}^*(x_{\tau_k})=\frac{T-\tau_{k+1}}{T-\tau_k}\beta(x_{\tau_k}).
\]

\subsection{Guided proposals}\label{guide}

For known $x_T$, \cite{vanDerMeulen2015} (see also \cite{Schauer_2014}) 
derive a bridge construct which they term a \emph{guided proposal} (GP). They take 
the SDE satisfied by the conditioned process $\{X_t,t\in[0,T]\}$ in (\ref{eqn:guip1}) 
and (\ref{eqn:guip2}) but replace the intractable $p(x_T|x_t)$ with the transition density associated with a 
class of linear processes $\{\hat{X}_t,t\in[0,T]\}$ satisfying 
\begin{equation}\label{eqn:linear}
d\hat{X}_t= B(t)\hat{X}_t\,dt+b(t)\,dt+\sqrt{\sigma(t)}\,dW_t, \quad \hat{X}_0=x_.
\end{equation}
Here, $B(t)$ and $\sigma(t)$ are $d\times d$ matrices and $b(t)$ is a $d$-vector. 
Note that the LNA (see Section~\ref{LNAsub}) satisfies 
(\ref{eqn:linear}) with $B(t)=H(\eta_t)$, $b(t)=\alpha(\eta_t)-H(\eta_t)\eta_t$ and 
$\sigma(t)=\beta(\eta_t)$.

The guided proposal can be extended to the 
Gaussian additive noise regime in (\ref{eqn:obs}) by noting 
that in this case, the drift in (\ref{eqn:guip2}) becomes 
\begin{equation}\label{eqn:guip3}
\tilde{\alpha}(X_t)=\alpha(X_t)+\beta(X_t)\nabla_{x_t}\log p(y_T|x_t).
\end{equation}
Given a tractable approximation of $p(y_T|x_t)$, the Euler-Maruyama approximation of (\ref{eqn:guip1}) 
can be applied over the discretisation of $[0,T]$ to give a proposal density of the form (\ref{eqn:qprop}) with
%\begin{equation}\label{eqn:guipProp}
%\[
%q_{\GP}(x_{\tau_{k+1}}| x_{\tau_{k}},y_{T})= 
%N\left(x_{\tau_{k+1}}\,;\, x_{\tau_k}+\mu_{\GP}(x_{\tau_k})\Delta\tau\,,\,\Psi_{\GP}(x_{\tau_k})\Delta\tau\right)
%\]
%\end{equation}
$\mu_{\GP}(x_{\tau_k})=~\tilde{\alpha}(x_{\tau_k})$ and $\Psi_{\GP}(x_{\tau_k})=\beta(x_{\tau_k})$. 

We will approximate $p(y_T|x_t)$ using the LNA. Using the 
partition $\hat{X}_t=\eta_t+\hat{R}_t$ and combining the transition 
density of $\hat{R}_t$ in (\ref{eqn:lna0}) with the observation 
regime defined in (\ref{eqn:obs}) gives
\[
\hat{p}(y_T|x_t)=
 N\left(y_T\,;\,F'\{\eta_T+P_{T|t}(x_t-\eta_t)\},F'P_{T|t}\psi_{T|t}P_{T|t}'F+\Sigma\right)
\]
where $P_{T|t}$ and $\psi_{T|t}$ are found by integrating the ODE system in 
(\ref{eqn:lna1}) and (\ref{eqn:lna2}) from $t$ to $T$ with 
$P_{t|t}=I_d$ and $\psi_{t|t}=0$. Hence the drift (\ref{eqn:guip3}) 
becomes
\begin{equation}
\tilde{\alpha}(X_t)=\alpha(X_t)+\beta(X_t)P_{T|t}'F(F'P_{T|t}\psi_{T|t}P_{T|t}'F+\Sigma)^{-1} 
\left\{y_T-F'(\eta_T+P_{T|t}[x_t-\eta_t])\right\}. \label{eqn:gpalpha}
\end{equation}
Note that a computationally efficient implementation of this approach is obtained by 
using the identities $P_{T|t}=P_{T}P_{t}^{-1}$ and $\psi_{T|t}=P_{t}(\psi_{T}-\psi_{t})P_{t}'$. 
Hence, the LNA ODEs in (\ref{eqn:dgode_eta}), (\ref{eqn:lna1}) and (\ref{eqn:lna2}) need only be 
integrated \emph{once} over the interval $[0,T]$. Unfortunately, we find that this approach 
does not work well in practice, unless the total measurement error $\tr(\Sigma)$ is large relative to 
the infinitesimal variance $\beta(\cdot)$. Note that the variance of $Y_T|x_t$ under the 
LNA is a function of the deterministic process $\eta_t$. If $\eta_t$ and 
$x_t$ diverge as $t$ is increased, the guiding term in (\ref{eqn:gpalpha}) will result in 
an over or under dispersed proposal mechanism (relative to the target conditioned process) 
at times close to $T$. The problem is exacerbated in the case of no measurement error, where the discrepancy between 
$x_t$ and $\eta_t$ can result in a singularity in the guiding term in (\ref{eqn:gpalpha}) at time $T$. 
This naive approach (henceforth referred to as GP-N) can be alleviated by integrating the ODE system given by 
(\ref{eqn:dgode_eta}), (\ref{eqn:lna1}) and (\ref{eqn:lna2}) \emph{for each interval} $[\tau_k, T]$, 
$k=0,1,\ldots,m-1$, with $\eta_{\tau_k}=x_{\tau_{k}}$. In this case, the drift (\ref{eqn:guip3}) 
is given by
\[
\tilde{\alpha}(X_t)=\alpha(X_t)+\beta(X_t)P_{T|t}'F(F'P_{T|t}\psi_{T|t}P_{T|t}'F+\Sigma)^{-1} 
\left(y_T-F'\eta_T\right). 
\]  

In the special case that $x_T$ is known, we have that $\Psi_{\GPN}^*(x_{\tau_k})=\Psi_{\GP}^*(x_{\tau_k})=\beta(x_{\tau_k})$, 
\[
\mu_{\GPN}^*(x_{\tau_k})=\alpha(x_{\tau_k})+\beta(x_{\tau_k})P_{T|\tau_k}'(P_{T|\tau_k}\psi_{T|\tau_k}P_{T|\tau_k}')^{-1}
\left\{x_T-[\eta_T+P_{T|\tau_k}(x_{\tau_k}-\eta_{\tau_k})]\right\}
\]
and 
\[
\mu_{\GP}^*(x_{\tau_k})=\alpha(x_{\tau_k})+\beta(x_{\tau_k})P_{T|\tau_k}'(P_{T|\tau_k}\psi_{T|\tau_k}P_{T|\tau_k}')^{-1}
\left(x_T-\eta_T\right).
\]
The limiting form of the acceptance rate in this case can be found 
in \cite{Schauer_2014}, who also remark that a key requirement for absolute continuity 
of the target and proposal process is that $\sigma(T)=\beta(x_T)$. For the LNA, 
we have $\sigma(t)=\beta(\eta_t)$. Again, we note that the naive implementation 
of the guided proposal (GP-N) will not meet this condition in general (when $x_T$ is known). 
Ensuring that $\sigma(t)\to\beta(x_T)$ as $t\to T$ by integrating (\ref{eqn:dgode_eta}), (\ref{eqn:lna1}) and (\ref{eqn:lna2}) 
for each $\tau_k$ is likely to be time consuming, unless the LNA ODE system is tractable. In the case of exact observations, 
a computationally less demanding approach is obtained in \cite{vanDerMeulen2015} by 
taking the transition density of (\ref{eqn:linear}) 
with $B(t)=0$ and $\sigma(t)=\beta(x_T)$ to construct the 
guided proposal. Setting $b(t)=\alpha(\eta_t)$ leads to 
a proposal density for the simplified guided proposal (GP-S) of the form (\ref{eqn:qprop}) 
%\[
%q_{\GPS}(x_{\tau_{k+1}}| x_{\tau_{k}},x_{T})= 
%N\left(x_{\tau_{k+1}}\,;\, x_{\tau_k}+\mu_{\GPS}^*(x_{\tau_k})\Delta\tau\,,\,\Psi_{\GPS}^*(x_{\tau_k})\Delta\tau\right)
%\]
with $\Psi_{\GPS}^*(x_{\tau_k})=\beta(x_{\tau_k})$ and 
\[
\mu_{\GPS}^*(x_{\tau_k})=\alpha(x_{\tau_k})+\beta(x_{\tau_k})\beta(x_T)^{-1}
\left\{\frac{x_T-x_{\tau_k}-(\eta_{T}-\eta_{\tau_k})}{T-\tau_k}\right\}.
\]
Further (example-dependent) methods for constructing guided proposals 
in the case of known $x_T$ can be found in \cite{vanDerMeulen2015}.

\subsubsection{Use of the MDB variance}\label{guipMDB}

Using the Euler-Maruyama approximation of (\ref{eqn:guip1}) 
gives the variance of $X_{\tau_{k+1}}|x_{\tau_{k}},y_T$ 
in the guided proposal process as 
$\Psi_{\GP}(x_{\tau_k})\Delta\tau=\beta(x_{\tau_{k}})\Delta\tau$. In 
Section~\ref{app} we investigate the effect of using the variance (\ref{eqn:mdb3})
of the modified diffusion bridge construct 
by taking $\Psi_{\GP}(x_{\tau_k})=\Psi_{\MDB}(x_{\tau_{k}})$. Although 
in this case, deriving the limiting form of the acceptance 
rate under the resulting proposal is problematic, we observe a 
worthwhile increase in empirical performance. In the case of 
known $x_T$, use of the MDB variance in place of 
$\beta(x_{\tau_k})\Delta\tau$ comes at almost no additional 
computational cost. We denote this construct GP-MDB.

\subsection{Computational considerations}\label{comp}

For the observation regime in (\ref{eqn:obs}), all bridge constructs 
(with the exception of the myopic approach) require the inversion of 
a $d_o\times d_o$ matrix at each intermediate time $\tau_{k}$, 
\hbox{$k=1,2\ldots,m-1$} and for each skeleton bridge required. For 
known $x_T$, the proposal densities associated with each construct 
simplify. In this case, only the LNA-based residual bridge 
and guided proposal require the inversion of a $d\times d$ matrix 
at each intermediate time.

The Lindstr\"{o}m  bridge and modified diffusion 
bridge have roughly the same computational 
cost. The bridges based on residual processes incur an additional 
computational cost of having to solve a system of either $d$ 
(when subtracting $\eta_t$) or order $d^2$ (when further subtracting 
$\rho_t$) coupled ODEs. However, we note that for known $x_0$, the 
ODE system need only be solved once, irrespective of the number 
of skeleton bridges required. This is also true of the naive 
and simplified guided proposals. However, we note that in the case of 
known $x_T$, the guided proposal requires solving order $d^2$ ODEs 
over each interval $[\tau_k,T]$, $k=0,1,\ldots,m-1$ for each 
simulated skeleton bridge, in order to maintain 
reasonable statistical efficiency (as measured by, for example, 
estimated acceptance rate of a Metropolis-Hastings independence sampler). 

\section{Applications}\label{app}

We now compare the accuracy and efficiency of the bridging methods discussed 
in the previous sections, by using them to make proposals inside a Metropolis-Hastings 
independence sampler. We consider three examples: a simple birth-death 
model in which the ODEs governing the LNA are tractable, 
a Lotka-Volterra system in which the use of numerical solvers are required, and a 
model of aphid growth inspired by real data taken from \cite{Matis_2008}. 
Generating discrete-time realisations from the SDE model of aphid growth 
is particularly challenging due to nonlinear dynamics, and an observation regime 
in which only one component is observed and is subject to additive Gaussian 
noise. 

In what follows, all results are based on 100K iterations of a Metropolis-Hastings 
independence sampler targeting either (\ref{eqn:target}) or (\ref{eqn:target2}), depending 
on the observation regime. We measure the statistical efficiency of each bridge via their 
empirical acceptance probability. \verb+R+ code for the implementation of the M-H scheme 
can be found at \emph{https://github.com/gawhitaker/bridges-apps}. The bridge constructs used in each example, 
together with their relative computational cost can be found in Table~\ref{tab times}. 
Note that in contrast to \cite{Lindstrom_2012}, we found that $\gamma\in[0.001,0.3]$ 
was required in order to find a near-optimal $\gamma$. 
Where LB is used, we only present results for the value 
of $\gamma$ that maximised empirical performance. 

\begin{table}
\begin{center}
\begin{tabular}{ll|ccc}
	\hline
&    & Birth-Death & Lotka-Volterra & Aphid \\
  \hline
              Myopic Euler-Maruyama &(EM)  & -- & -- & 1.0 \\  
         Modified diffusion bridge &(MDB)  & 1.0 & 1.0 & -- \\
                 Lindstr\"om bridge &(LB)  & 1.1 & 1.1 &  -- \\
Residual bridge, subtract $\eta_t$  &(RB)  & 1.0 & 1.0 & 7.3  \\ 
  RB, further subtract $\rho_t$ &(RB$^-$)  & 1.0 & 1.0 & 7.9  \\
                    Guided proposal &(GP)  & 1.2 & 30.7 & 7.1 \\ %194.7 old time for solving each bridge
	   GP with MDB variance &(GP-MDB)  & 1.3 & 31.0 & 7.9  \\ %199.5 old time for solving each bridge
                         Naive GP &(GP-N)  & 1.2 & -- & -- \\
	            Simplified GP &(GP-S)  & 1.1 & -- & --  \\
	\hline
\end{tabular}
\caption{Example and bridge specific relative CPU cost for 100K iterations of a Metropolis-Hastings independence sampler. Due to well known poor performance in the case of known $x_T$, EM is not implemented for the first two examples. Likewise, due to poor performance, we omit results based on  GP-N and GP-S in the second example, and results based on MDB and LB in the final example.} \label{tab times}
\end{center}
\end{table} 

\subsection{Birth-death}\label{bd}

We consider a simple birth-death process with birth rate $\theta_1$ 
and death rate $\theta_2$, characterised by the SDE
%\begin{eqnarray*}
%\textrm{\bf{Reaction 1:}} & \mathcal{X} & \overset{\lambda}\rightarrow \quad 2\mathcal{X} \\
%\textrm{\bf{Reaction 2:}} & \mathcal{X} & \overset{\mu}\rightarrow \quad \emptyset
%\end{eqnarray*}
%can be approximated through the SDE
\begin{equation}\label{eqn:bd}
dX_t = (\theta_1-\theta_2)X_t\,dt + \sqrt{(\theta_1+\theta_2)X_t}\,dW_t,\quad X_0=x_0
\end{equation}
which can be seen as a degenerate case of a Feller square-root diffusion \citep{feller52}. 
The ODE system (\eqref{eqn:dgode_eta}, (\ref{eqn:lna1}) and (\ref{eqn:lna2})) 
governing the linear noise approximation of (\ref{eqn:bd}) is tractable, and we 
obtain $\eta_t=x_0e^{(\theta_1-\theta_2)t}$, $P_t=e^{(\theta_1-\theta_2)t}$ and 
\[
\psi_t = \dfrac{\theta_1+\theta_2}{\theta_1-\theta_2}\left(1-e^{-(\theta_1-\theta_2)t}\right)x_0 .
\]

In this example we assume that $x_T$ is known and, to adequately assess the 
performance of each bridge construct, we take $x_T$ to be either the 5\%, 50\% or 95\% quantile 
(denoted by $x_{T,(5)}$, $x_{T,(50)}$ and $x_{T,(95)}$ respectively) 
of $X_T|X_0=~x_0$, found by repeatedly applying the Euler-Maruyama approximation 
to (\ref{eqn:bd}) with a small time-step. To allow for different inter-observation 
intervals, we take $T\in\{1,2\}$. An initial condition of 
$x_0=50$ and parameter values $\theta=(0.1,0.8)'$ gives 
$(x_{1,(5)},x_{1,(50)},x_{1,(95)})=(18.49,24.62,31.68)$ and 
$(x_{2,(5)},x_{2,(50)},x_{2,(95)})=(6.97,12.00,18.35)$.   

\begin{comment}
We compare the performance of all bridging strategies 
discussed in Sections~\ref{sde} and \ref{bridge}, except the 
myopic approach, which is well known to behave poorly in the 
absence of measurement error. Specifically, we apply the 
modified diffusion bridge (MDB), Lindstr\"om bridge (LB), 
residual bridge based on both subtraction of $\eta_t$ (RB) 
and further subtraction of the expected residual under 
the LNA (RB$^-$), the guided proposal with variance 
based on both the Euler-Maruyama approximation (GP) and MDB 
(GP-MDB), and the simplified approach (GP-S). We also 
investigate the performance of a naive implementation 
of GP, where the ODE system governing the 
approximating transition density is only integrated once 
over $[0,T]$. We denote this construct by GP-N.   
 
A Metropolis-Hastings independence sampler targeting 
(\ref{eqn:target2}) was run for 100K iterations with each 
bridge construct used as a proposal mechanism. 
\end{comment}

Since the ODE system governing the LNA is tractable 
for this example, there is little difference in CPU 
cost between the bridges (see Table~\ref{tab times}). 
Therefore, we use statistical efficiency (as measured by 
empirical Metropolis-Hastings acceptance probablity) 
as a proxy for overall efficiency of each bridge, with 
higher probabilities preferred.  

\begin{figure}[t]
\begin{center}
\begin{minipage}[b]{0.32\linewidth}
        \centering
				\caption*{\qquad $x_T=x_{T,(5)}$}\vspace{-0.25cm}
        \includegraphics[scale=0.285]{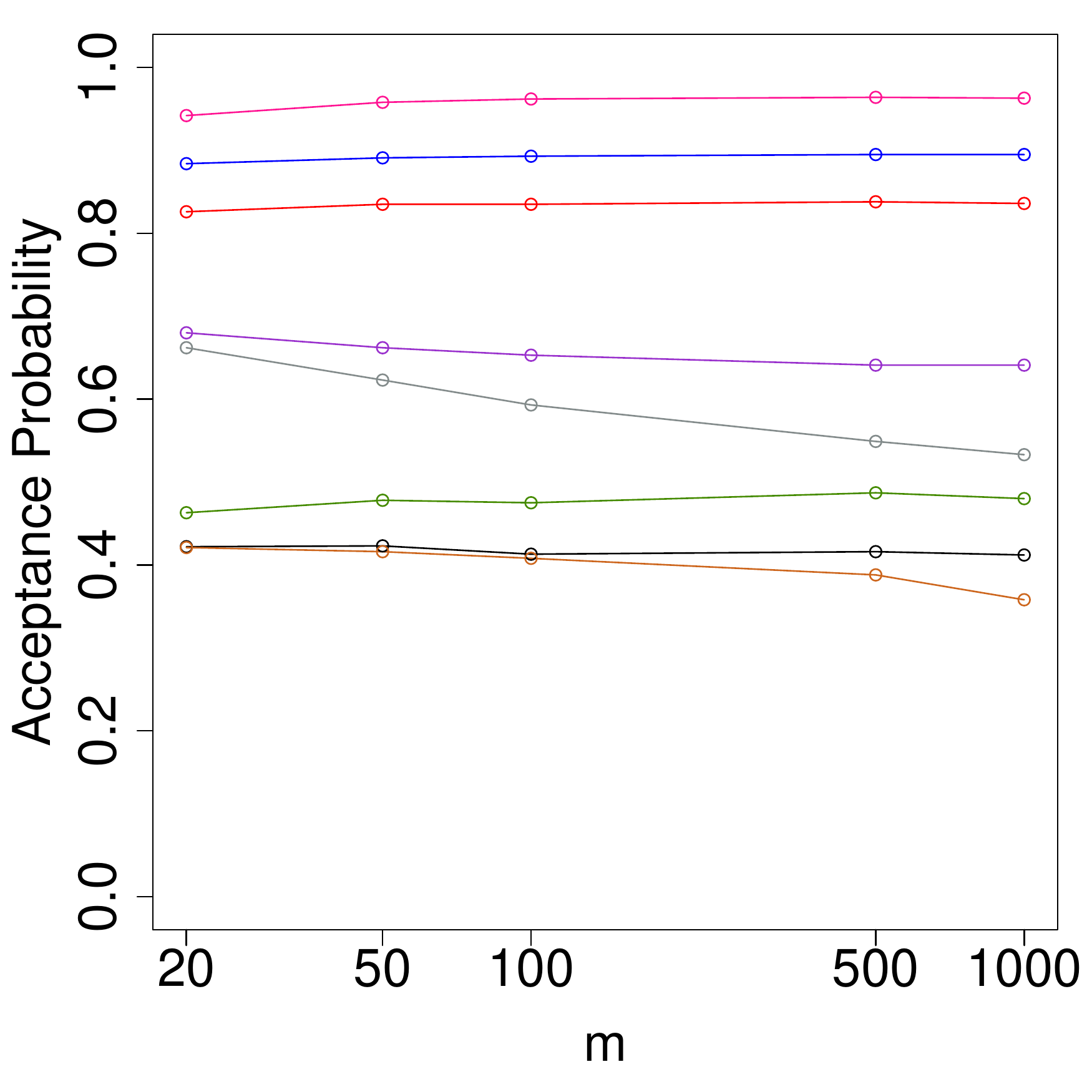}
\end{minipage} 
\begin{minipage}[b]{0.32\linewidth}
        \centering
				\caption*{\qquad $x_T=x_{T,(50)}$}\vspace{-0.25cm}
        \includegraphics[scale=0.285]{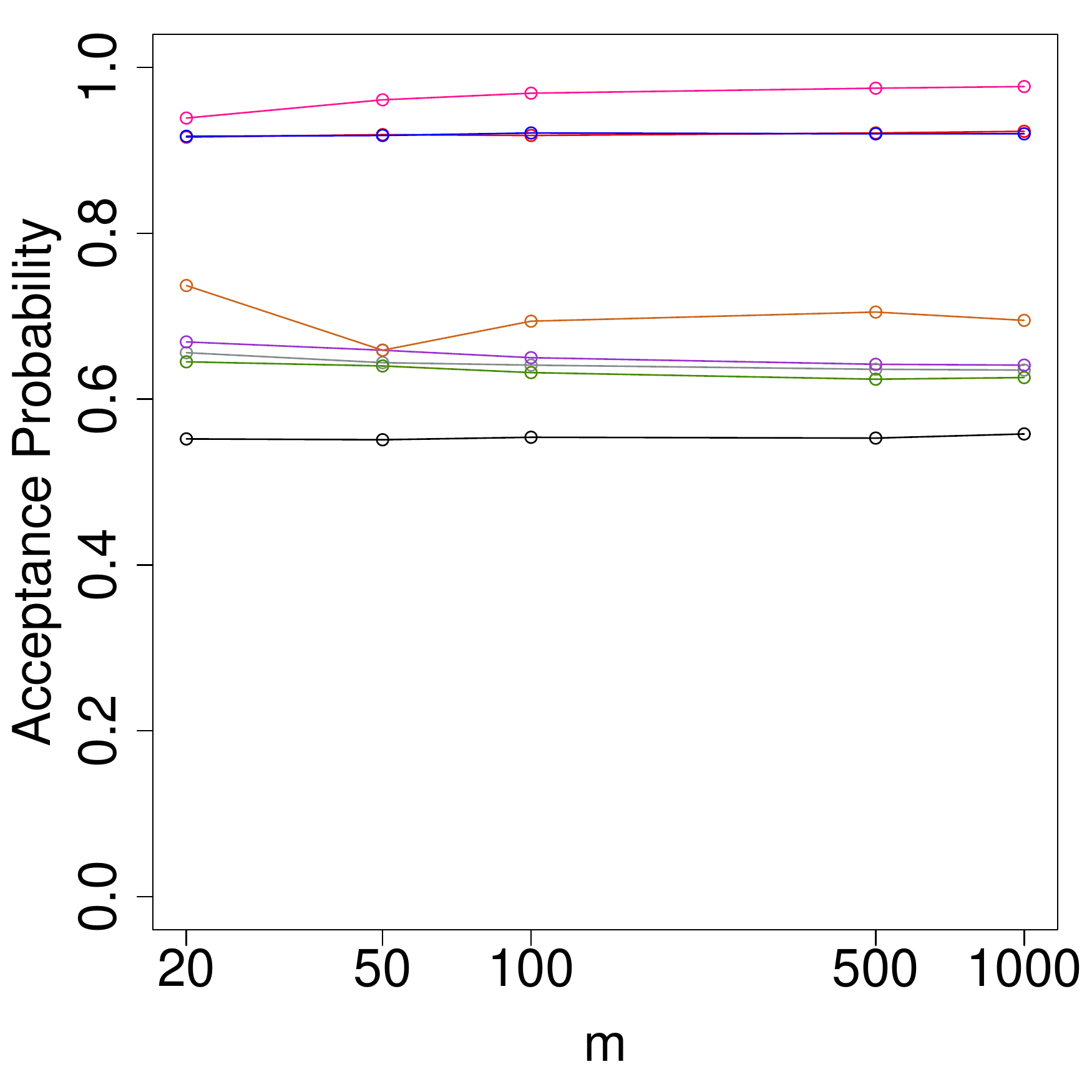}
\end{minipage}
\begin{minipage}[b]{0.32\linewidth}
				\centering
				\caption*{\qquad $x_T=x_{T,(95)}$}\vspace{-0.25cm}
        \includegraphics[scale=0.285]{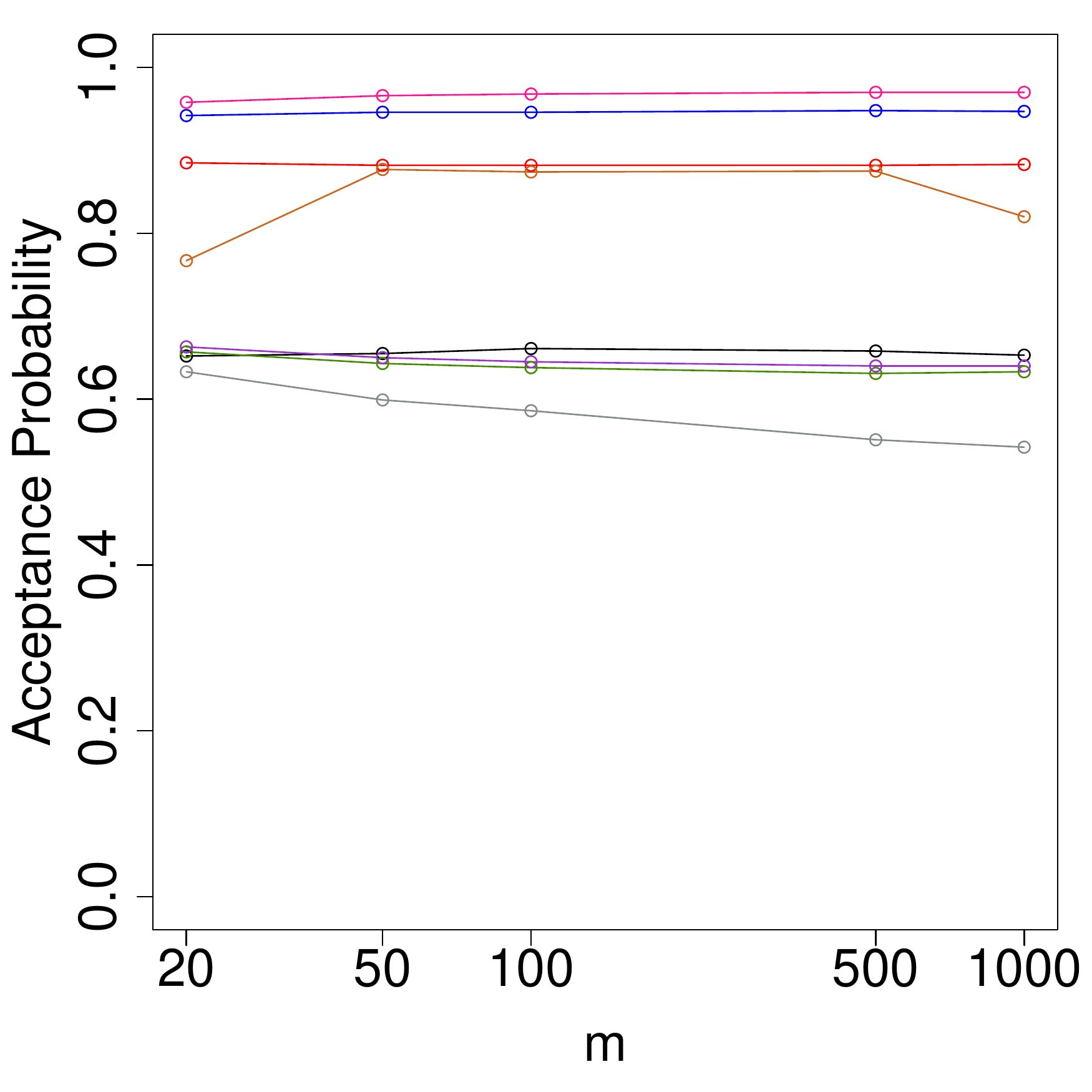}
\end{minipage}\\
\vspace{0.2cm}
\begin{minipage}[b]{0.32\linewidth}
				\centering
        \includegraphics[scale=0.285]{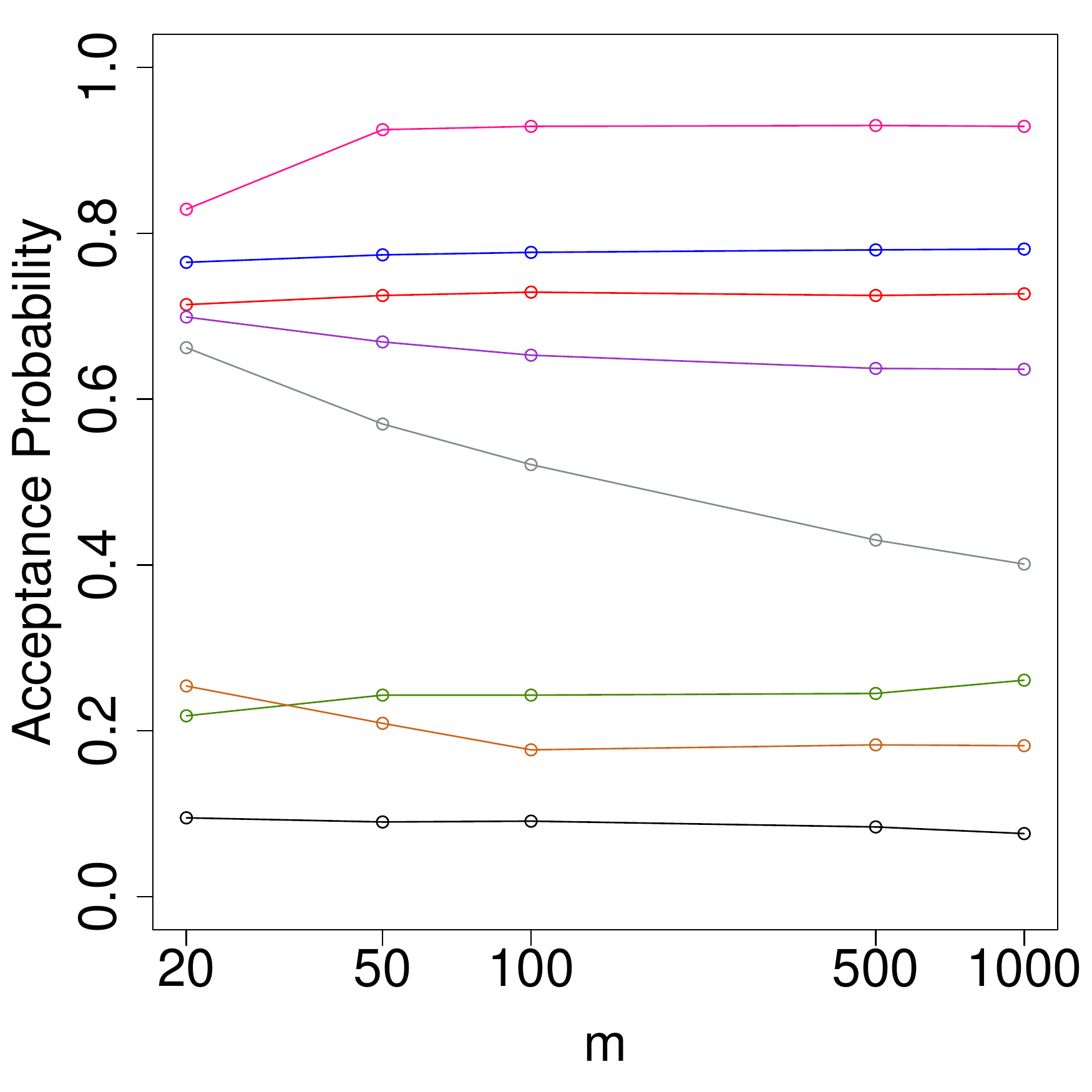}
\end{minipage}
\begin{minipage}[b]{0.32\linewidth}
        \centering
        \includegraphics[scale=0.285]{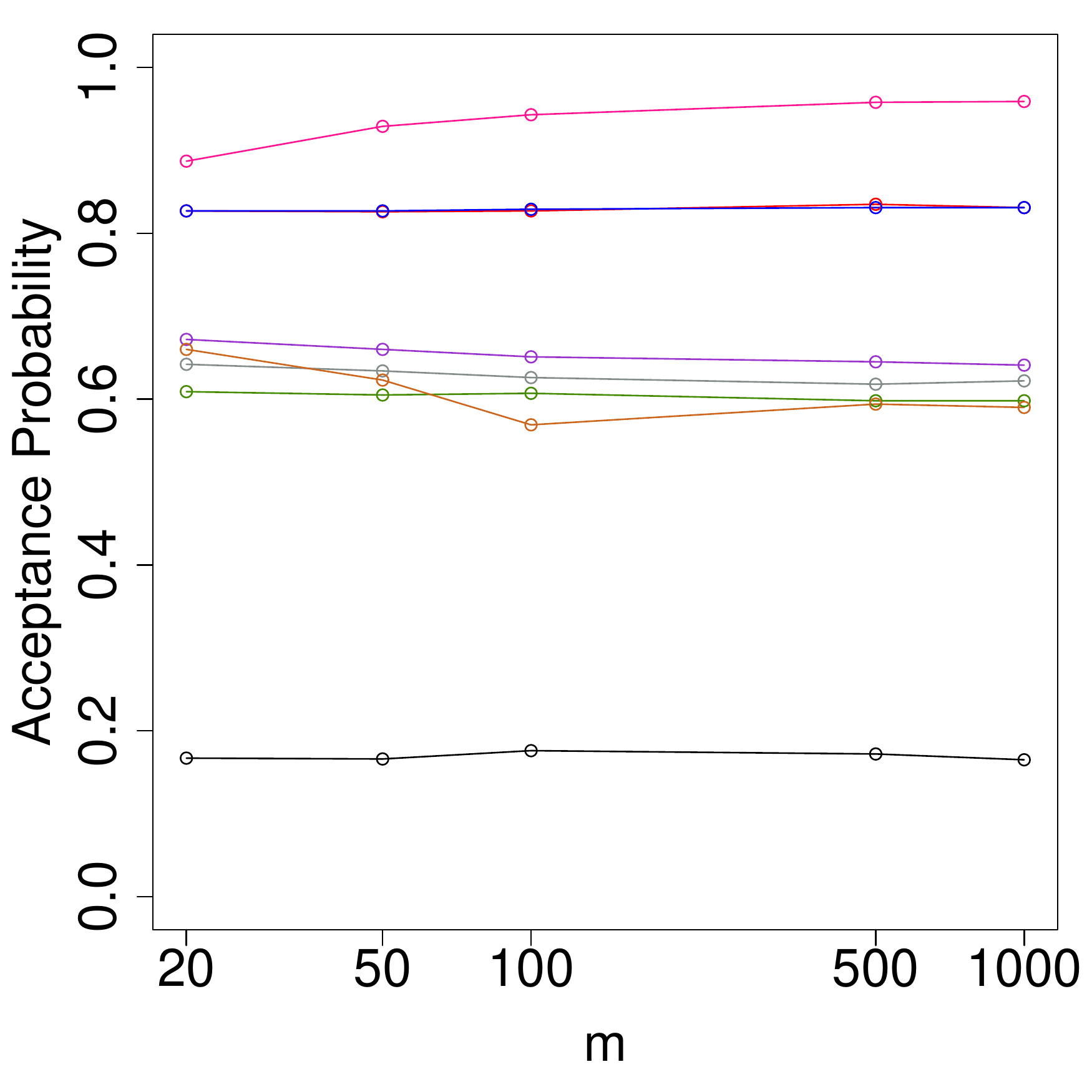}
\end{minipage}
\begin{minipage}[b]{0.32\linewidth}
        \centering
        \includegraphics[scale=0.285]{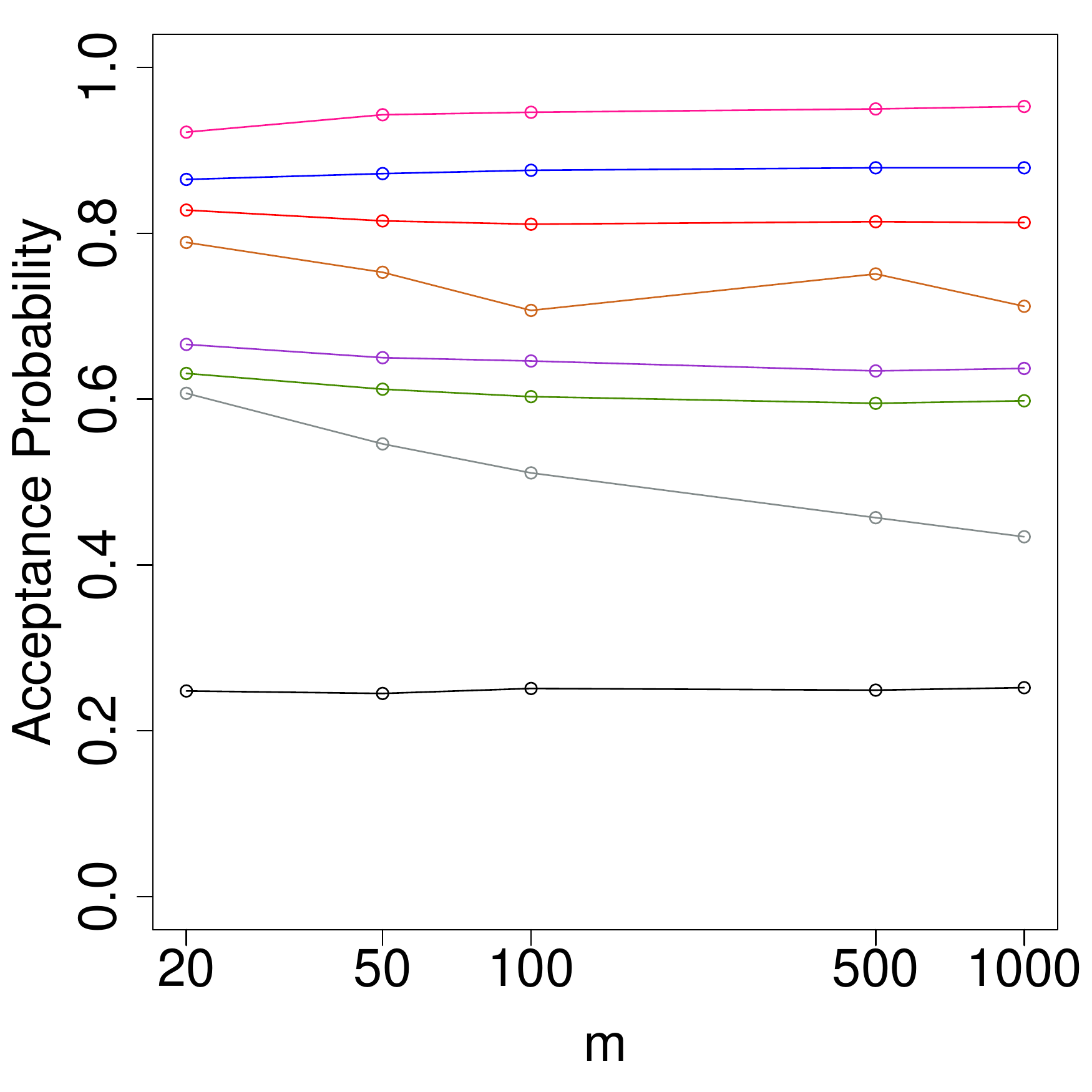}
\end{minipage}
\caption{Birth-death model. Empirical acceptance probability against $m$ with $T=1$ 
(1$^{\textrm{st}}$ row) and $T=2$ (2$^{\textrm{nd}}$ row). The results are based on 100K iterations of a Metropolis-Hastings independence 
sampler. Black: MDB. Brown: LB. Red: RB. Blue: RB$^-$. Grey: GP-N. Green: GP-S. Purple:~GP. Pink: GP-MDB.} \label{bd fig_ap}
\end{center}
\end{figure}

\begin{figure}[t]
\begin{center}
\begin{minipage}[b]{0.32\linewidth}
        \centering
				\caption*{\qquad MDB}\vspace{-0.25cm}
        \includegraphics[scale=0.285]{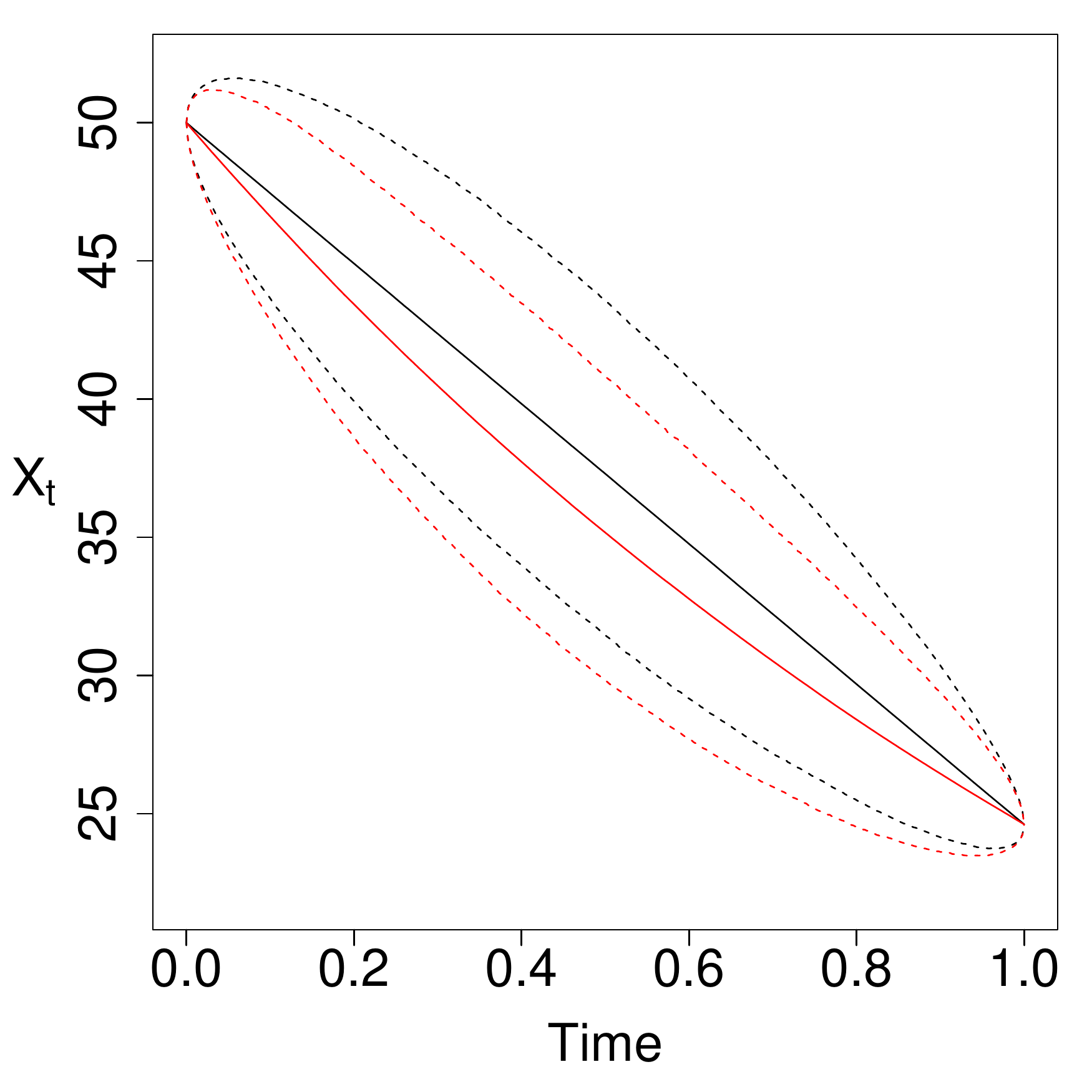}
\end{minipage} 
\begin{minipage}[b]{0.32\linewidth}
        \centering
				\caption*{\qquad RB}\vspace{-0.25cm}
        \includegraphics[scale=0.285]{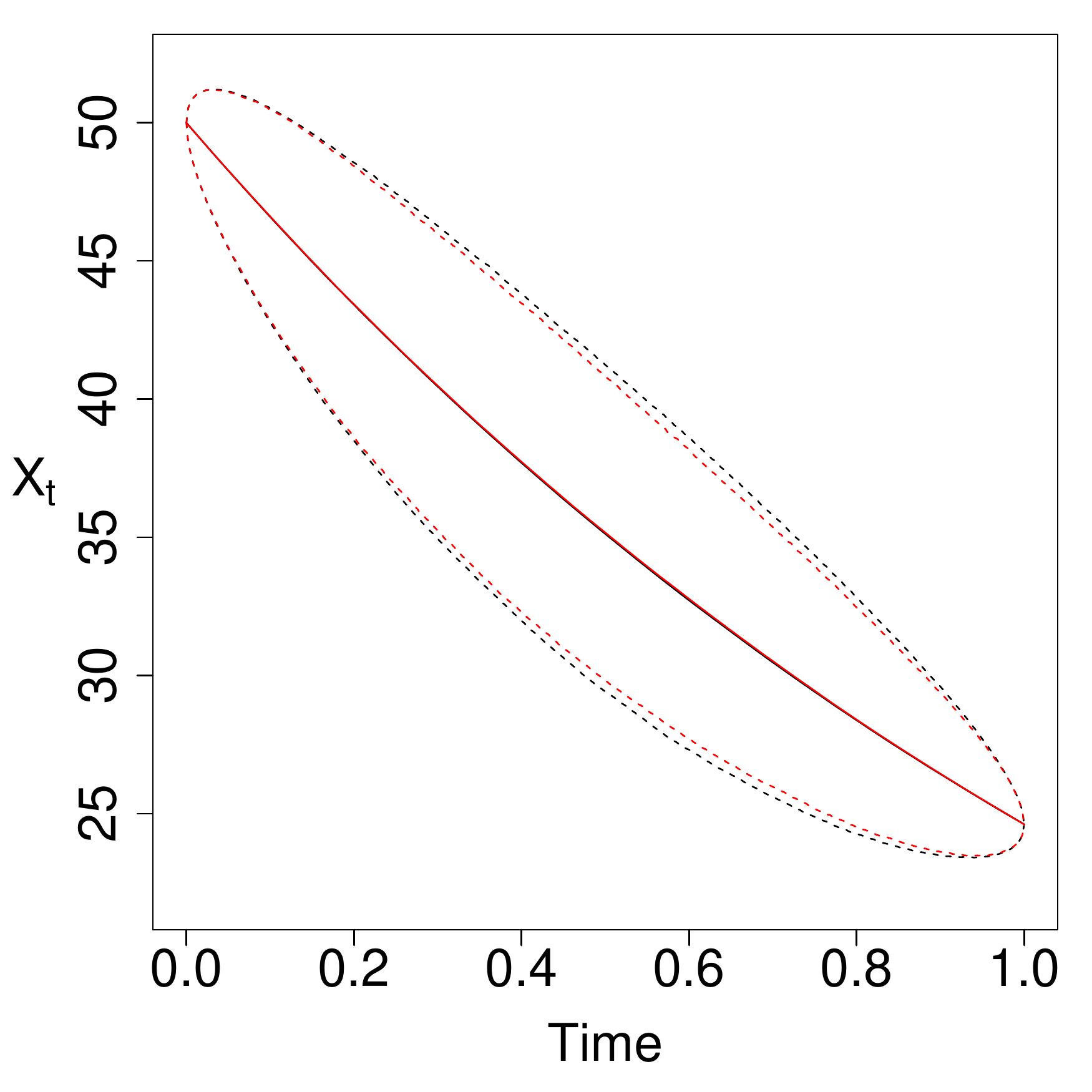}
\end{minipage}
\begin{minipage}[b]{0.32\linewidth}
				\centering
				\caption*{\qquad LB, $\gamma=0.0025$}\vspace{-0.25cm}
        \includegraphics[scale=0.285]{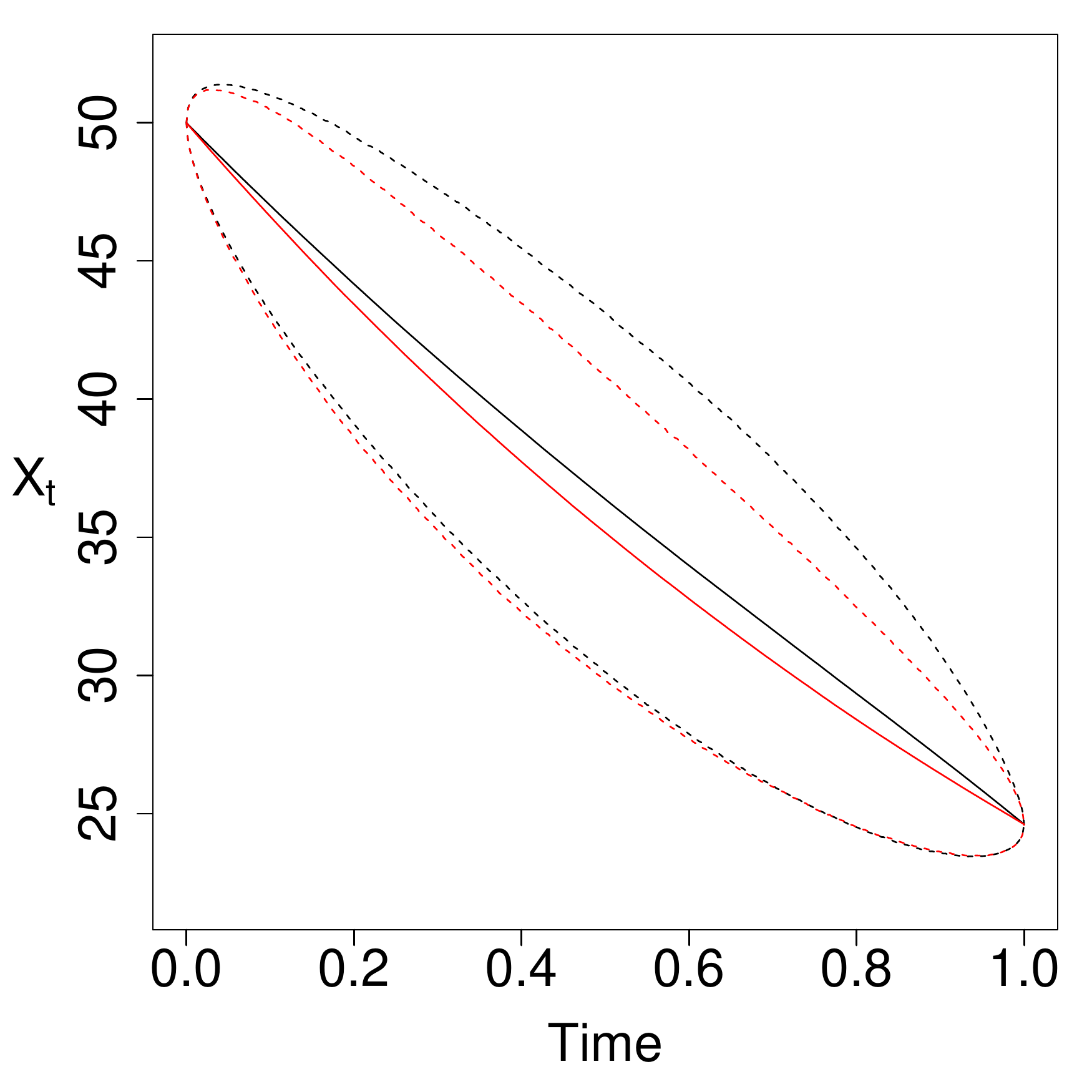}
\end{minipage}\\
\vspace{0.2cm} 
\begin{minipage}[b]{0.32\linewidth}
				\centering
				\caption*{\qquad GP-N}\vspace{-0.25cm}
        \includegraphics[scale=0.285]{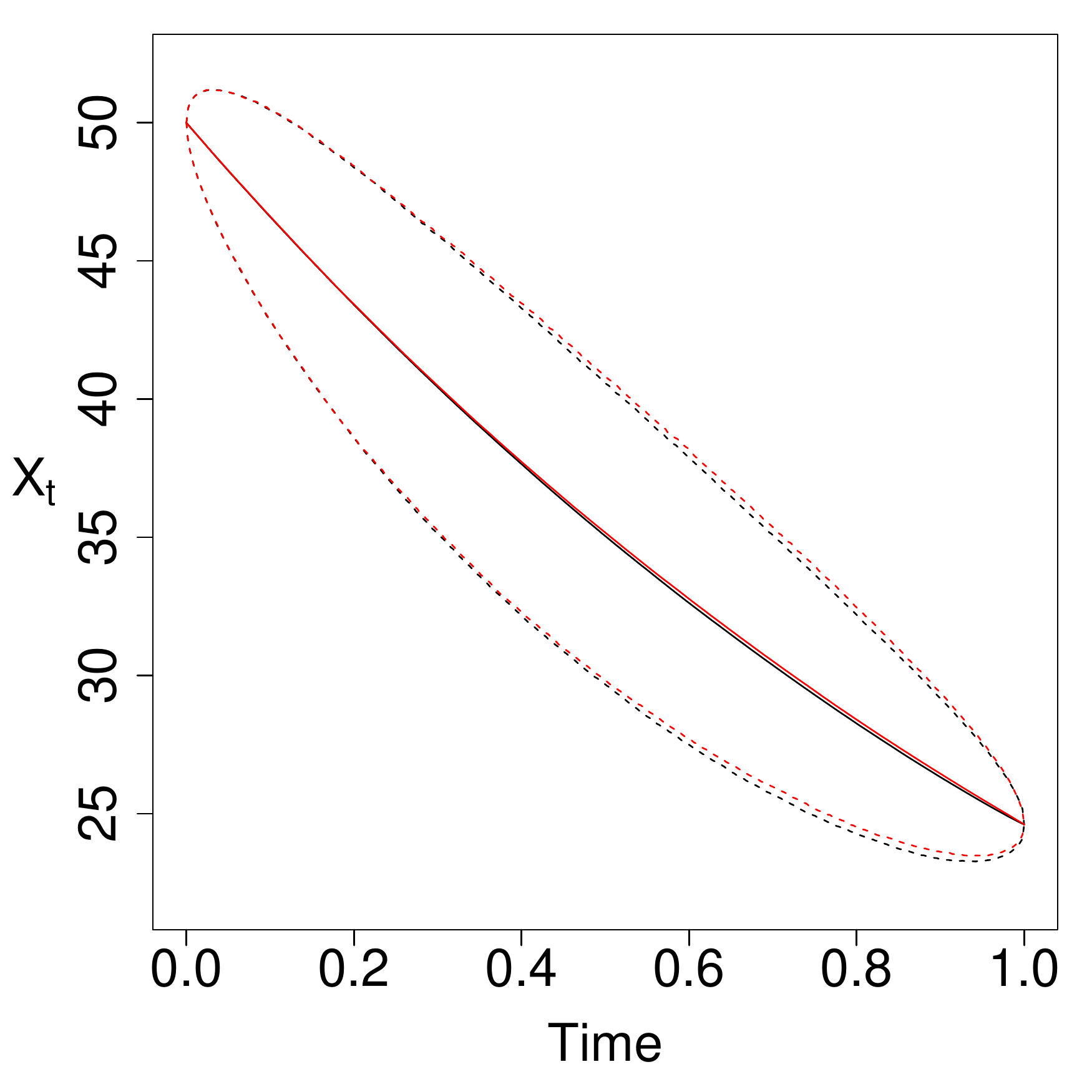}
\end{minipage}
\begin{minipage}[b]{0.32\linewidth}
        \centering
				\caption*{\qquad GP-S}\vspace{-0.25cm}
        \includegraphics[scale=0.285]{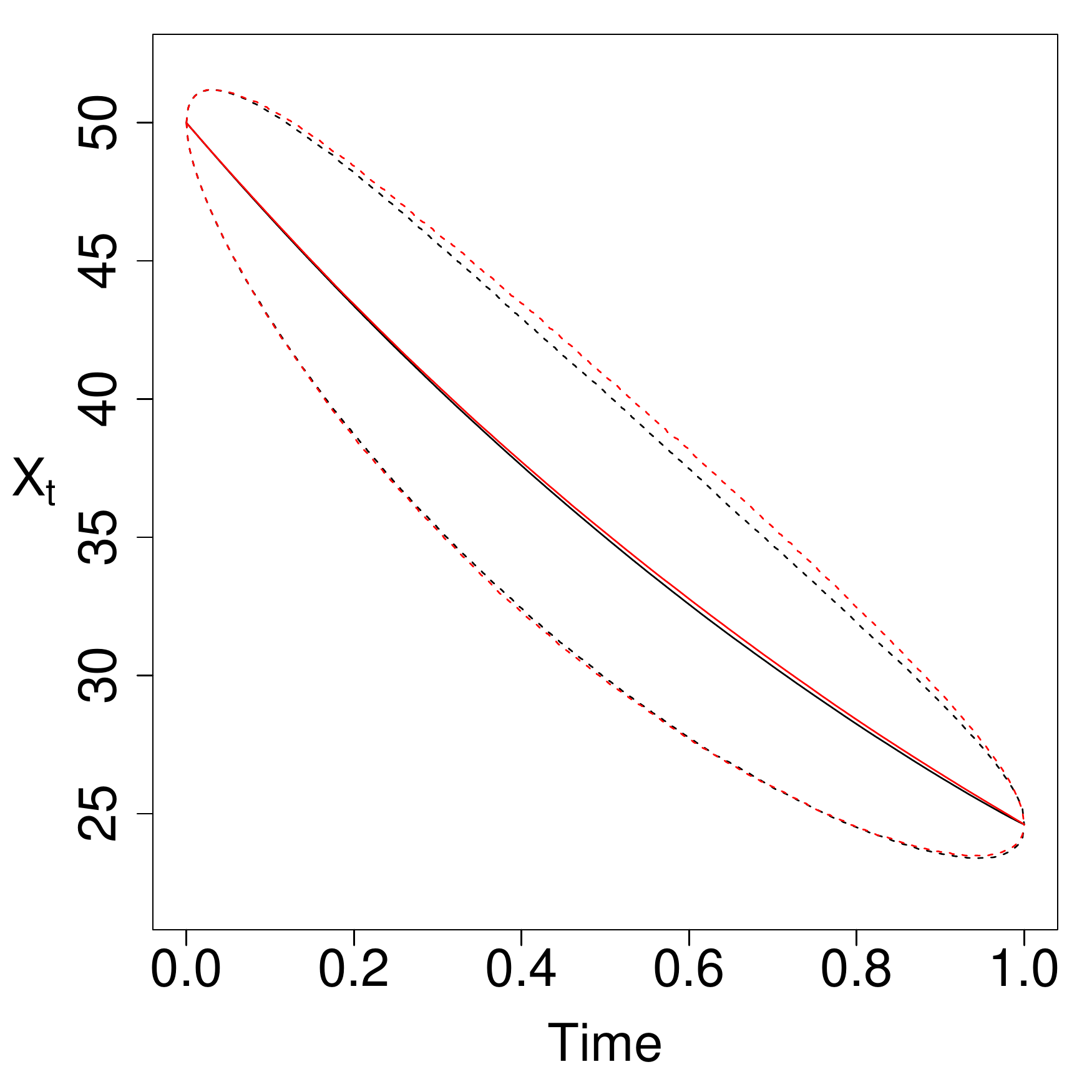}
\end{minipage}
\begin{minipage}[b]{0.32\linewidth}
        \centering
				\caption*{\qquad GP}\vspace{-0.25cm}
        \includegraphics[scale=0.285]{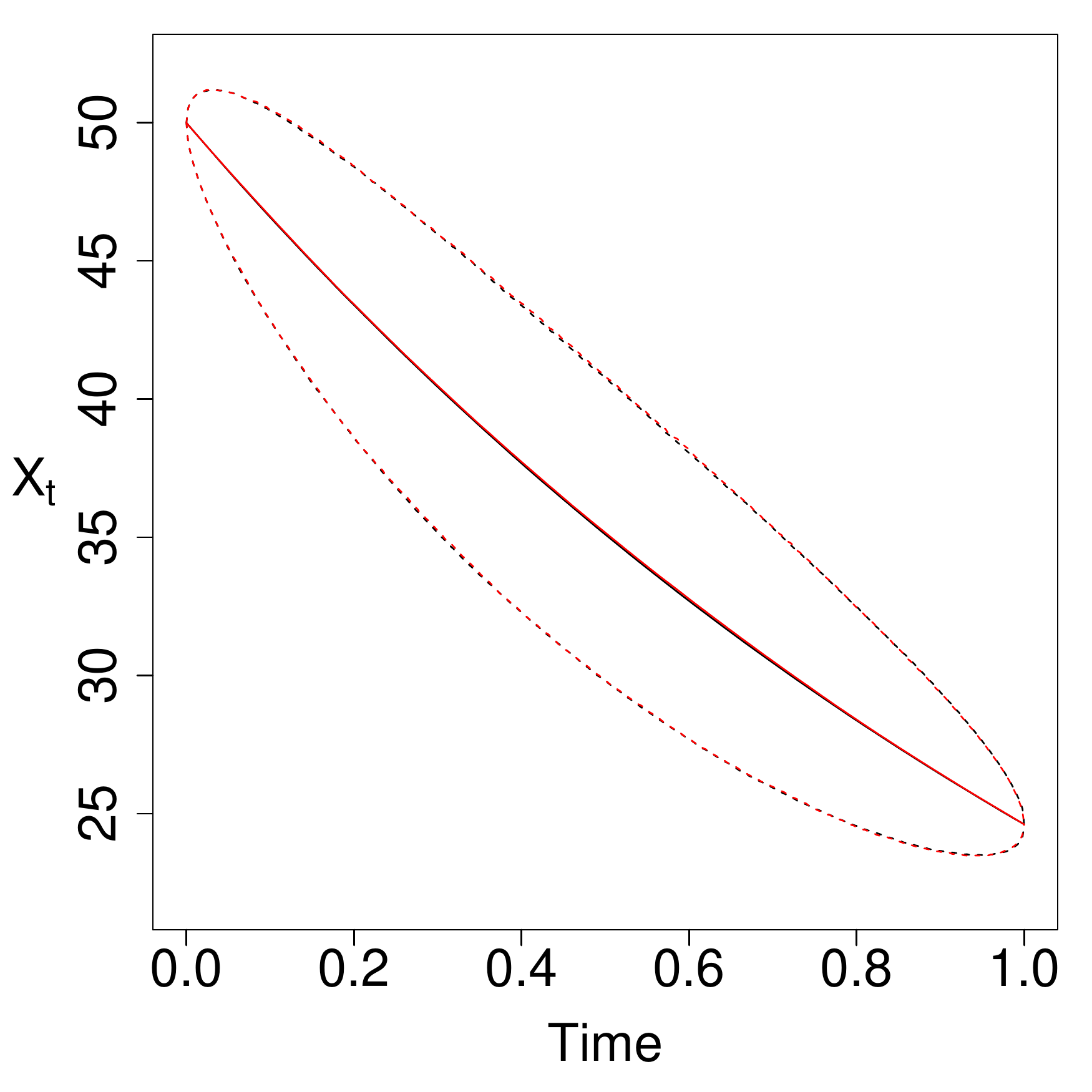}
\end{minipage}
\caption{Birth-death model. 95\% credible region (dashed line) and mean (solid line) of the true conditioned 
process (red) and various bridge constructs (black) using $x_T=x_{1,(50)}$.} \label{bd fig_prop_t1}
\end{center}
\end{figure}

\begin{figure}[t]
\begin{center}
\begin{minipage}[b]{0.32\linewidth}
        \centering
				\caption*{\qquad MDB}\vspace{-0.25cm}
        \includegraphics[scale=0.285]{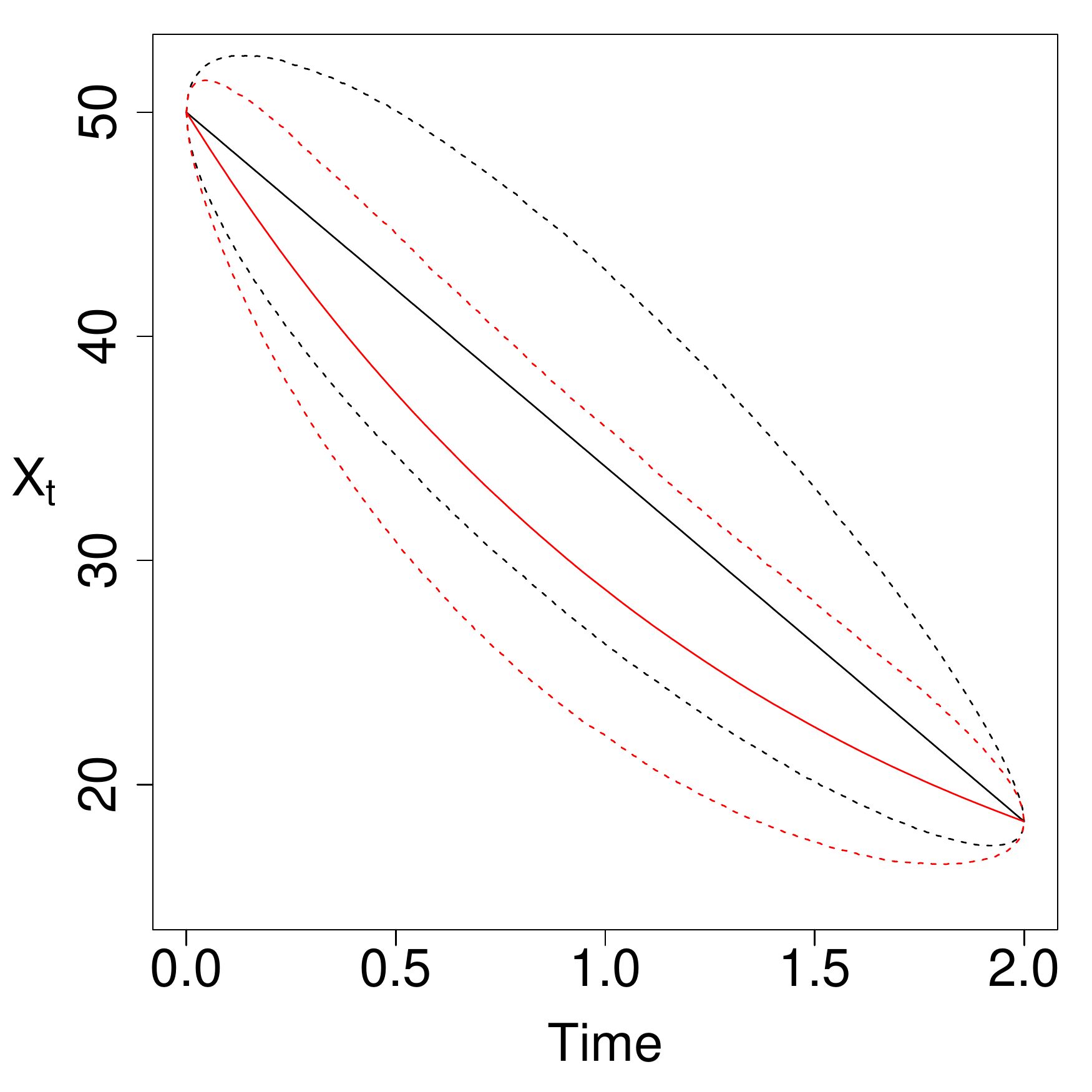}
\end{minipage} 
\begin{minipage}[b]{0.32\linewidth}
        \centering
				\caption*{\qquad RB}\vspace{-0.25cm}
        \includegraphics[scale=0.285]{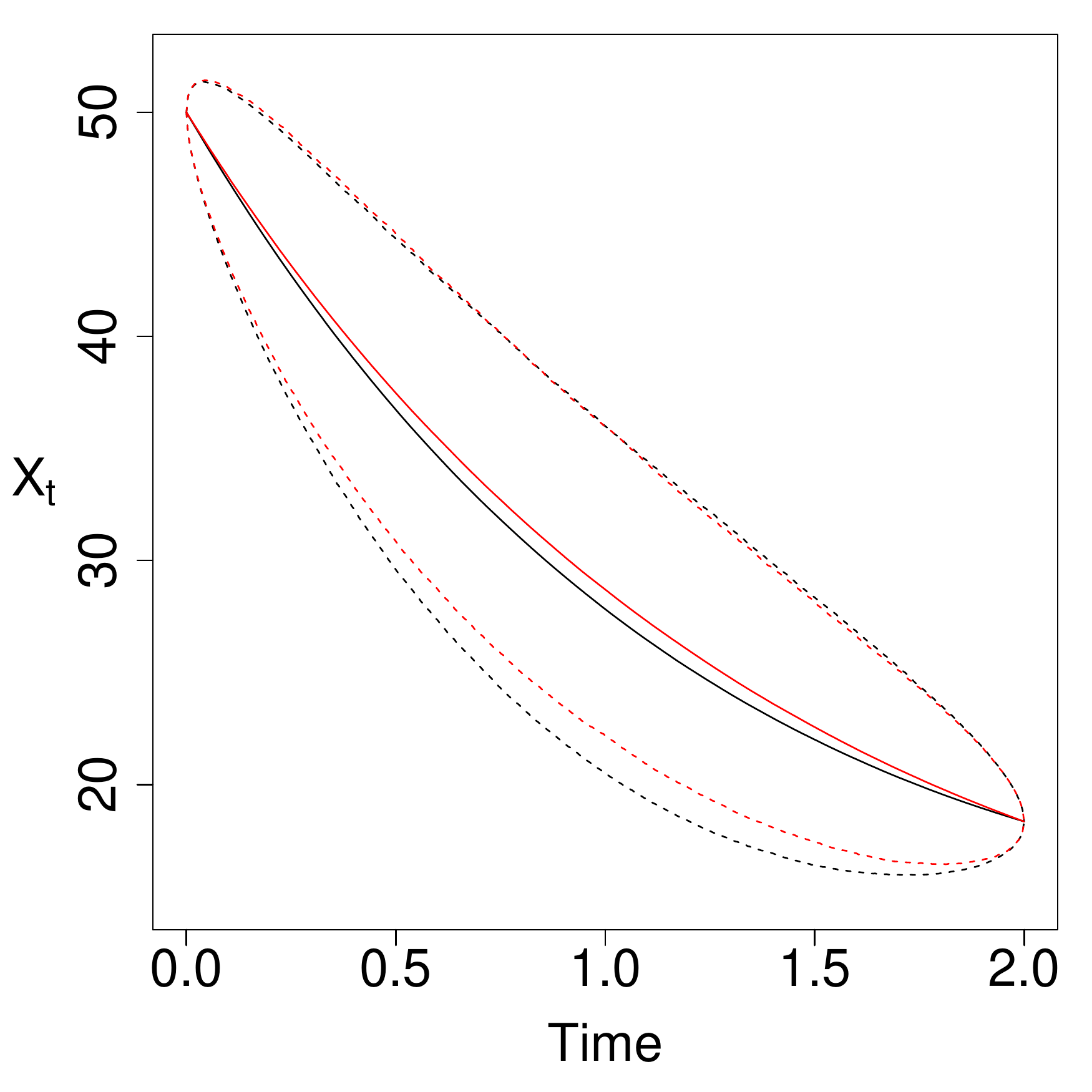}
\end{minipage} 
\begin{minipage}[b]{0.32\linewidth}
				\centering
				\caption*{\qquad LB, $\gamma=0.0025$}\vspace{-0.25cm}
        \includegraphics[scale=0.285]{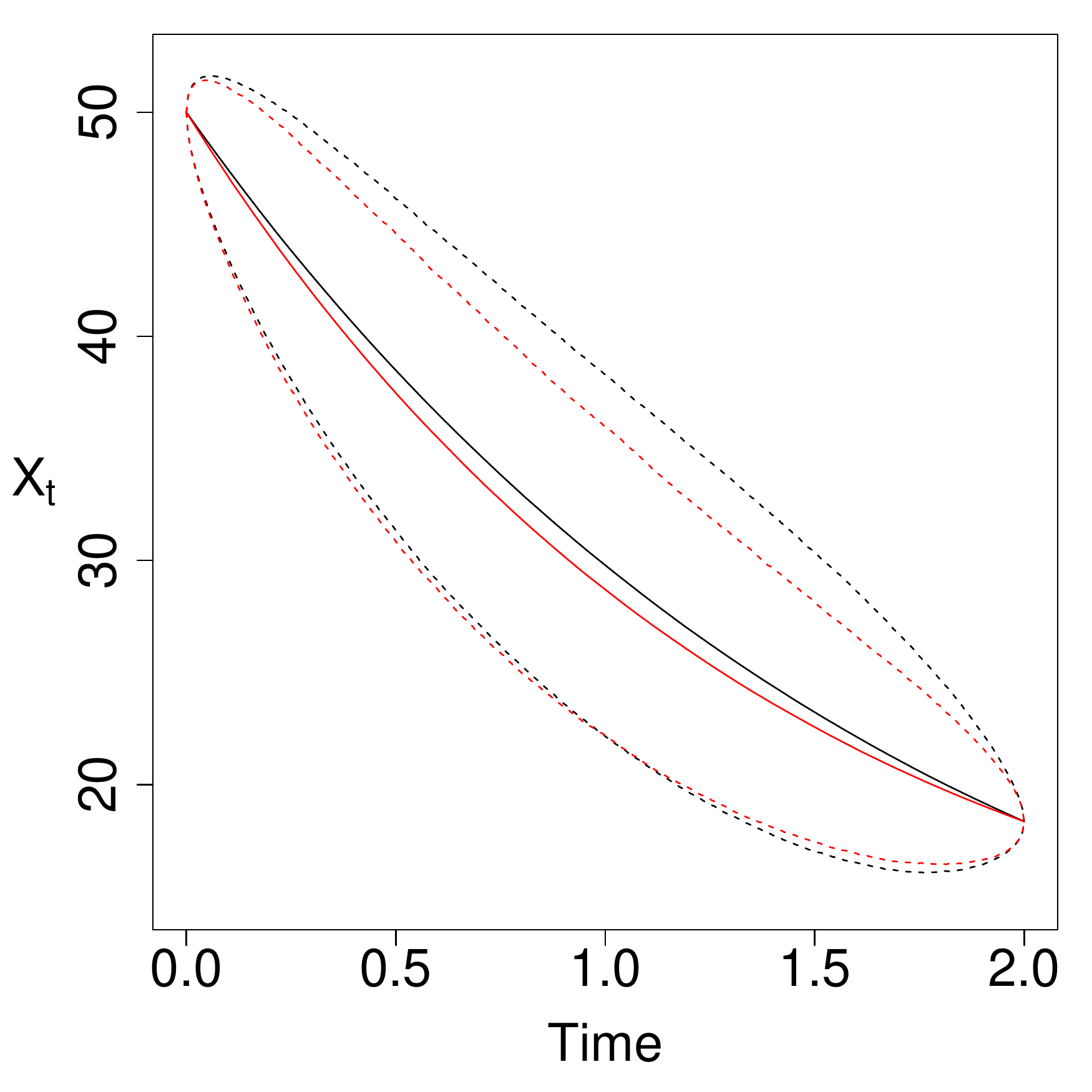}
\end{minipage}\\
\vspace{0.2cm} 
\begin{minipage}[b]{0.32\linewidth}
				\centering
				\caption*{\qquad GP-N}\vspace{-0.25cm}
        \includegraphics[scale=0.285]{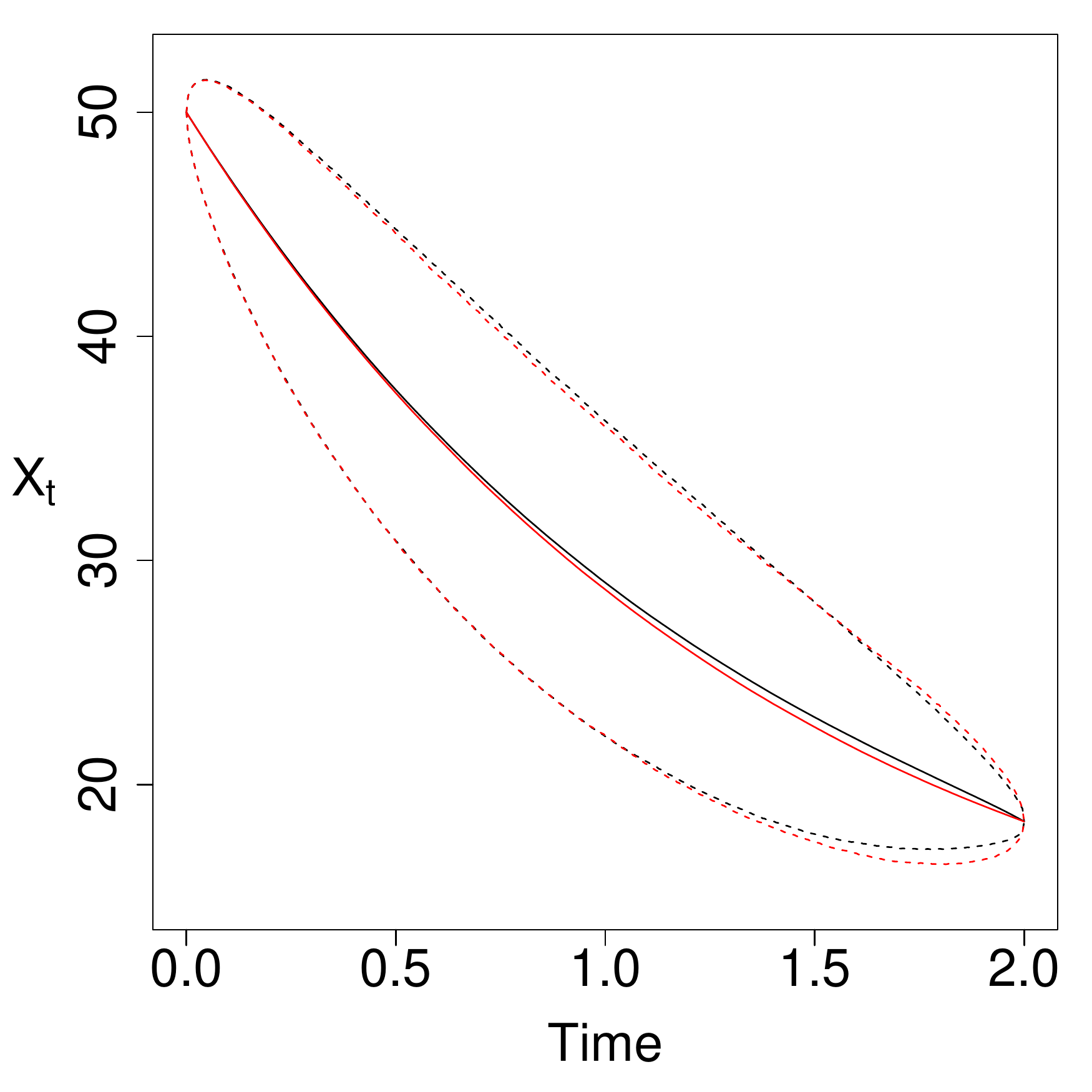}
\end{minipage}
\begin{minipage}[b]{0.32\linewidth}
        \centering
				\caption*{\qquad GP-S}\vspace{-0.25cm}
        \includegraphics[scale=0.285]{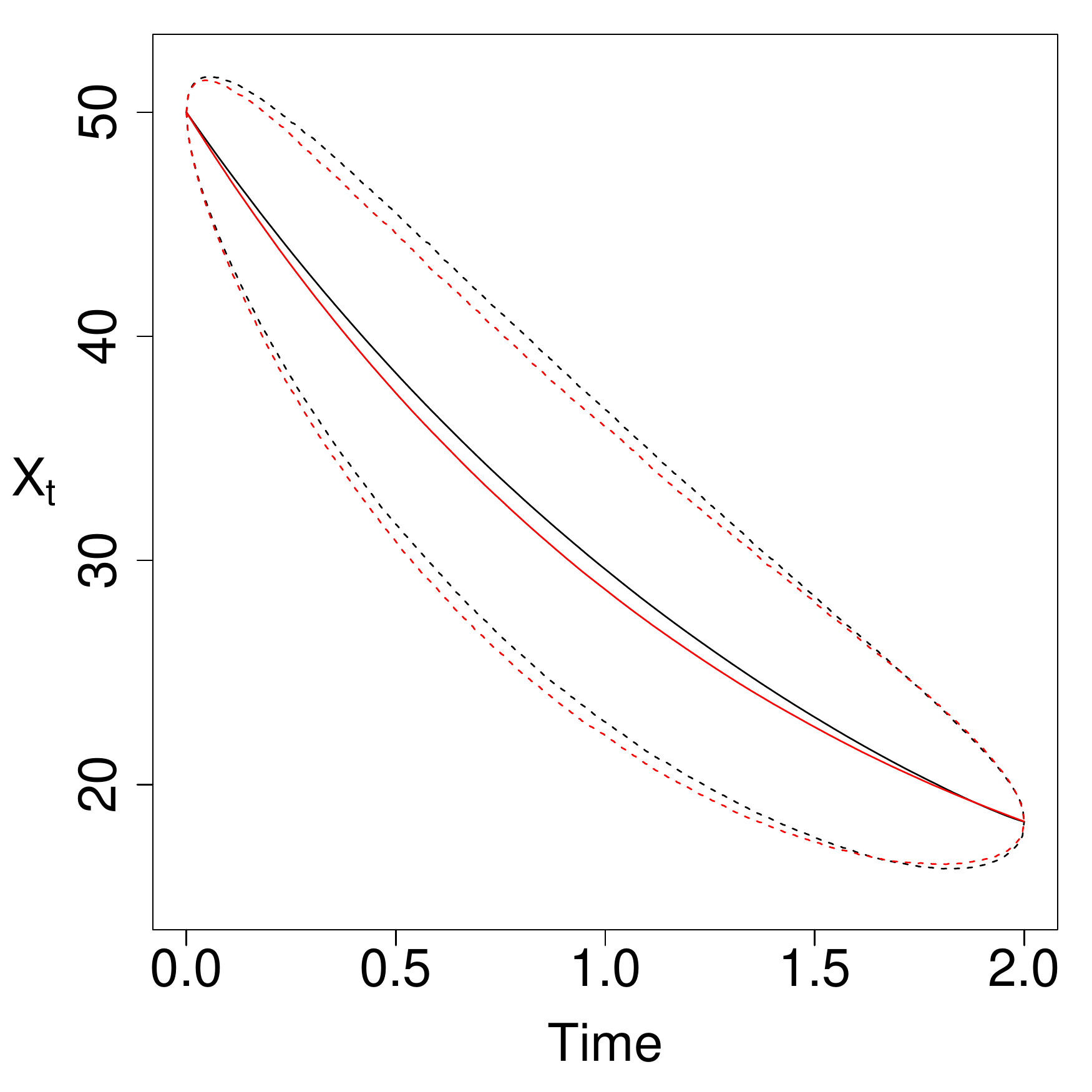}
\end{minipage}
\begin{minipage}[b]{0.32\linewidth}
        \centering
				\caption*{\qquad GP}\vspace{-0.25cm}
        \includegraphics[scale=0.285]{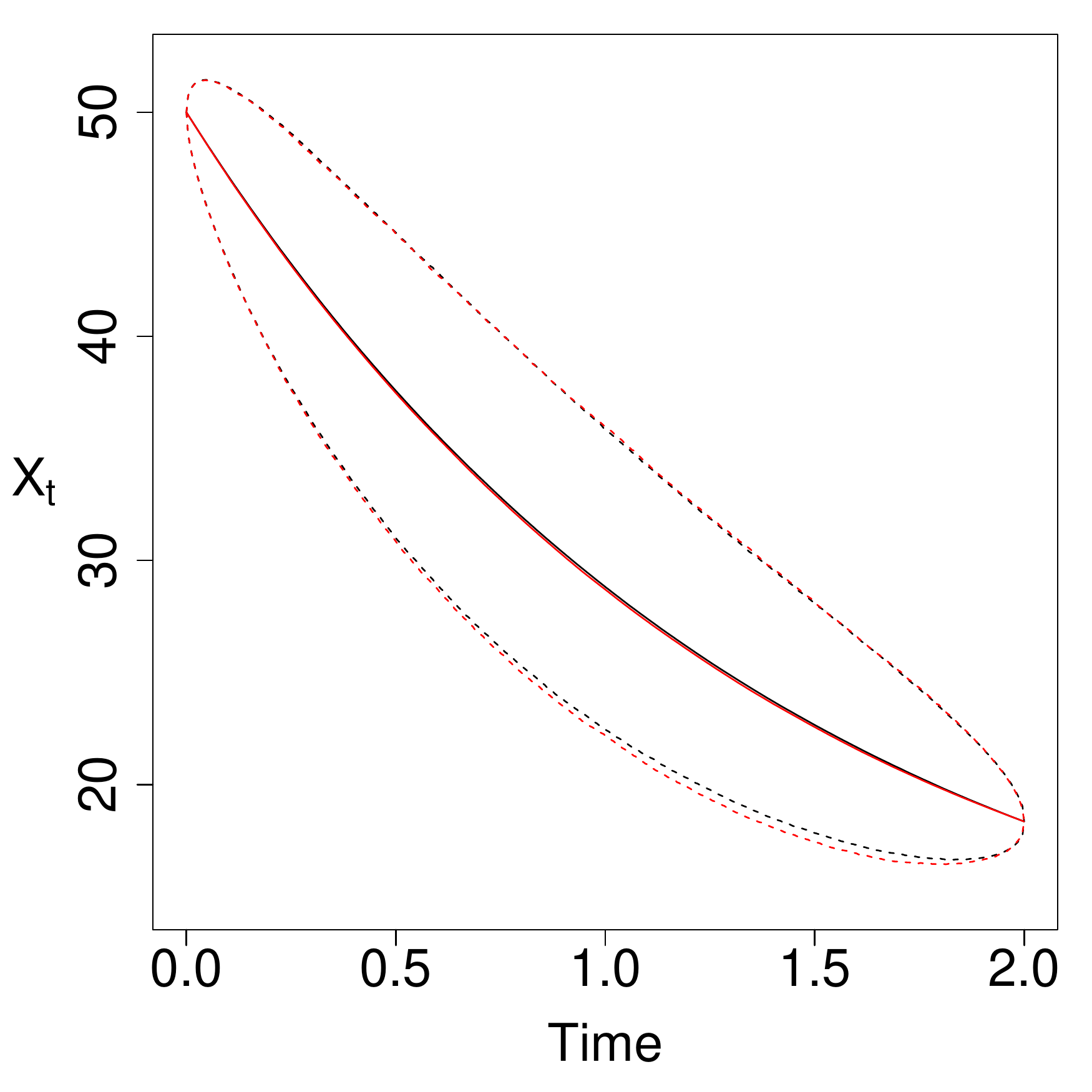}
\end{minipage}
\caption{Birth-death model. 95\% credible region (dashed line) and mean (solid line) of the true conditioned 
process (red) and various bridge constructs (black) using $x_T=x_{2,(95)}$.}\label{bd fig_prop_t2}
\end{center}
\end{figure}

Figure~\ref{bd fig_ap} shows empirical acceptance 
probabilities against the number of 
sub-intervals $m$ for each bridge and each $x_T$. 
Figures~\ref{bd fig_prop_t1} and \ref{bd fig_prop_t2} compare 
95\% credible regions of the proposal under various bridging strategies 
with the true conditioned process (obtained from the output of 
the Metropolis-Hastings independence sampler). It is clear from the 
figures that as $T$ is increased, the MDB fails to 
adequately account for the nonlinear behaviour of the 
conditioned process. Indeed, in terms of empirical 
acceptance rate, MDB is outperformed by all other 
bridges for $T=2$. As $m$ is increased so that the 
discretisation gets finer, the acceptance rates under 
all bridges (with the exception of GP-N) stay roughly 
constant. For GP-N, the acceptance rates decrease with 
$m$ when $x_T$ is either the $5\%$ or $95\%$ quantile 
of $X_T|X_0=50$. In this case, the variance associated 
with the approximate transition 
density either overestimates (when $x_T$ is the $5\%$ quantile) 
or underestimates (when $x_T$ is the $95\%$ quantile) the 
true variance at the end-point. For example, when $x_T$ is the 
$95\%$ quantile, this results (see Figure~\ref{bd fig_prop_t2}) 
in a `tapering in' of the proposal relative to 
the true conditioned process. \hbox{GP-S}, GP and LB give 
similar performance, although we note that GP-S and LB 
perform particularly poorly when $x_T$ is the $5\%$ quantile. 
Moreover, LB requires the specification of a tuning parameter 
$\gamma$ and we found that the acceptance rate was fairly 
sensitive to the choice of $\gamma$. In all scenarios, 
RB, RB$^-$ and GP-MDB comprehensively outperform all other 
bridge constructs. When $x_T$ is the median of $X_T|X_0=50$, 
we see that RB and RB$^-$ (red and blue lines in 
Figure~\ref{bd fig_ap}) give near identical performance, 
with $\eta_t$ adequately accounting for the observed 
nonlinear dynamics. In terms of statistical efficiency, 
GP-MDB outperforms both RB and RB$^-$ in all scenarios, 
although the relative difference is small. 

\subsection{Lotka-Volterra}\label{lv}

In this example we consider a Lotka-Volterra model of pred\-ator-prey 
dynamics. We denote the system state at time $t$ by $X_t=(X_{1,t},X_{2,t})'$, 
ordered as prey, predators. The mass-action SDE representation of system 
dynamics takes the form  
%\begin{eqnarray*}
%\textrm{\bf{Reaction 1:}} & \mathcal{X}_1 & \overset{\theta_1}\rightarrow \quad 2\mathcal{X}_1 \\
%\textrm{\bf{Reaction 2:}} & \mathcal{X}_1+\mathcal{X}_2 & \overset{\theta_2}\rightarrow \quad 2\mathcal{X}_2 \\
%\textrm{\bf{Reaction 3:}} & \mathcal{X}_2 & \overset{\theta_3}\rightarrow \quad \emptyset.
%\end{eqnarray*}
\begin{equation}\label{eqn:lv}
dX_t= \begin{pmatrix}
\theta_1X_{1,t}-\theta_2X_{1,t}X_{2,t} \\
\theta_2X_{1,t}X_{2,t}-\theta_3X_{2,t} \\ \end{pmatrix} dt 
+ \begin{pmatrix}
\theta_1X_{1,t}+\theta_2X_{1,t}X_{2,t} &  -\theta_2X_{1,t}X_{2,t}  \\
-\theta_2X_{1,t}X_{2,t} & \theta_3X_{2,t}+\theta_2X_{1,t}X_{2,t} \\ \end{pmatrix}^{\frac{1}{2}}dW_t. 
\end{equation}

\begin{table}[!h]
\begin{center}
\begin{tabular}{c|cccc}
	\hline
   & $T=1$ & $T=2$ & $T=3$ & $T=4$ \\
  \hline
  $x_{T,(5)}$ & (82.47,62.78) & (107.35,57.95) & (142.00,60.02) & (185.04,71.23) \\
  $x_{T,(50)}$ & (96.82,71.93) & (133.35,70.75) & (182.64,77.36) & (242.08,97.23) \\ 
	$x_{T,(95)}$ & (112.13,81.58) & (162.28,84.63) & (228.82,97.12) & (308.58,128.76) \\
	\hline
\end{tabular}
\caption{Lotka-Volterra model. Quantiles of $X_T|X_0=(71,79)'$ found by repeatedly simulating from the Euler-Maruyama 
approximation of (\ref{eqn:lv}) with $\theta=(0.5,0.0025,0.3)'$.} \label{tab lvdata}
\end{center}
\end{table}
\noindent The components of $\theta=(\theta_1,\theta_2,\theta_3)'$ can be interpreted as 
prey reproduction rate, prey death and predator reproduction rate, and predator 
death. Note that the ODE system (\eqref{eqn:dgode_eta}, (\ref{eqn:lna1}) and (\ref{eqn:lna2})) 
governing the linear noise approximation of (\ref{eqn:lv}) is intractable 
and we therefore use the \verb+R+ package \verb+lsoda+ to numerically solve the system 
when necessary.
%\[
%H(\eta_t)= \begin{pmatrix}
%\theta_1 - \theta_2\eta_{2,t}  &  -\theta_2\eta_{1,t} \\
%\theta_2\eta_{2,t}  &  \theta_2\eta_{1,t} - \theta_3 \end{pmatrix}.
%\]

Following \cite{Boys_Wilkinson_Kirkwood_2008} we impose the parameter values 
$\theta=(\theta_1,\theta_2,\theta_3)'=(0.5,0.0025,0.3)'$ and let $x_0=(71,79)'$. 
We assume that $x_T$ is known and generate a number of challenging scenarios 
by taking $x_T$ as either the 5\%, 50\% or 95\% marginal quantiles of $X_T|X_0=(71,79)'$ 
for $T\in\{1,2,3,4\}$. These quantiles are shown in Table~\ref{tab lvdata}. Note that 
for this parameter choice, the expectation of $X_t|X_0=(71,79)'$ is approximately 
periodic with a period around 17.

\begin{comment}
\begin{figure}
\begin{center}
				\centering
        \includegraphics[scale=0.37]{10augpaper2_lv_data.pdf}
\caption{Simulated data for a Lotka-Volterra model with $\theta_1=0.5$, $\theta_2=0.0025$, $\theta_3=0.3$ and $x_0=(71,79)'$ (denoted by a cross). The 5\%, 50\% and 95\% quantiles are represented by triangles, circles and diamonds respectively for times $t=1,2,3,4$. Black: Prey ($X_1$). Red: Predator ($X_2$).} \label{fig:lvdata}
\end{center}
\end{figure}
\end{comment}

\begin{figure}
\begin{center}
\begin{minipage}[b]{0.32\linewidth}
        \centering
				\caption*{\qquad $x_T=x_{T,(5)}$}\vspace{-0.25cm}
        \includegraphics[scale=0.285]{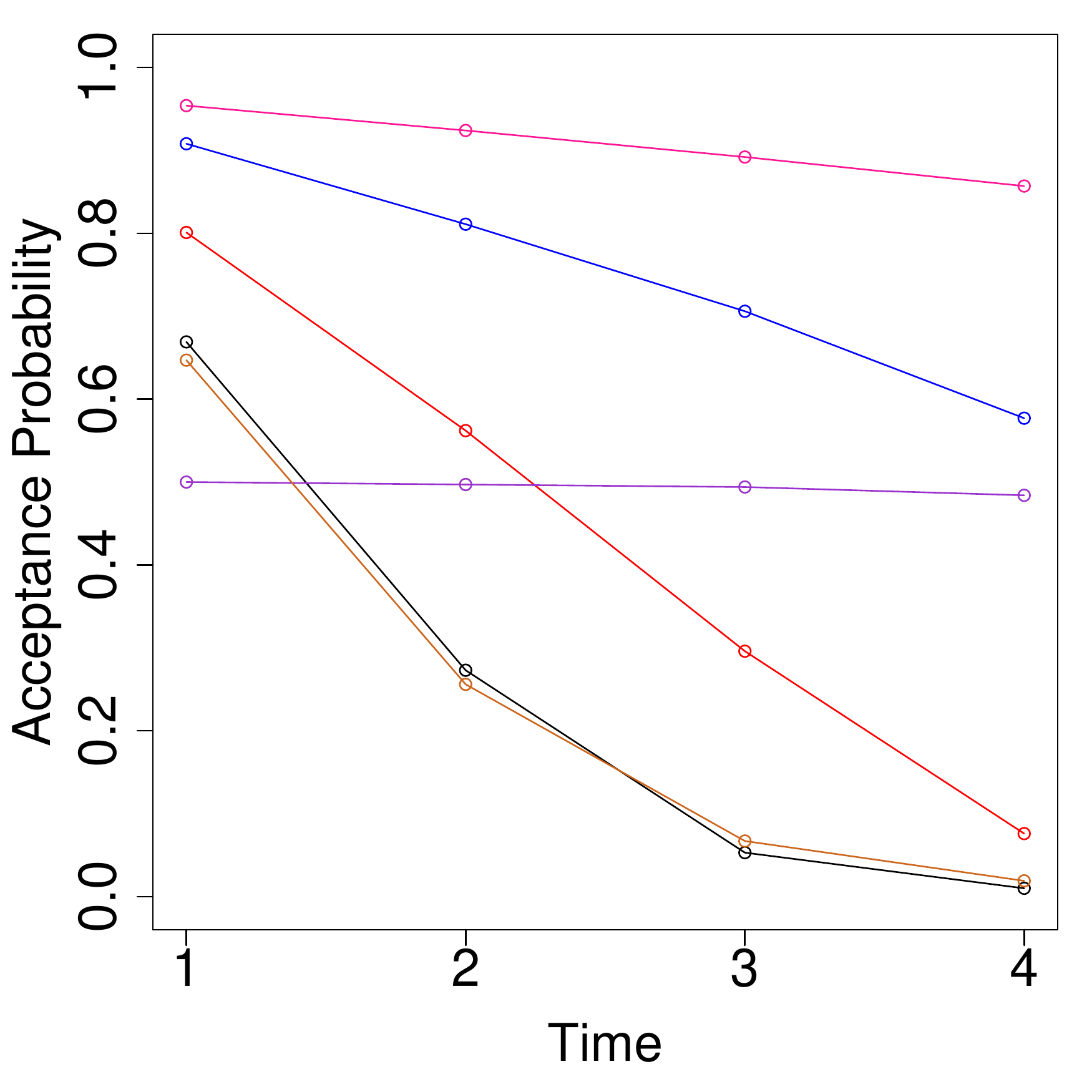}
\end{minipage} 
\begin{minipage}[b]{0.32\linewidth}
        \centering
				\caption*{\qquad $x_T=x_{T,(50)}$}\vspace{-0.25cm}
        \includegraphics[scale=0.285]{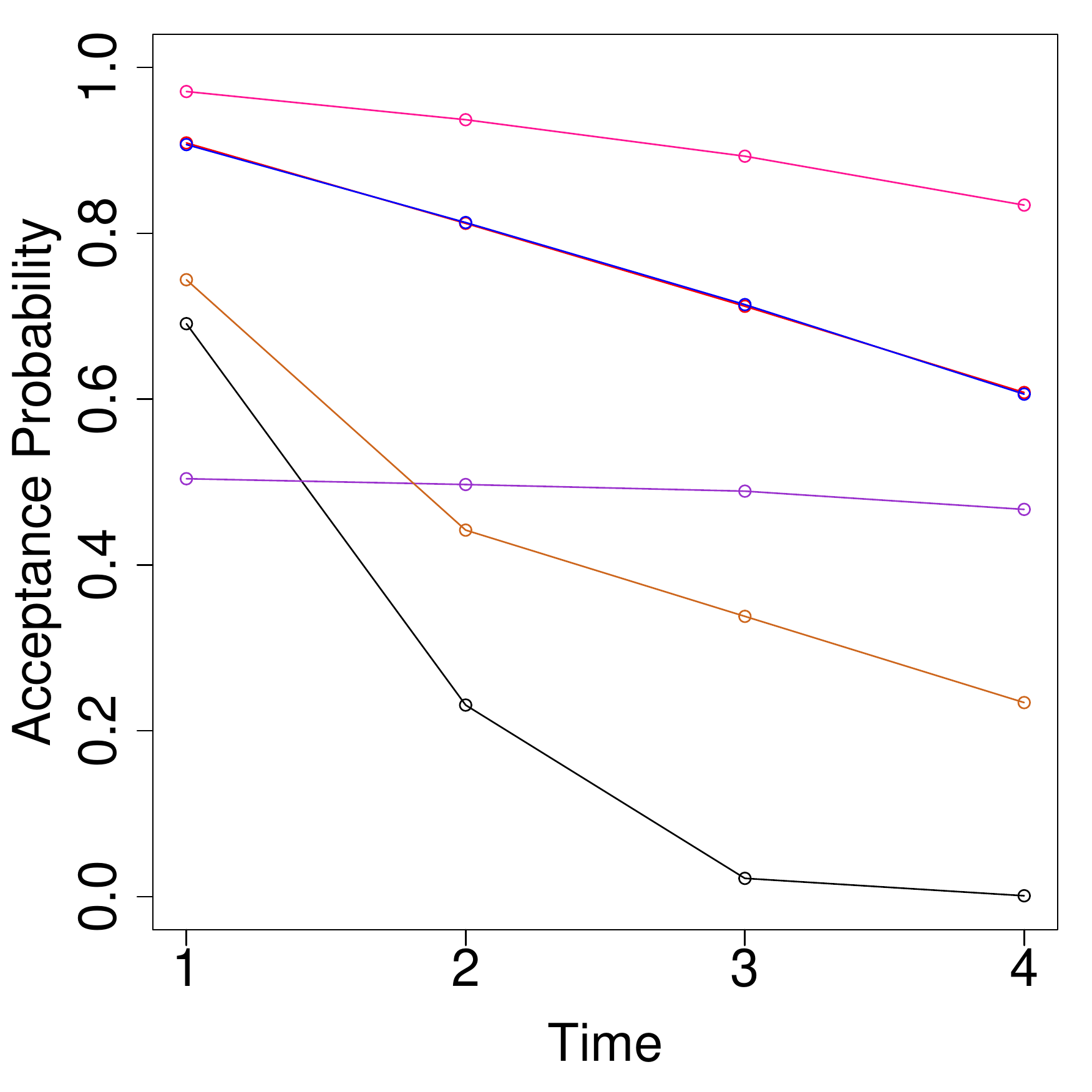}
\end{minipage} 
\begin{minipage}[b]{0.32\linewidth}
        \centering
				\caption*{\qquad $x_T=x_{T,(95)}$}\vspace{-0.25cm}
        \includegraphics[scale=0.285]{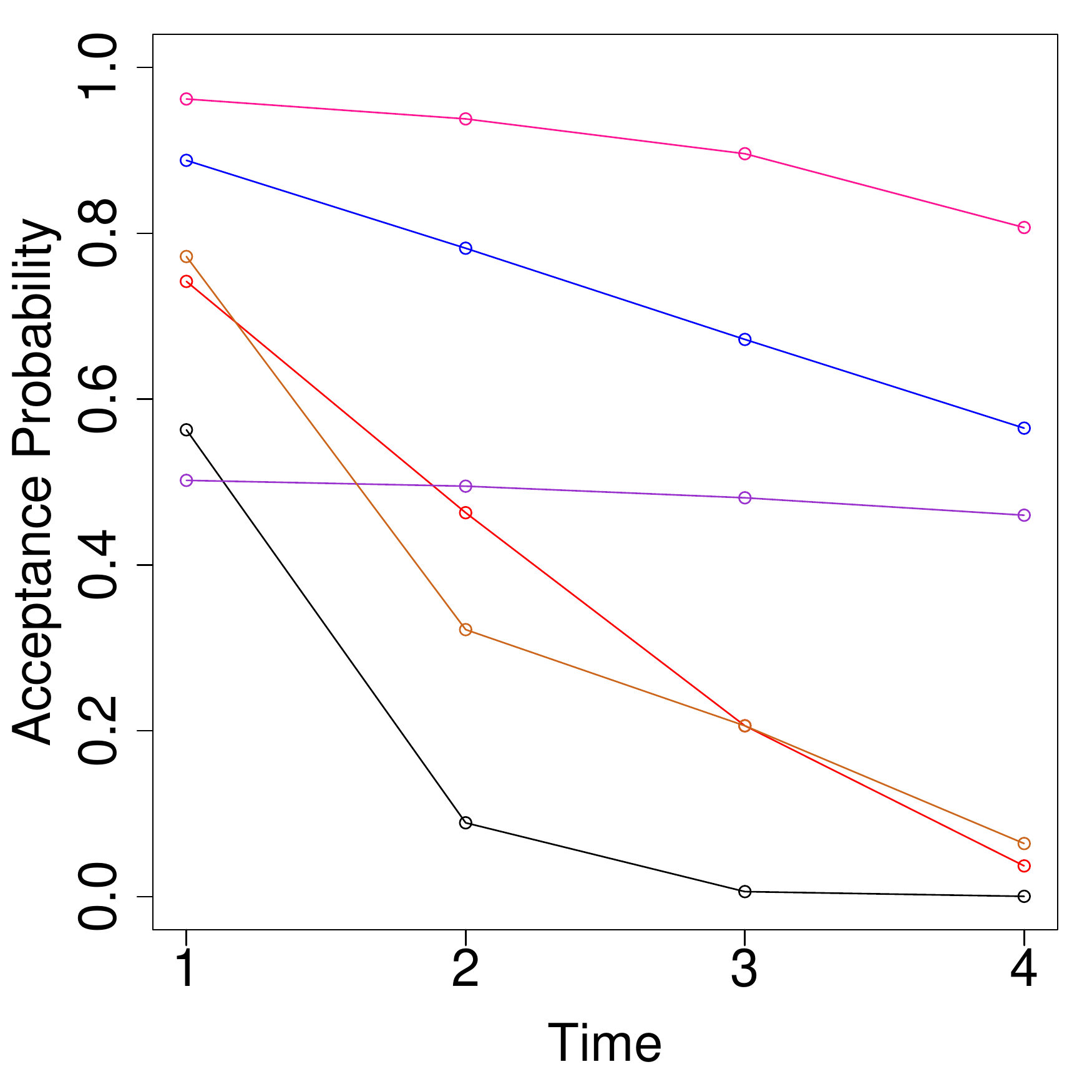}
\end{minipage} 
\caption{Lotka-Volterra model. Empirical acceptance probabilities against $T$. The results are based on 100K iterations 
of a Metropolis-Hastings independence sampler. Black: MDB. Brown:~LB. Red: RB. Blue: RB$^-$. Purple: GP. Pink: GP-MDB.} \label{lv fig_ap}
\end{center}
\end{figure}

\begin{figure}
\begin{center}
\begin{minipage}[b]{0.32\linewidth}
        \centering
				\caption*{\qquad RB}\vspace{-0.25cm}
        \includegraphics[scale=0.285]{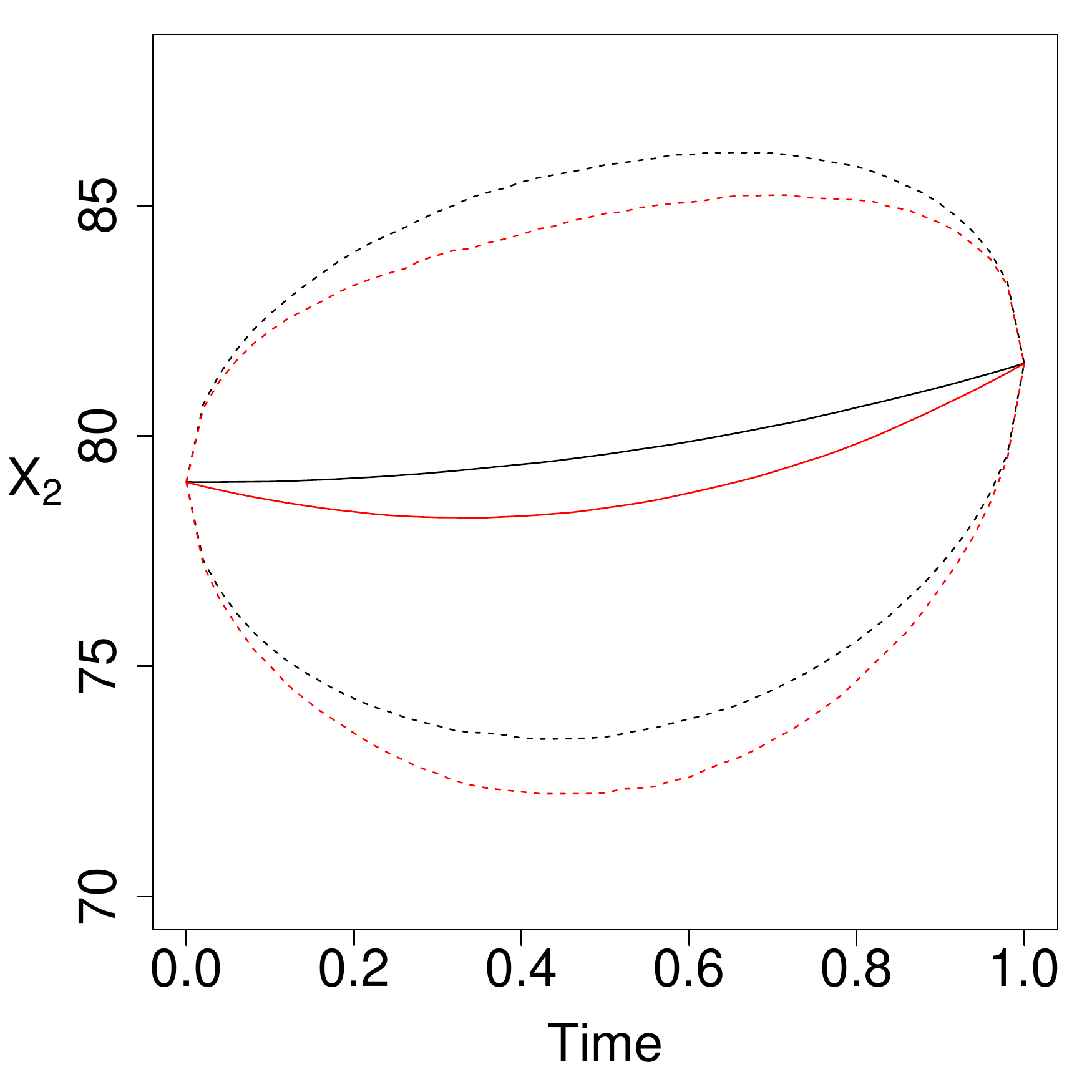}
\end{minipage} 
\begin{minipage}[b]{0.32\linewidth}
        \centering
				\caption*{\qquad RB$^-$}\vspace{-0.25cm}
        \includegraphics[scale=0.285]{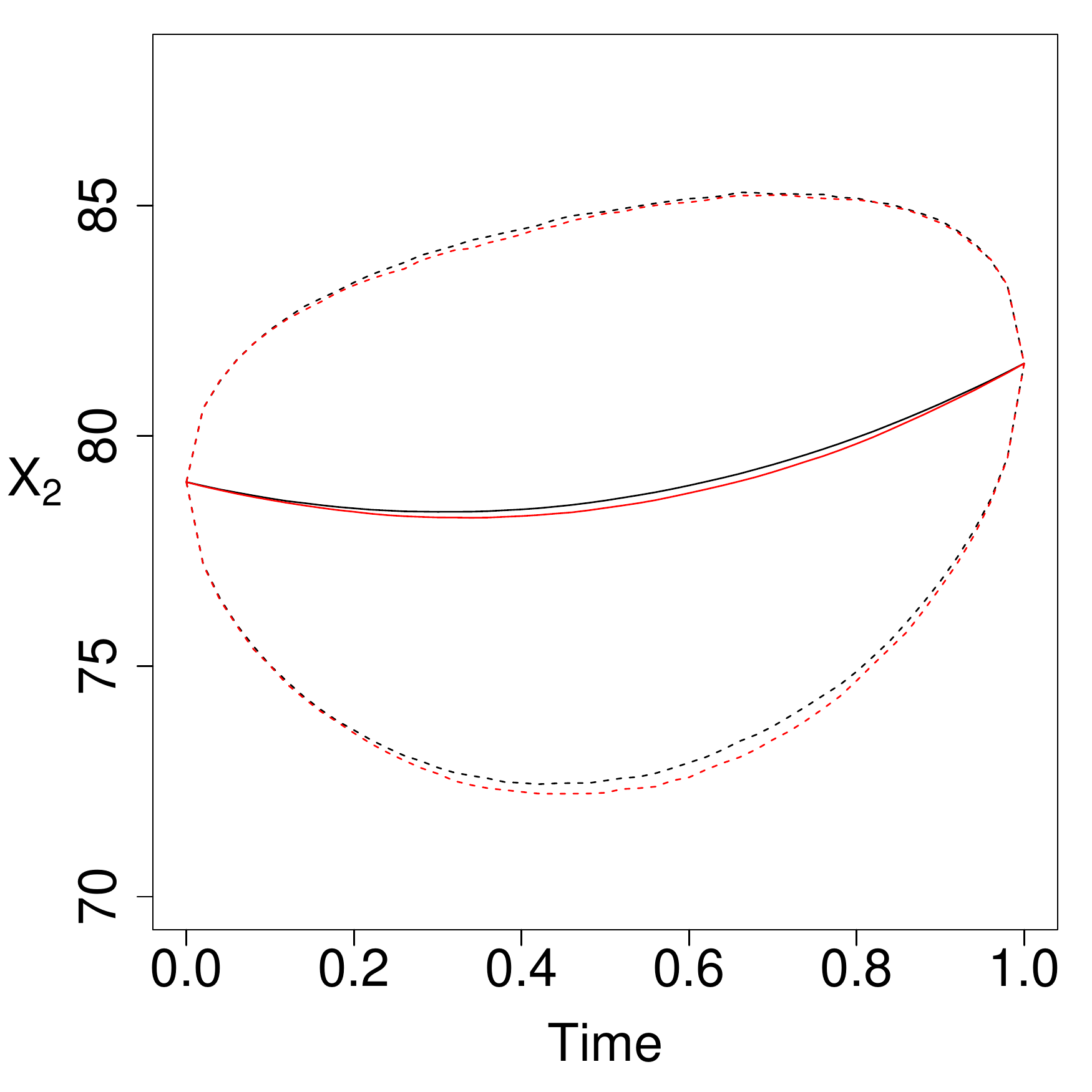}
\end{minipage}
\begin{minipage}[b]{0.32\linewidth}
				\centering
				\caption*{\qquad LB, $\gamma=0.01$}\vspace{-0.25cm}
        \includegraphics[scale=0.285]{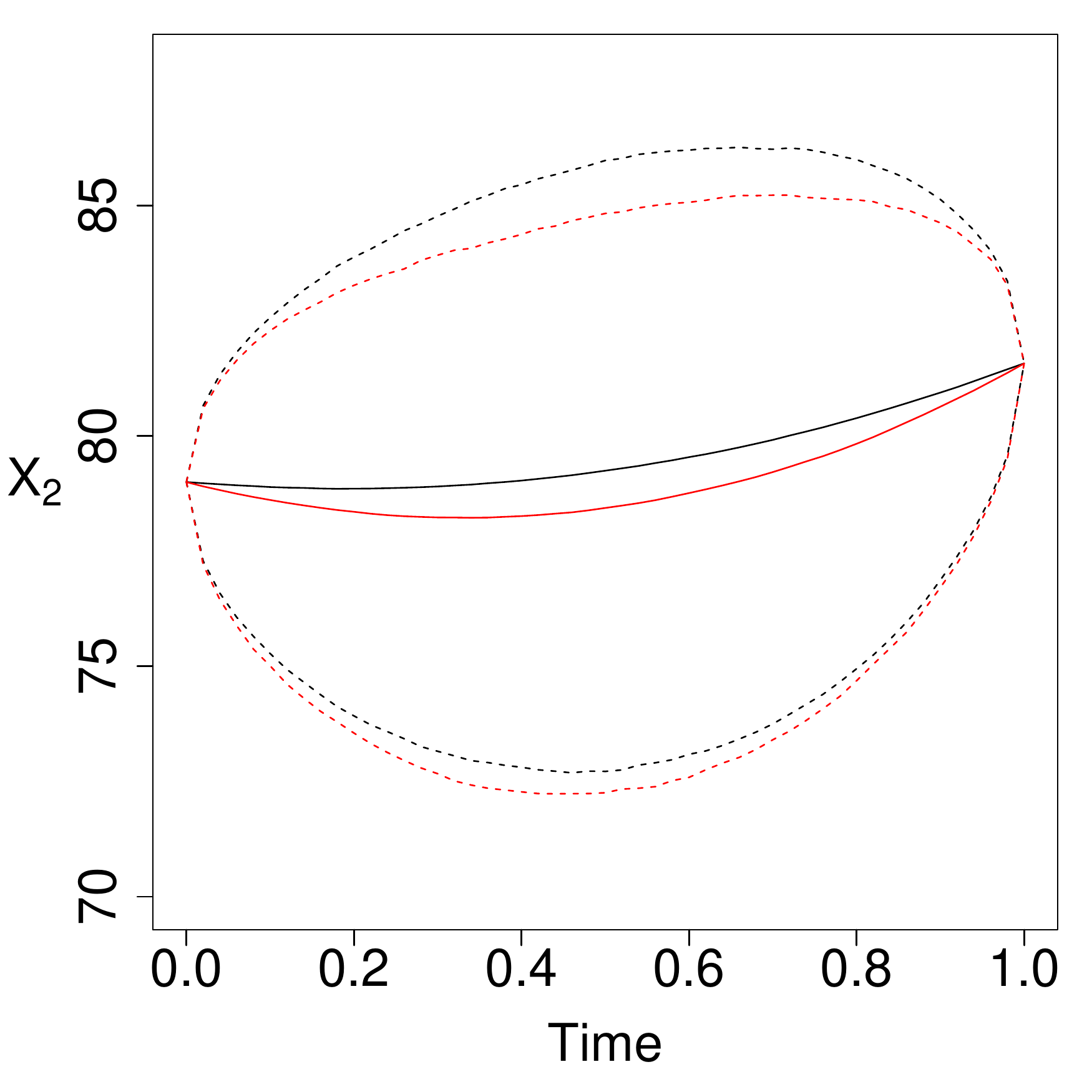}
\end{minipage}\\
\vspace{0.2cm} 
\begin{minipage}[b]{0.32\linewidth}
				\centering
        \includegraphics[scale=0.285]{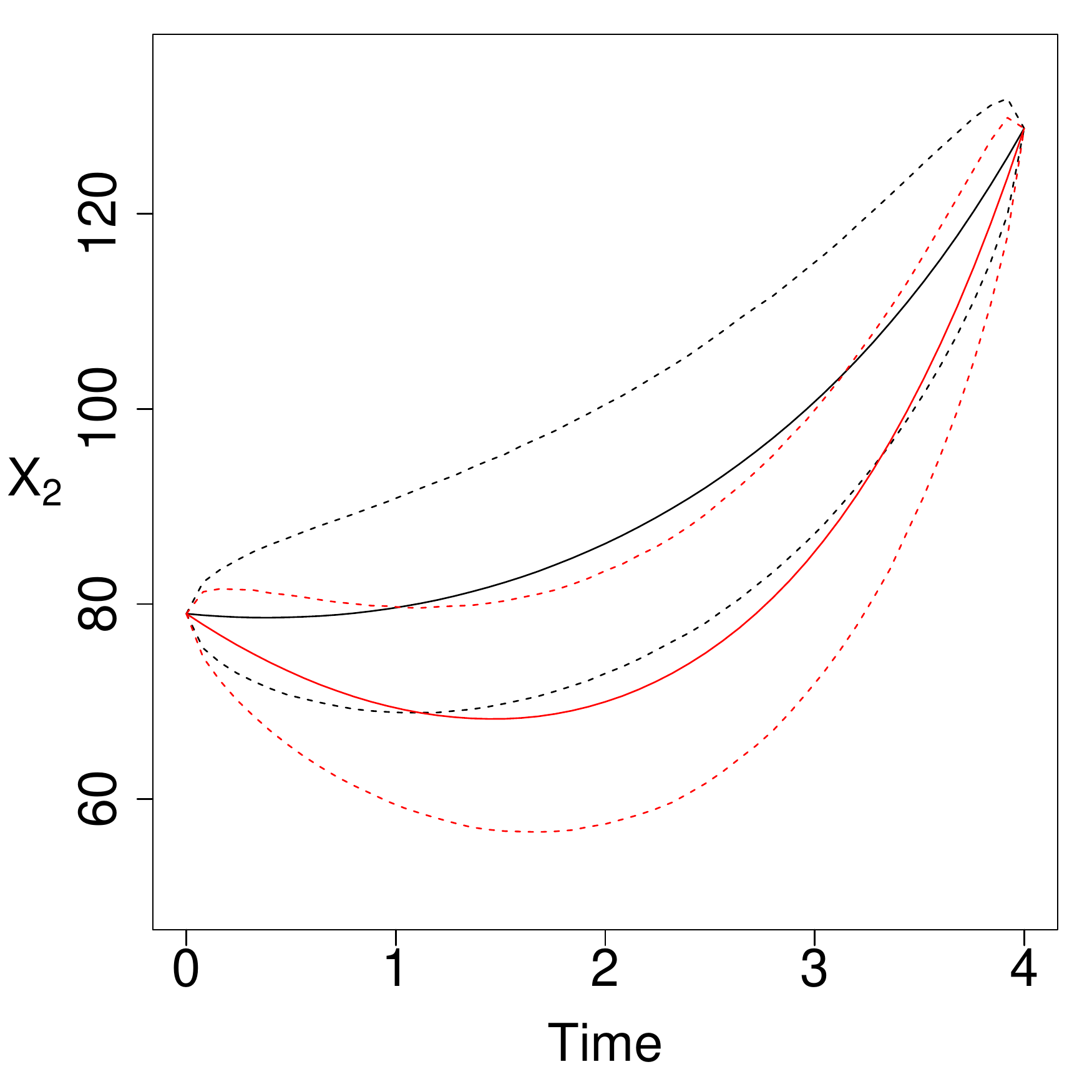}
\end{minipage}
\begin{minipage}[b]{0.32\linewidth}
        \centering
        \includegraphics[scale=0.285]{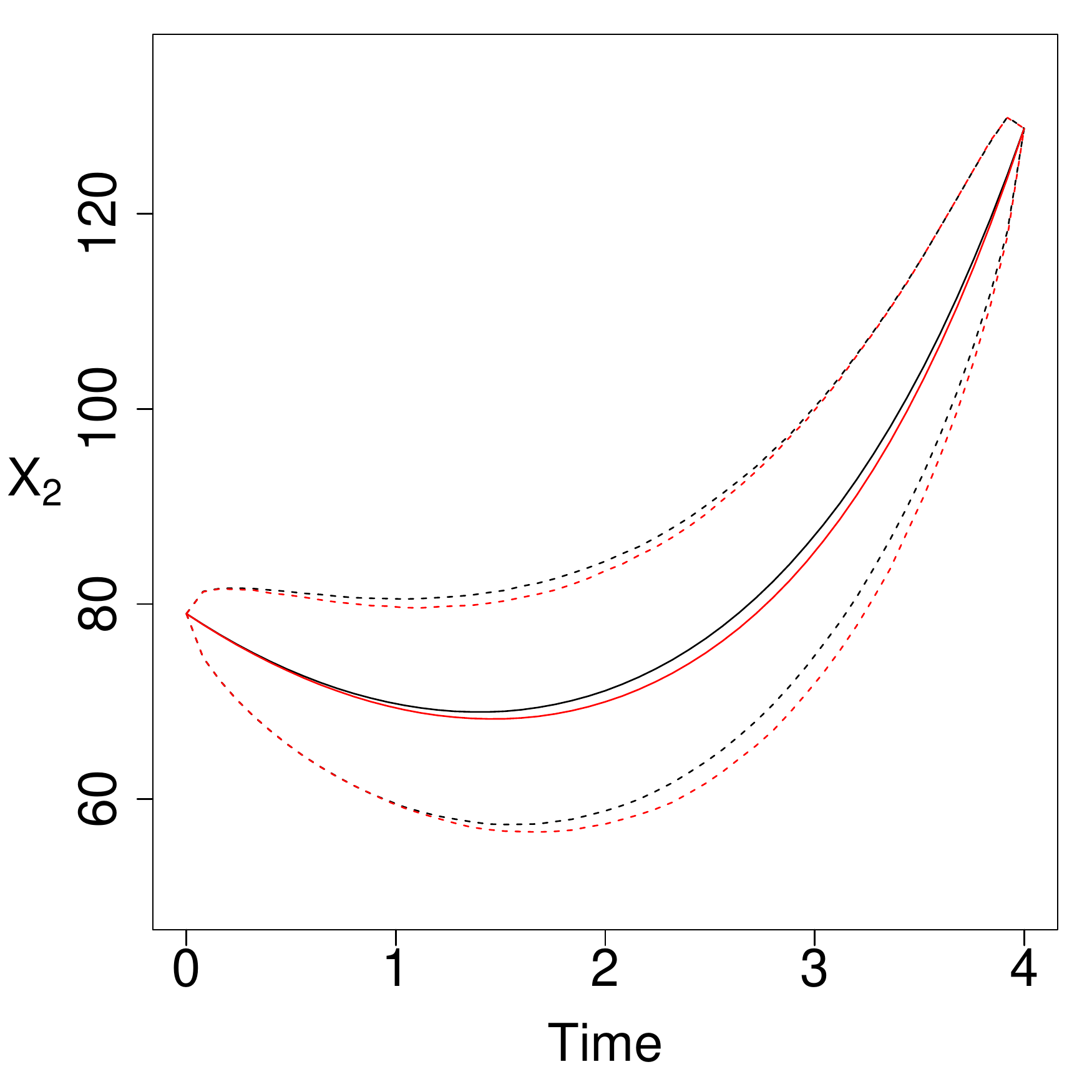}
\end{minipage}
\begin{minipage}[b]{0.32\linewidth}
        \centering
				\caption*{\qquad $\gamma=0.1$}\vspace{-0.25cm}
        \includegraphics[scale=0.285]{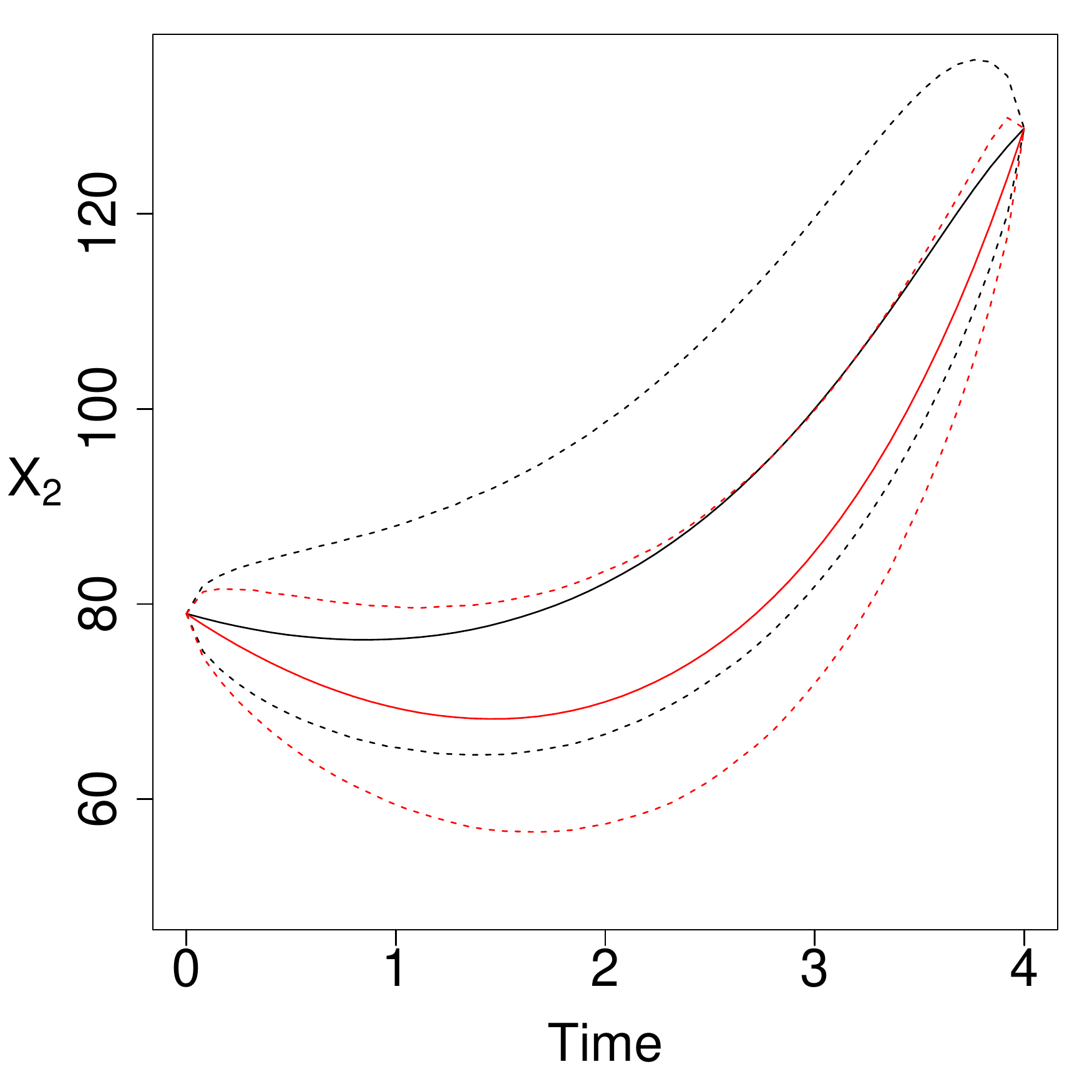}
\end{minipage}
\caption{Lotka-Volterra model. 95\% credible region (dashed line) and mean (solid line) of the true conditioned 
predator component $X_{2,t}|x_0,x_T$ (red) and various bridge constructs (black) using $x_T=x_{T,(95)}$ with 
$T=1$ (1$^{\textrm{st}}$ row) and $T=4$ (2$^{\textrm{nd}}$ row).} \label{lv fig_prop} 
\end{center}
\end{figure}

We fixed the discretisation by taking $m=50$, but note 
no appreciable difference in results for finer discretisations 
(e.g.~$m=1000$). As in the 
previous example, GP-N and \hbox{GP-S} perform relatively 
poorly, therefore in what follows we omit these bridges from 
the results. Note that we include MDB for reference. 
Figure~\ref{lv fig_ap} shows empirical acceptance 
probabilities against $T$ for each bridge and each $x_T$. 
Figure~\ref{lv fig_prop} compares 95\% credible regions of the proposal 
under various bridging strategies with the true 
conditioned process (obtained from the output of the 
Metropolis-Hastings independence sampler). 

Unsurprisingly, as $T$ is increased, MDB fails to 
adequately account for the nonlinear behaviour of the 
conditioned process. LB offers a modest improvement 
(except when $x_T=x_{T,(5)}$) but is generally 
outperformed by the other bridge constructs. We found 
that as $T$ was increased, LB required larger values 
of $\gamma$, reflecting the need for more weight to be placed 
on the myopic component of the construct. As for the previous 
example, unless $x_T$ is the median of $X_T|x_0$, RB is 
comprehensively outperformed 
by RB$^-$ (see Figure~\ref{lv fig_prop} for the effect 
of increasing $T$ on RB and RB$^-$). 
However, we see that the acceptance probabilities 
are decreasing in $T$ for both constructs. As noted by 
\cite{Fearnhead_2014}, the LNA can become poor as $T$ increases, 
with the implication here being that the approximation of the 
expected residual (as used in RB$^-$) degrades with $T$. 

We note that the estimated acceptance probabilities are roughly 
constant for GP and (to a lesser extent) GP-MDB, and in terms of statistical 
efficiency for a fixed number of iterations, GP-MDB 
should be preferred over all other algorithms considered in this article. 
However, the difference in estimated acceptance probabilities 
between GP-MDB and RB$^-$ is fairly small, even when $T=4$ 
(e.g. 0.857 vs 0.577 when $x_T=x_{T,(5)}$ and 0.834 vs 0.606 
when $x_T=x_{T,(50)}$). We also note that a Metropolis-Hastings scheme that 
uses RB or RB$^-$ is some 30 times faster than a scheme with GP or GP-MDB, 
since the latter require solving the LNA ODE system for each sub-interval 
$[\tau_{k},T]$ to maintain reasonable statistical efficiency 
for a given $m$. Therefore, we further compare RB, RB$^-$, GP and GP-MDB by 
computing the minimum effective sample size (ESS) at time $T/2$ 
(where the minimum is over each component of $X_{T/2}$) divided by 
CPU cost (in seconds). We denote this measure of overall efficiency by ESS/s.   
When $x_T=x_{T,(5)}$ and $T=1$, ESS/s scales roughly as $1:3:56:83$ for 
GP : GP-MDB : RB : RB$^-$. When $T=4$, ESS/s scales roughly as $1:3:1:17$. Hence, 
for this example, RB$^-$ is to be preferred in terms of overall efficiency, 
although the relative difference between RB$^-$ and GP-MDB appears 
to decrease as $T$ is increased, consistent with the behaviour of the empirical 
acceptance rates observed in Figure~\ref{lv fig_ap}.

\subsection{Aphid growth}\label{aphid}

\cite{Matis_2008} describe a stochastic model for aphid 
dynamics in terms of population size ($N_t$) and
cumulative population size ($C_t$). The diffusion 
approximation of their model is given by
\begin{equation}\label{eqn:aphidSDE}
\begin{pmatrix}
dN_t\\
dC_t\\ \end{pmatrix}= \begin{pmatrix}
\theta_1 N_t-\theta_2 N_tC_t \\
\theta_1 N_t \\ \end{pmatrix}dt 
+ \begin{pmatrix}
\theta_1 N_t+\theta_2 N_tC_t & ~\theta_1 N_t  \\
\theta_1 N_t & ~\theta_1 N_t \\ \end{pmatrix}^\half dW_t 
\end{equation}
where the components of $\theta=(\theta_1,\theta_2)'$ characterise 
the birth and death rate respectively. \cite{Matis_2008} 
also provide a dataset consisting of cotton aphid counts 
recorded at times \hbox{$t=0, 1.14, 2.29, 3.57$} and $4.57$ weeks, 
and collected for 27 different treatment block combinations. 
The analysis of these data via a stochastic differential 
mixed-effects model driven by (\ref{eqn:aphidSDE}) is the 
focus of \cite{Whitaker_2015}.  
 
Driven by the real data of \cite{Matis_2008} and 
to illustrate the proposed methodology in a challenging 
partial observation scenario, we assume that $X_T$ cannot 
be measured exactly. Rather, we observe  
\[
Y_T = F' X_T + \epsilon_T, \qquad \epsilon_T|\Sigma\sim N(0,\Sigma),    
\]
where $\Sigma=\sigma^2$ and $F=(1,0)'$ so that only noisy observation 
of $N_T$ is possible, and $C_T$ is not observed at all. We consider 
a single treatment-block combination and consider the dynamics of the 
process over an observation time interval $[2.29,3.57]$, over which nonlinear 
dynamics are typically observed. We fix $\theta$ and $x_{2.29}$ at their marginal 
posterior means found by \cite{Whitaker_2015}, that is, at $\theta=(1.45,0.0009)'$ 
and $x_{2.29}=(347.55,398.94)'$. We generate various end-point conditioned 
scenarios by taking $y_{3.57}$ to be either the 5\%, 50\% or 95\% 
quantile of \hbox{$Y_{3.57}|X_{2.29}=(347.55,398.94)',\sigma$.} To investigate the 
effect of measurement error, we further take $\sigma\in\{5,10,50\}$. The 
resulting quantiles are shown in Table~\ref{tab aphiddata}. As 
with the previous example, the ODE system 
governing the linear noise approximation of 
(\ref{eqn:aphidSDE}) is intractable and we again 
use the \verb+lsoda+ package to numerically 
solve the system when necessary. 
%Again we implement the LNA through \eqref{eqn:dgode_eta}, (\ref{eqn:lna1}) and (\ref{eqn:lna2}) with
%\[
%H_t= \begin{pmatrix}
%\theta_1 - \theta_2\eta_{C,t}  &  -\theta_2\eta_{N,t} \\
%\theta_1  &  0 \end{pmatrix}.
%\]
%As above in Section~\ref{lv} the ODE system is intractable and as such we 
%numerically solve the system. 

\begin{table}
\begin{center}
\begin{tabular}{c|ccc}
	\hline
    & $\sigma=5$ & $\sigma=10$ & $\sigma=50$ \\
  \hline
  $y_{3.57,(5)}$  & 726.75 & 724.57 & 762.36 \\
  $y_{3.57,(50)}$  & 786.09 & 815.51 & 774.41 \\ 
	$y_{3.57,(95)}$  & 841.82 & 856.36 & 910.86 \\
	\hline
\end{tabular}
\caption{Aphid growth model. Quantiles of $Y_{3.57}|X_{2.29}=(347.55,398.94)'$ found by repeatedly simulating from the Euler-Maruyama 
approximation of (\ref{eqn:aphidSDE}) with $\theta=(1.45,0.0009)'$, and corrupting $N_{3.57}$ with additive $N(0,\sigma^2)$ noise.} \label{tab aphiddata}
\end{center}
\end{table} 

\begin{figure}
\begin{center}
\begin{minipage}[b]{0.32\linewidth}
        \centering
				\caption*{\qquad $y_{3.57}=y_{3.57,(5)}$}\vspace{-0.25cm}
        \includegraphics[scale=0.285]{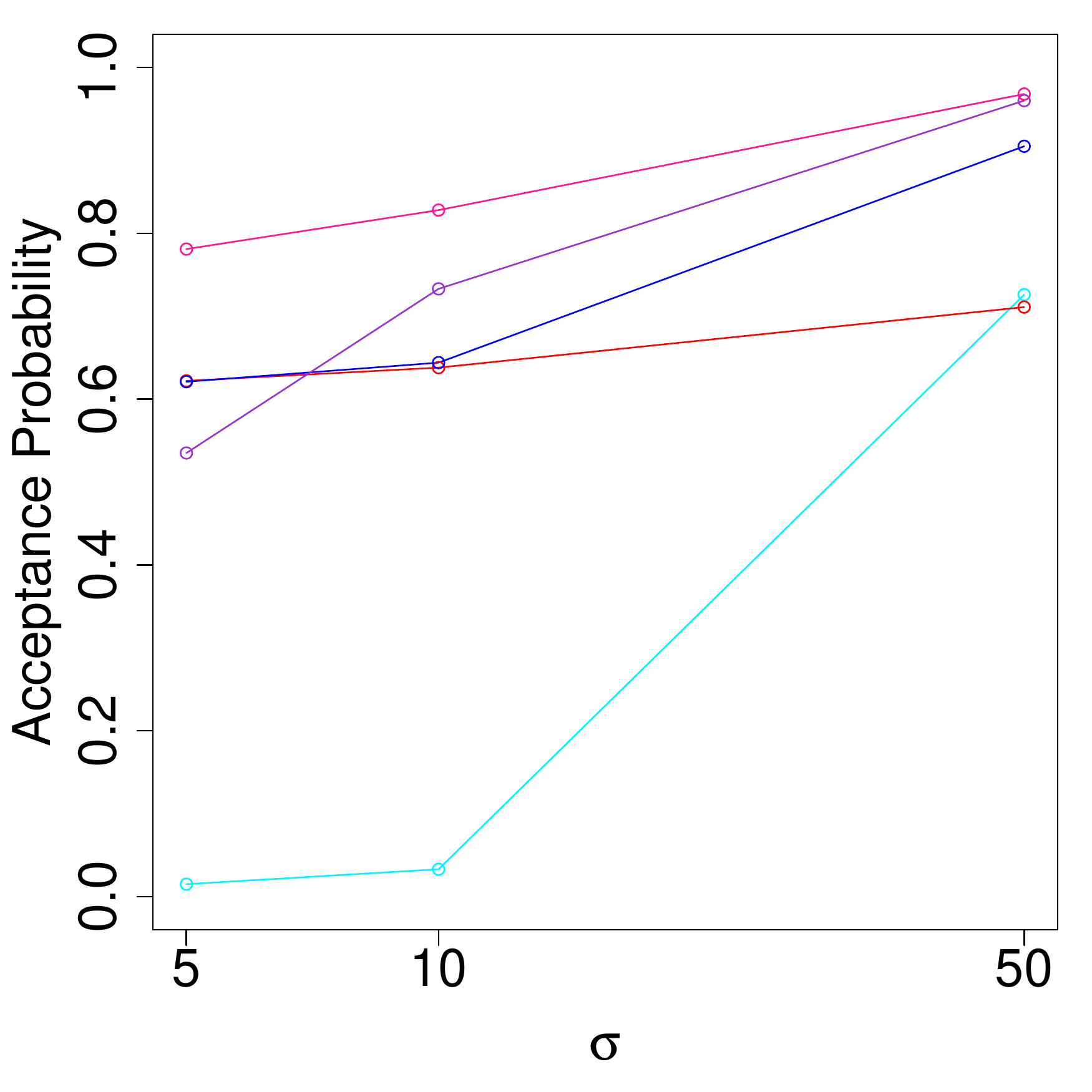}
\end{minipage} 
\begin{minipage}[b]{0.32\linewidth}
        \centering
				\caption*{\qquad $y_{3.57}=y_{3.57,(50)}$}\vspace{-0.25cm}
        \includegraphics[scale=0.285]{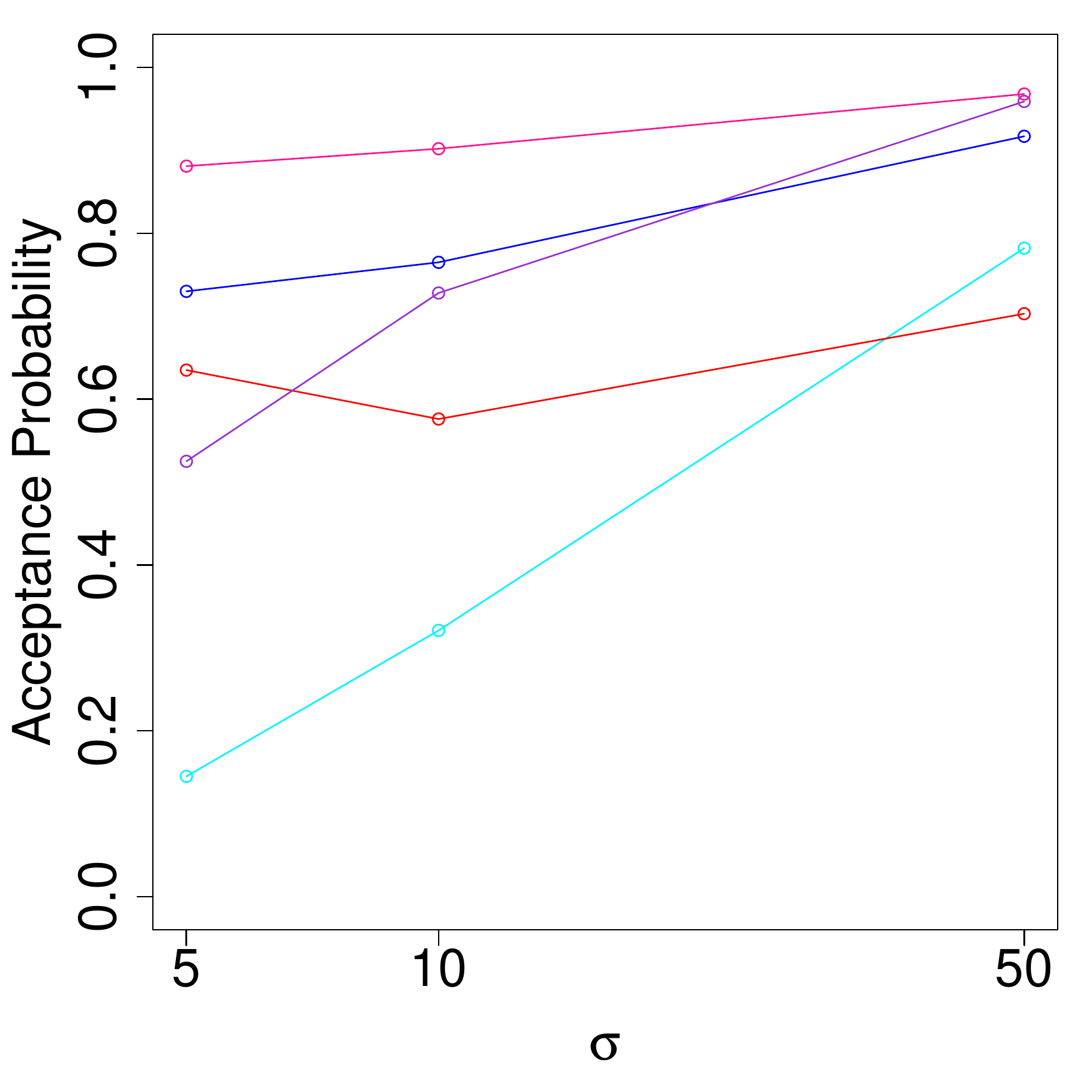}
\end{minipage} 
\begin{minipage}[b]{0.32\linewidth}
        \centering
				\caption*{\qquad $y_{3.57}=y_{3.57,(95)}$}\vspace{-0.25cm}
        \includegraphics[scale=0.285]{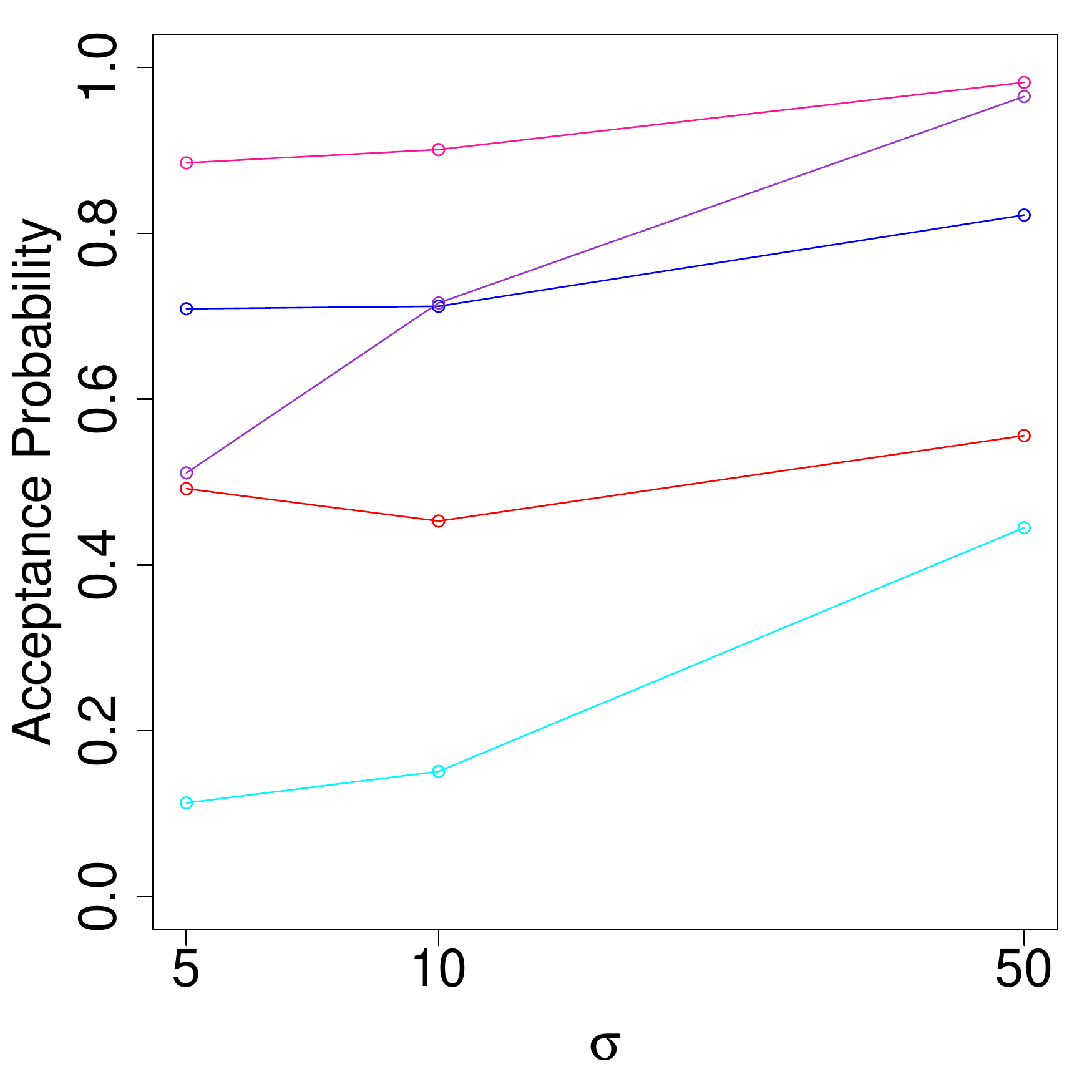}
\end{minipage} 
\caption{Aphid growth model. Empirical acceptance probabilities against $\sigma$. The results are based on 100K iterations 
of a Metropolis-Hastings independence sampler. Turquoise: EM. Red:~RB. Blue:~RB$^-$. Purple: GP. Pink: GP-MDB.} \label{aphid fig_ap}
\end{center}
\end{figure}

\begin{figure}
\begin{center}
\begin{minipage}[b]{0.32\linewidth}
        \centering
				\caption*{\qquad EM}\vspace{-0.25cm}
        \includegraphics[scale=0.285]{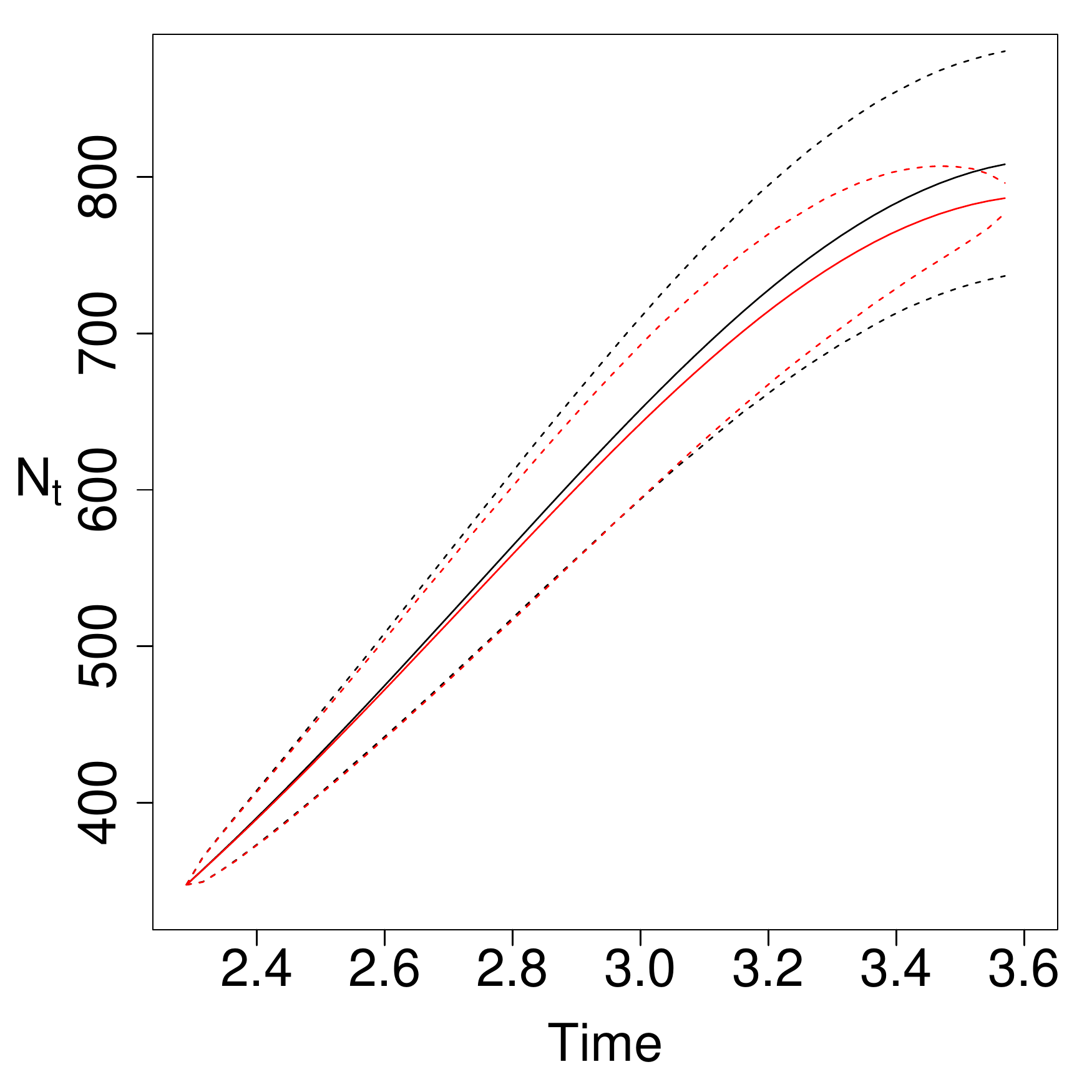}
\end{minipage} 
\begin{minipage}[b]{0.32\linewidth}
        \centering
				\caption*{\qquad GP-MDB}\vspace{-0.25cm}
        \includegraphics[scale=0.285]{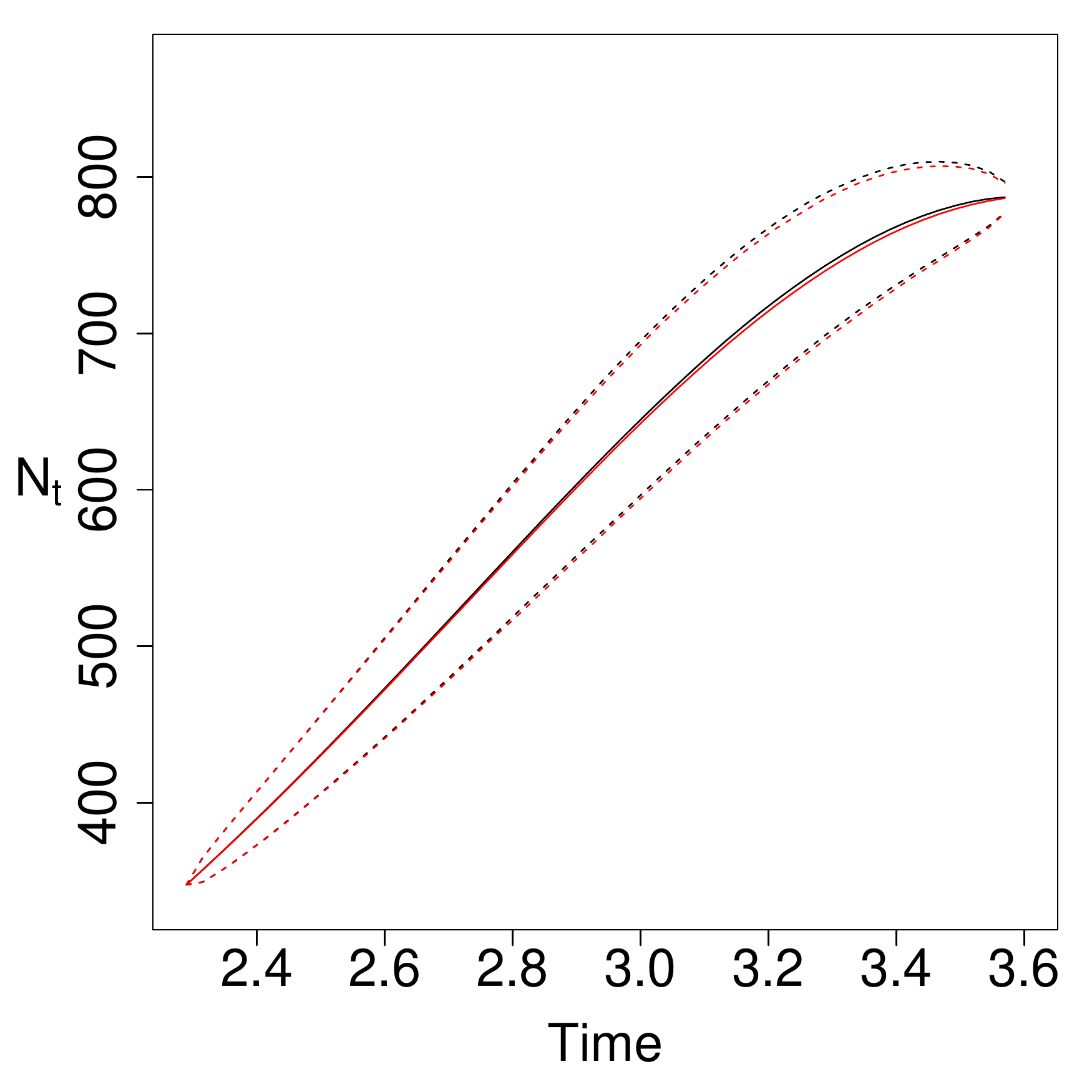}
\end{minipage} 
\begin{minipage}[b]{0.32\linewidth}
				\centering
				\caption*{\qquad RB$^-$}\vspace{-0.25cm}
        \includegraphics[scale=0.285]{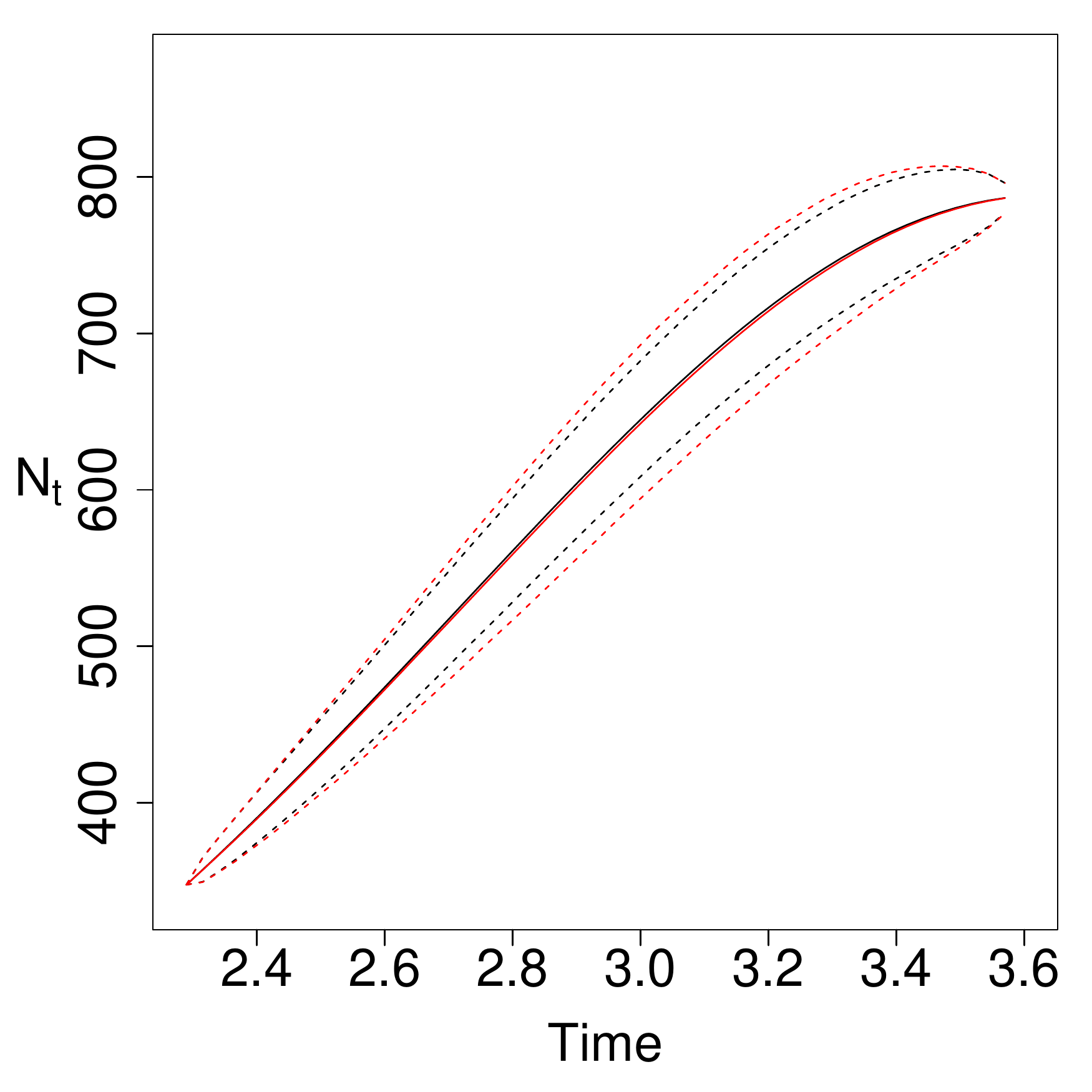}
\end{minipage}\\
\vspace{0.2cm}
\begin{minipage}[b]{0.32\linewidth}
				\centering
        \includegraphics[scale=0.285]{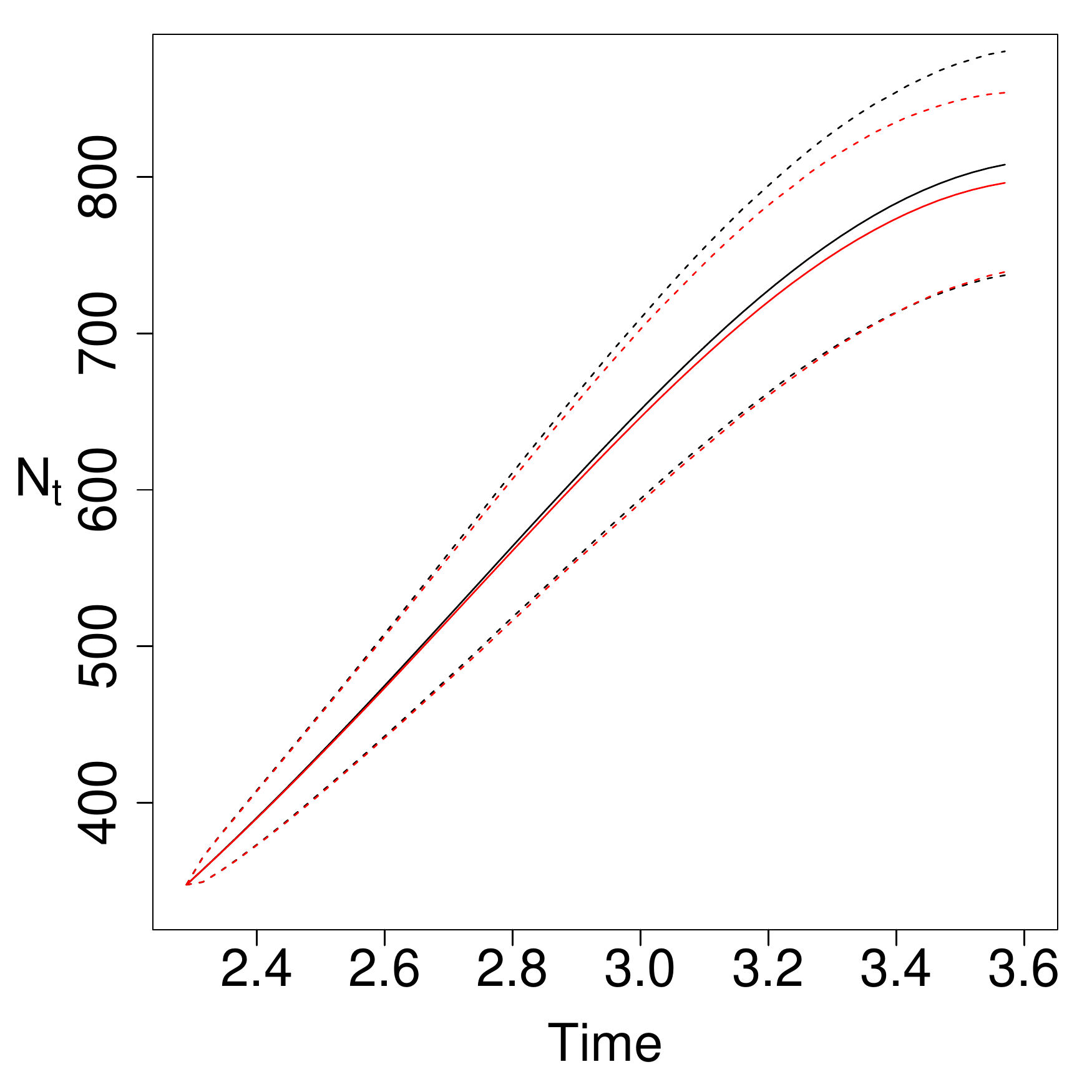}
\end{minipage}
\begin{minipage}[b]{0.32\linewidth}
        \centering
        \includegraphics[scale=0.285]{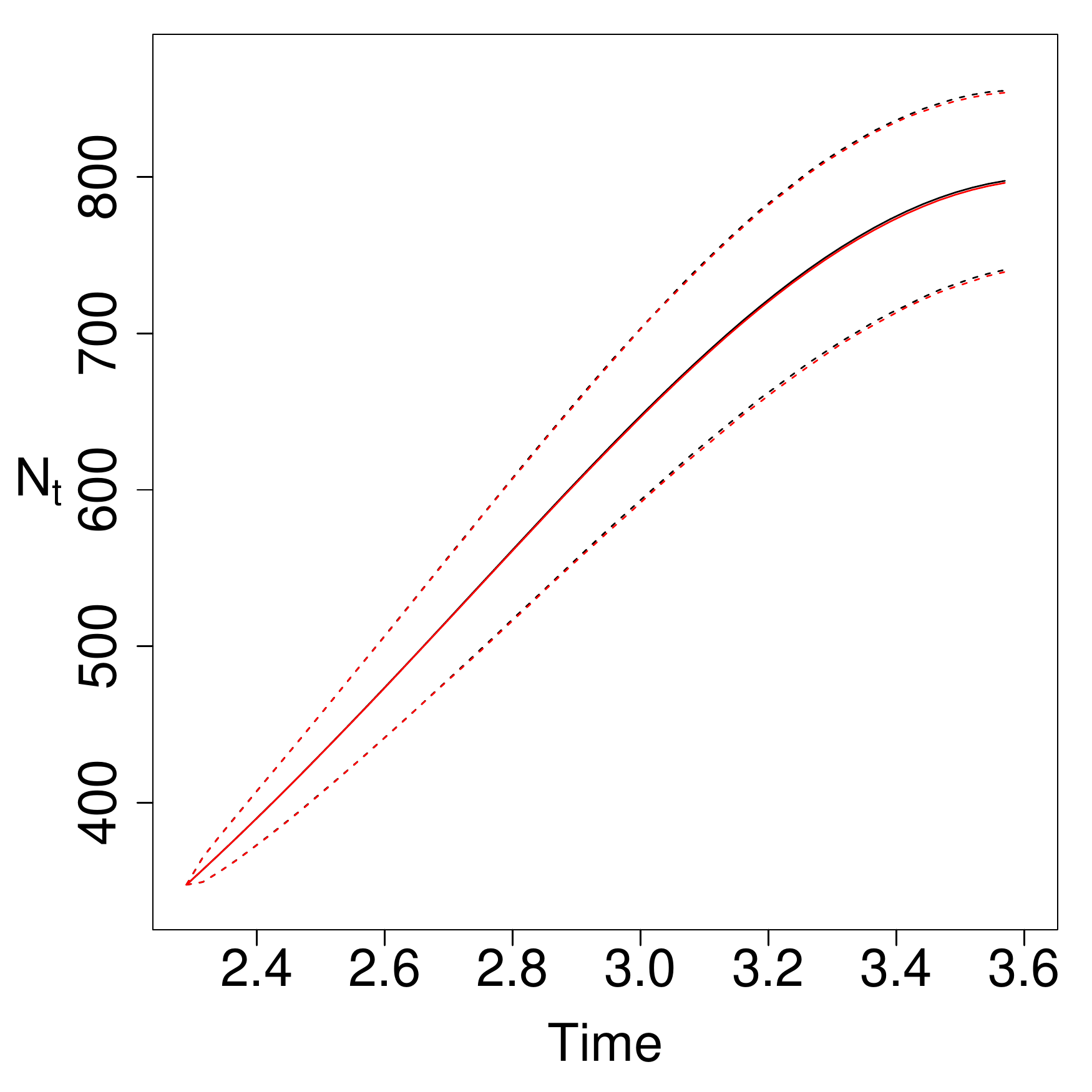}
\end{minipage}
\begin{minipage}[b]{0.32\linewidth}
        \centering
        \includegraphics[scale=0.285]{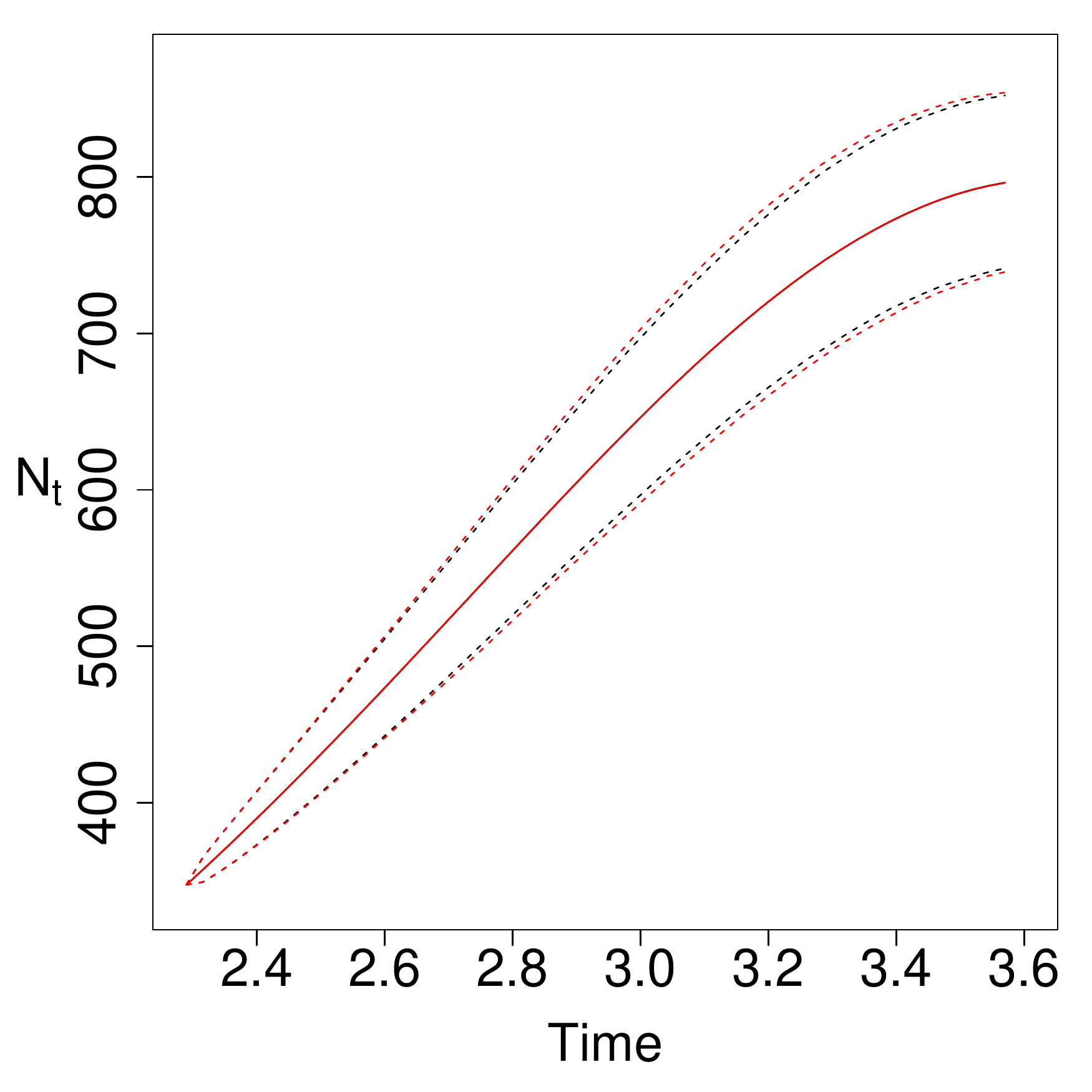}
\end{minipage}
\caption{Aphid growth model. 95\% credible region (dashed line) and mean (solid line) of the true conditioned 
aphid population component $N_t|x_{2.29},y_{3.57}$ (red) and various bridge constructs (black) using $y_{3.57}=y_{3.57,(50)}$ with 
$\sigma=5$ (1$^{\textrm{st}}$ row) and $\sigma=50$ (2$^{\textrm{nd}}$ row).} \label{aphid fig_n_prop}
\end{center}
\end{figure}

Figure~\ref{aphid fig_ap} shows empirical acceptance 
probabilities against $\sigma$ for EM, RB, RB$^-$, GP and GP-MDB. 
Figure~\ref{aphid fig_n_prop} compares 95\% credible regions 
for a selection of bridges with the true 
conditioned process (obtained from the output of the independence sampler). 
All results are based on $m=50$ (but note that no discernible 
difference in output was obtained for finer discretisations). As illustrated 
by both figures, the myopic sampler (EM) performs poorly (in terms of statistical 
efficiency, as measured by empirical acceptance probability) 
when the measurement error variance is relatively small 
($\sigma=5$). For $\sigma=50$, the performance of EM 
is comparable with the other bridge constructs. In fact, as 
$\sigma$ increases, the bridge constructs coincide with the 
Euler-Maruyama approximation of the target process. The gain in 
statistical performance of RB$^-$ over RB is clear. Likewise, 
GP-MDB outperforms GP, although the difference is very small 
for $\sigma=~50$ and again we note that as $\sigma$ increases, 
the variance under GP-MDB, $\Psi_{\MDB}(x_{\tau_{k}})$, 
approaches the Euler-Maruyama variance, as used in GP.   
  
The relative computational cost of each scheme can be found in 
Table~\ref{tab times}. EM is particularly cheap to implement, 
given the simple form of the construct and the M-H acceptance 
probability. However, this approach cannot be recommended 
in this example for $\sigma < 10$, due to its dire 
statistical efficiency. The computational cost of RB, RB$^-$, 
GP and GP-M is roughly the same, since for the guided proposals, 
we found that a naive implementation that only solves the LNA ODEs 
once, gave no appreciable difference in empirical acceptance probability 
as obtained when repeatedly solving the ODE system for each sub-interval 
$[\tau_{k},T]$ (as is required in the case of no measurement error). 
Consequently, in this example, GP-MDB outperforms RB$^-$ in terms of overall efficiency.

\begin{comment}
Although, as with the Lotka-Volterra model, 
in terms of statistical efficiency for a fixed number of 
iterations, RB$^-$ is outperformed by GP-MDB, the difference 
is generally fairly small (with acceptance rates for RB- never lower 
than 70\% of those of GP-MDB) and decreases as $\sigma$ decreases. Moreover, as with the 
Lotka-Volterra model, each iteration of RB$^-$ is much quicker than 
an iteration of GP-MDB, here taking approximately 
4\% of the CPU time.
\end{comment}

\section{Discussion}\label{disc}

We have presented a novel class of bridge constructs that are 
both computationally and statistically efficient, and can 
be readily applied in situations where only noisy and partial 
observation of the process is possible. Our approach is 
straightforward to implement and is based on a partition 
of the process into a deterministic part that accounts 
for forward dynamics, and a residual stochastic process. The 
intractable end-point conditioned residual SDE is approximated 
using the modified diffusion bridge of \cite{Durham_2001}. Using 
three examples, we have investigated the empirical performance 
of two variants of the residual bridge. The first constructs 
the residual SDE by subtraction of a deterministic process 
based on the drift governing the target process (denoted RB). The second 
variant further subtracts the linear noise approximation (LNA) 
of the expected conditioned residual process (denoted RB$^-$). 
Our examples included a scenario in which the LNA system is tractable, 
and another where the system must be solved numerically. An example 
that considers partial and noisy observation of the process at a future 
time was also presented.  

\subsection{Choice of residual bridge}

We find that for all examples considered, the residual bridge 
that further subtracts the LNA mean 
results in improved statistical efficiency (over the simple implementation 
based on the drift subtraction only) at 
the expense of having to solve a larger ODE system consisting 
of order $d^2$ equations (as opposed to just $d$ when using 
the simpler variant). For a known initial time-point 
$x_0$, the ODE system need only be solved once, irrespective of the 
number of skeleton bridges required. Taking the Lotka-Volterra diffusion (described 
in Section~\ref{lv}) as an example, overall 
efficiency (as measured by minimum effective sample size per second, ESS/s, at time 
$T/2$) of RB$^-$ is 1.5 times that of RB when $T=1$ and $x_T$ is either the 
$5\%$ or $95\%$ quantile of $X_T|x_0$. This factor increases to 
17 when $T=4$. However, for unknown 
$x_0$, as would typically be the case when performing parameter inference, 
the ODE solution will be required for each skeleton 
bridge, and the difference in computational cost between the 
two approaches is likely to be important, especially as the dimension of the 
state space increases. For the Lotka-Volterra example, the computational 
cost for solving the ODE system \emph{for each bridge} scales as $1:2.8$ for 
RB : RB$^-$. Therefore, the relative difference in ESS/s would reduce 
to a factor of roughly 0.5 when $T=1$ (so that RB would be preferred) 
and 6 when $T=4$. We therefore anticipate that in problems where $x_0$ 
is unknown, the simple residual bridge is to be preferred, unless the ODE 
system governing the LNA is tractable, or the dimension $d$ of $X_t$ is 
relatively small, say $d<5$.

\subsection{Residual bridge or guided proposal?} 

We have compared the performance of our approach to several 
existing bridge constructs (adapting where necessary to the 
case of noisy and partial observation). These include the 
modified diffusion bridge \citep{Durham_2001}, Lindstr\"{o}m bridge \citep{Lindstrom_2012} and 
guided proposal \citep{Schauer_2014}. Our implementation of the latter uses 
the LNA to guide the proposal. We find that a further modification that replaces the 
Euler-Maruyama variance with the MDB variance gives 
a particularly effective bridge, outperforming 
all others considered here, in terms of statistical efficiency. 
We find that for fixed $x_0$ and noisy observation of $x_T$, 
an efficient implementation of the guided proposal is possible, 
where the ODE system governing the LNA need only be solved once. 
In this case, the guided proposal outperforms both implementations 
of the residual bridge in terms of overall efficiency. However, we 
found that in the case of no measurement error (so that $x_T$ 
is known exactly), the guided proposal required that the ODEs governing 
the LNA be re-integrated for each intermediate time-point and 
for each skeleton bridge required. Unless the ODE system can be solved 
analytically, we find that when combining statistical and computational efficiency, 
the guided proposal is outperformed by both implementations of the residual bridge. 

\subsection{Extensions}

Our work can be extended in a number of ways. For example, it may 
be possible to improve the statistical performance of the residual 
bridges by replacing the Euler-Maruyama approximation of the 
variance of $Y_T|X_0$ with that obtained under the LNA. This approach 
could also be combined with the Lindstr\"{o}m sampler to avoid 
specification of a tuning parameter. Deriving the limiting (as $\Delta\tau\to 0$) 
forms of the Metropolis-Hastings acceptance rates associated with the 
residual bridges would be problematic due to the time dependent terms 
entering the variance of the constructs. Nevertheless, this merits 
further research. Interest also lies in the comparison of the bridge 
constructs for SDEs that exhibit multimodal behaviour, although we anticipate that 
further modification of the constructs will be required to efficiently deal with 
such a scenario.
\\

\noindent\textbf{Acknowledgements} The authors would like to thank the
associate editor and three anonymous referees for their suggestions for
improving this paper.

\bibliographystyle{apalike}
\bibliography{paper2}

\end{document}